\documentclass[amsmath,amssymb,10pt,aps,twocolumn,superscriptaddress]{revtex4-2} 
\usepackage[colorlinks=true]{hyperref}
\usepackage{comment}
\usepackage{mathrsfs}
 \usepackage{lipsum}
 \usepackage{graphicx}
 \usepackage{textcomp}
\usepackage[dvipsnames]{xcolor}
\usepackage{latexsym}
\usepackage{amsmath}
\usepackage{amsthm}
\usepackage{amssymb}
\usepackage{epstopdf}
\usepackage{enumitem}
\usepackage{setspace} % controllable line spacing
\usepackage{dcolumn} % Align table columns on decimal point
\usepackage{bm} % bold math
\usepackage{setspace} % controllable line spacing
\usepackage{slashed}
\usepackage{color}
\usepackage{xcolor}
\usepackage{youngtab}
\usepackage{tikz}
\usepackage{tikz-3dplot}
\usepackage{braket}
\usepackage{mhchem}
\usepackage{bm}
\usepackage{dsfont}
\usepackage{physics}
\usepackage{mathrsfs}

\usetikzlibrary{shapes,snakes,arrows,chains,matrix,positioning,scopes,calc}
\usetikzlibrary{decorations.markings}

\usepackage[tabbotcap]{subfigure}

\allowdisplaybreaks

\usepackage{lipsum}

\newlength{\diamondrulelength}
\newlength{\diamondrulethickness}
\begin{document}
\title{Chiral superconductors from parent states with nonuniform Berry curvature: Momentum-space vortices, Bogoliubov–de Gennes topology, and thermal Hall conductivity}
\author{L. David Le Nir \footnotemark[1]} 
\thanks{These authors contributed equally.}
\affiliation{Department of Physics, University of Toronto}
\author{Asimpunya Mitra \footnotemark[1]} 
\thanks{These authors contributed equally.}
\affiliation{Department of Physics, University of Toronto}
\author{Yong Baek Kim} 
\affiliation{Department of Physics, University of Toronto, Toronto, Ontario M5S 1A7, Canada}
\affiliation{Canadian Institute for Advanced Research, Toronto, Ontario M5G 1M1, Canada}

\date{\today}   

\begin{abstract}
We investigate chiral superconductivity emerging from parent electronic states with non-uniform Berry curvature, motivated by recent experiments in rhombohedral graphene multilayers. Using the continuum $\lambda_N$-model—a tunable platform with independently controllable Berry curvature profiles—we solve the full BCS gap equation on a continuum Chern band beyond the weak-coupling limit. We find that a non-uniform Berry curvature of the parent band enriches the superconducting order parameter, leading to the formation of momentum-space vortices in the gap function away from high-symmetry points. By tuning the Berry curvature profile, we identify distinct regimes associated with vortex nucleation and vortex number saturation, and show that the nucleation of momentum-space vortices tends to lower the condensation energy. We then show analytically that the parent band Chern number constrains the number of momentum-space vortices that can nucleate in the gap—independent of details of the $\lambda_N$-model. We also provide a gauge-invariant formulation for computing the Bogoliubov-de Gennes (BdG) Berry curvature for continuum models, and find that it is determined by a momentum-space phase current. The winding of this current around vortices in the occupied region in turn determines the BdG Chern number. Finally, we discuss how thermal Hall measurements can be used to probe the formation of momentum-space vortices. Our results highlight the crucial role of Berry curvature in shaping chiral superconductivity, and offer guiding principles for its identification in systems such as rhombohedral graphene.

\end{abstract}

\maketitle

\section{Introduction}

Chiral superconducting states are characterized by spontaneously broken time-reversal symmetry and a finite orbital angular momentum carried by Cooper pairs \cite{Kallin2016}. Such states naturally arise in systems where crystal symmetries permit degenerate multi-component gap functions \cite{Mineev1999}. In this situation, any linear combination of the degenerate components is, in principle, allowed. However, the system selects the configuration that minimizes its condensation energy—defined as the free-energy difference between the superconducting and normal states. In the weak-coupling limit, this corresponds to maximizing the Fermi-surface average of $|\Delta(\mathbf{k})|^2$ (here $\Delta(\mathbf{k})$ is the superconducting gap), which favors fully gapped states. As a result, chiral combinations such as $p_x \pm i p_y$ are energetically preferred, leading to spontaneous time-reversal symmetry breaking.

Chiral superconductivity has been proposed in a variety of systems, including the kagome metals \ce{AV3Sb5} \cite{DiSante2026,Wilson2024,Wu_2021,schultz2025,mitra2025}, \ce{MoTe2} \cite{xu2025MoTe2,Xu_MoTe2_2025,Chen2026}, and \ce{UTe2} \cite{Aoki2019,Jiao2020,Aoki2022}, although its existence in these materials remains under active debate \cite{Theuss2024}. To date, the most established realization is the A-phase of superfluid $^3$He, which exhibits chiral $p+ip$ pairing \cite{Volovik2003,Vollhardt2013,Ikegami2013,Ikegami2015,Kasai_2018}.
Recent experiments on rhombohedral graphene multilayers have reinvigorated interest in this topic. Superconductivity has been observed across a wide variety of multilayered graphene platforms \cite{Zhou2021half,Choi2025superconductivity,Li2024tunable,Park2021tunable,Hao2021electric,Zhang2023enhanced,Zhou2022isospin,Oh2021evidence,Cao2021pauli,Yankowitz2019tuning,Cao2018unconventional,kumar2025}, motivating extensive theoretical efforts to understand the pairing mechanism and order parameter symmetry in these systems \cite{xu2018topological,Isobe2018,you2019superconductivity,lee2019theory,lian2019twisted,Yuan2019,wu2018theory,Ghazaryan2021,khalaf2021charged,chou2021correlation,chou2021acoustic,dong2024superconductivity,szabo2022metals,chou2022acousticbernal,chou2022acoustic,Cea2022,christos2022correlated,Lu2022,Ghazaryan2023,dong2023superconductivity,JimenoPozo2023,Li2023,chou2024topological,Qin2025,ji2025phonon}. Most recently, signatures consistent with chiral superconductivity have been reported in non-moiré tetralayer rhombohedral graphene \cite{Han2025}. Under electron doping, the system exhibits a large spontaneous anomalous Hall effect in the normal state and magnetic hysteresis in the resistance below $T_c$, suggesting the emergence of superconductivity from a spin-valley polarized “quarter-metal” phase near a single $K$ valley. Complementary measurements in rhombohedral hexalayer graphene report anisotropic transport within the superconducting state \cite{qin2026,nguyen2025}. Together, these observations motivate a closer examination of mechanisms that can stabilize chiral superconductivity.

In rhombohedral graphene multilayers, the relevant electronic states are characterized by non-uniform Berry curvature. Hence, the nature of chiral superconductors arising from such parent states and its key experimental signatures are important issues. In conventional type-II superconductors, an external magnetic field in the range $H_{c1} < H < H_{c2}$ induces vortices in the real-space order parameter \cite{Tinkham2004}. These vortices, arranged in an Abrikosov lattice, carry quantized flux $\Phi_0 = h/2e$ and are associated with a $2\pi$ phase winding of the gap. An intriguing analogy arises in momentum space: Berry curvature acts as an effective magnetic field, coupling to the orbital motion of electrons and generating orbital magnetization in bands with nonzero Chern number. This raises a natural question: can Berry curvature similarly seed vortex-like structures in the superconducting gap, but in momentum space? Understanding how such momentum space vortices emerge and what the corresponding experimental signatures are would serve as useful guiding principles for the identification  of chiral superconductors in these systems.

In this work, we investigate how momentum-space vortices emerge in chiral superconducting states descending from parent states with non-uniform Berry curvature, and how these vortices determine the topology and thermal Hall response of the resulting BdG quasiparticles. In a conventional weak-coupling chiral $p$-wave superconductor, the gap function takes the form $\Delta(\mathbf{k}) \sim k_x + i k_y = |\mathbf{k}| e^{i \phi_{\mathbf{k}}}$, corresponding to a momentum-space vortex of winding $\ell = 1$ at the origin. Higher-angular-momentum states exhibit analogous structures; for instance, chiral $f+if$ pairing yields $\Delta(\mathbf{k}) \sim |\mathbf{k}|^3 e^{i 3\phi_{\mathbf{k}}}$ with $\ell = 3$. In the presence of nontrivial Berry curvature, however, we show that it becomes possible for vortices to nucleate away from the origin, reflecting the underlying topology of the normal-state band. Understanding the formation, stability, and topology of such momentum-space vortices is the central goal of this work.

To this end, we employ the $\lambda_N$-model introduced in Ref.~\cite{desrochers2025}, which provides a tunable platform for studying electronic states in bands with non-uniform Berry curvature. The model allows independent control over both the momentum-space distribution of Berry curvature and the total enclosed Berry flux. Originally motivated as a minimal description of spin-valley polarized $N$-layer rhombohedral graphene, it captures Berry curvature effects through band-projected form factors that are decoupled from the dispersion. We focus on superconductivity within the lowest band, which carries a nonzero continuum Chern number.
Previous studies of pairing in Chern bands have typically assumed a uniform Berry curvature, often modeled with the lowest-Landau-level form factors \cite{maymann2025,Wang2024,jahin2025,Guerci2025}. In contrast, the $\lambda_N$-model enables us to explore the impact of strongly nonuniform Berry curvature profiles, analogous to spatially varying magnetic fields in real space. Moreover, prior work has largely relied on weak-coupling approaches that restrict pairing to states near the Fermi surface \cite{jahin2025,maymann2025,Shavit2025,Geier2025,Parra2025,christos2025}. In such treatments, the topology of the superconducting state is determined solely by the winding of the gap at the origin. Here, we instead solve the full nonlinear BCS gap equation, allowing pairing across a large patch in momentum space. In this regime, electrons probe the full Berry curvature landscape, enabling the emergence of momentum-space vortices away from high-symmetry points.
Momentum-space vortices have recently been reported in realistic models of spin-valley polarized rhombohedral graphene \cite{patri2025}, where they were found to nucleate along rings of maximal Berry curvature. However, in those systems, Berry curvature is intrinsically tied to the band structure, making it difficult to isolate its role. The $\lambda_N$-model overcomes this limitation by allowing systematic tuning of the Berry curvature profile independently.

In the first part of the paper, we use this controlled setting to study the formation and energetics of momentum-space vortices. We find, by solving the BCS gap equation, that strong interactions promote momentum-space vortex formation by enabling pairing across a broader region of the band, while sharply peaked Berry curvature profiles enhance vortex nucleation despite locally suppressing the gap magnitude. By tracking the condensation energy and vortex positions, we identify distinct regimes of vortex nucleation and vortex-number saturation. We further extend the analysis to annular Fermi surface geometries—relevant for rhombohedral graphene at large displacement fields—where we show that vortices away from inversion-invariant momenta can qualitatively change the topology of the superconducting state.

The second part of the paper explains why these vortices control the topological response. A central difficulty in band-projected superconductivity is that the projected gap is not an ordinary scalar function. It is defined using a gauge-dependent Bloch state and therefore transforms under parent-band gauge transformations. Guided by gauge invariance, we formulate band-projected superconductivity in an orbital basis, where the order parameter is intrinsically gauge invariant. In this formulation, the momentum-space analogue of the superfluid velocity coupled to an external vector potential is the difference between two connections: the singular connection, constructed from the gap-phase gradient, and the pair Berry connection inherited from the parent band. This structure reveals that parent-band topology forces vorticity into the projected gap. The BdG Chern number, however, is not determined by the total vorticity alone. Instead, we show that the BdG Berry curvature is controlled by the coherence-factor-weighted phase velocity, which plays the role of a momentum-space analogue of the Meissner current. The superconducting coherence factor suppresses vortices outside the occupied region, so only the vortices in the occupied region of the Fermi sea contribute to the BdG Chern number. In this sense, the BdG Chern number is a momentum-space analogue of flux quantization in a type-II superconductor. This leads to the result
\begin{align}
    C_{\text{BdG}}=\mathcal{Q},
\end{align}
where $\mathcal{Q}$ is the signed sum of vortex windings enclosed within the occupied region. Under mild regularity assumptions, we show that $\mathcal{Q}$ provides the canonical occupied-vortex representative of the class-D $\mathbb{Z}$ invariant for a single-band projected superconductor. This directly fixes the thermal Hall response and provides a unique experimental signature for the presence of momentum-space vortices.

The paper is organized as follows. In Sec.~\ref{sec:recap}, we review the $\lambda_N$-model. In Sec.~\ref{sec:band_proj_SC}, we introduce the formalism for superconducting pairing. In Sec.~\ref{sec:results_gap_eqn}, we present solutions of the full gap equation and analyze the emergence of momentum-space vortices and their impact on the condensation energy. In Sec.~\ref{sec_BdG Topology}, we detail a mathematically rigorous way of computing the topology of the superconducting state, and discuss how it is tied to the parent band Chern number. We then discuss a thermal Hall measurement that can be used as a probe for the presence of these momentum space vortices. We conclude our discussion in Sec.~\ref{sec:discussions}.

\section{Recap of the $\lambda_N$  model}\label{sec:recap}
In this section, we review the essential features of the $\lambda_N$-jellium model, introduced in Ref. \cite{desrochers2025}. The $\lambda_N$-model features an isolated spinless band with tunable quantum geometry. This model is defined over the momentum continuum $\mathbf{k}\in \mathbb{R}^2$, and encloses a quantized Berry flux through the momentum plane. Since the lowest band of the model extends naturally to the one-point compactification of the momentum plane, it may be regarded as a Chern band defined over the compactified momentum-space sphere. The $\lambda_N$-model interpolates between two limiting cases, the $\lambda$-jellium model $(N = 2)$ \cite{Soejima2025} and the infinite Chern band (ICB) model $(N\rightarrow \infty)$ \cite{Tan2024}, both of which have been extensively studied in the context of electronic crystallization (Refs. \cite{Tan2024,TanDevakul2025WavefunctionAHC,desrochers2025}). The $\lambda_2$ model is the simplest model with non-uniform Berry curvature peaked near the origin. By contrast, the ICB model has uniform Berry curvature, making it the momentum-space analogue of an electron in a uniform magnetic field. The $\lambda_N$ model therefore provides a useful platform to study the effect of both uniform and nonuniform Berry curvature on superconducting pairing.

\subsection{The model}

The $\lambda_N$-jellium model is a non-interacting continuum model with $N$ internal degrees of freedom \cite{desrochers2025}. These internal degrees of freedom, which we refer to as orbitals, are labeled by $\ket{\alpha}$ and $\ket{\beta}$, with $\alpha,\beta=0,\dots,N-1$. Spin degrees of freedom are neglected. Physically, the $\lambda_N$-model was motivated by spin-valley-polarized N-layer rhombohedral graphene (RNG) with layer dependent couplings and Fermi velocities (see Ref.~\cite{desrochers2025,han2025_theory}). The point $\vb{k}=0$ of the model is identified with one of the $K$-valley points of RNG. For this reason, the momentum $\vb{k}$ is a relative momentum measured from one of the $K$-valleys of RNG. In what follows, we will generically treat the $\lambda_N$ model as a standalone model, referring to the valley point $K$ as the origin. The Hamiltonian takes the form
\begin{gather} 
    H^{\lambda_N}_{\alpha \beta}(\mathbf{k})=(\epsilon_{\vb{k}}-\mu)\delta_{\alpha \beta}+JD^{N}_{\alpha \beta}(\vb{k}),\label{eq:Hamiltonian_l_N}
\end{gather}
where $\epsilon_{\mathbf{k}}$ is the kinetic energy, $\mu$ is the chemical potential, and $J\geq0$ sets the separation between adjacent bands.  The operator $D^{N}(\vb{k})$, shown in Appendix~\ref{sec_app:LambdaNModelDetails}, endows the otherwise trivial electron gas with a tunable Berry curvature. For this reason, the corresponding single-particle eigenstates can be written as $e^{i\vb{k}\cdot \vb{r}}\ket{s_{n\vb{k}}}$. Here $\ket{s_{n\vb{k}}}$ plays the formal role of a cell-periodic Bloch state: it is the $\vb{k}$-dependent internal spinor from which the Berry connection and Berry curvature are computed. Unlike in a lattice model, however, $\ket{s_{n\vb{k}}}$ carries no real-space periodic dependence, and is instead an eigenstate of the operator $D^{N}(\vb{k})$. This state should be physically thought of as a local representation of the true Bloch state of RNG in a small region around the valley point. In the orbital basis, we may represent the Bloch state as the linear superposition 
\begin{align}
    \ket{s_{n\vb{k}}}=\sum_{\alpha=0}^{N-1}U_{\alpha n}(\vb{k})\ket{\alpha}.
\end{align}
The functions $U_{\alpha n}(\vb{k})$ are the orbital components of the Bloch states near the valley, and may be assembled into the unitary matrix
\begin{align}
    U(\vb{k})=\begin{pmatrix}
        \ket{s_{0\vb{k}}},\cdots , \ket{s_{N-1\vb{k}}}
    \end{pmatrix}.
\end{align}
As we shall see in Sec.~\ref{sec_BdG Topology}, the orbital degrees of freedom play an important role in band projected superconductivity, which is why we emphasize this structure here. We diagonalize Eq.~\eqref{eq:Hamiltonian_l_N} to obtain band energies of the form
\begin{align}
    E_{n\vb{k}}=\epsilon_{\vb{k}}-\mu+J\varepsilon_{n\vb{k}}.
\end{align}
For the lowest band $(n=0)$, it can be shown that  $D^{N}(\vb{k})$ has a zero mode, i.e. $\varepsilon_{0\vb{k}}=0$ (see Appendix~\ref{sec_app:LambdaNModelDetails}). Therefore, the dispersion of the lowest band $\ket{s_{0\mathbf{k}}}$ is simply $E_{0\vb{k}}=\epsilon_{\vb{k}}-\mu$, independent of the Bloch eigenstates and their associated Berry curvature. 

When studying superconductivity, we shall project all interactions onto this lowest active band. This is justified as we assume that $J$ is large so that all inter-band processes are negligible. For R4G it is known that the pairing gap $|\Delta|\sim 300 \text{ mK }\sim 0.025$ meV \cite{Han2025}, whereas the separation from the higher bands in R3G and R4G is $J\sim 10-300$ meV \cite{maymann2025,Zhang_2010}. Thus, $J\gg |\Delta|$ supporting the band-projected approximation.

\subsection{Quantum Geometry}\label{GeometryofLambda}
We now discuss the quantum geometry of the lowest band of the $\lambda_N$ model. The quantum geometry of an isolated band is characterized by its Berry curvature \cite{BerryOriginal}, defined as
\begin{align}
    \Omega(\vb{k})=i(\braket{\partial_{x}s_{0\vb{k}}}{\partial_{y}s_{0\vb{k}}}-\braket{\partial_{y}s_{0\vb{k}}}{\partial_{x}s_{0\vb{k}}}),
\end{align}
and its quantum metric tensor \cite{ProvostVallee1980}
\begin{align}
    g_{\mu\nu}(\vb{k})=\text{Re}\left[\braket{\partial_{\mu}s_{0\vb{k}}}{\partial_{\nu}s_{0\vb{k}}}-\braket{\partial_{\mu}s_{0\vb{k}}}{s_{0\vb{k}}}\braket{s_{0\vb{k}}}{\partial_{\nu}s_{0\vb{k}}}\right].
\end{align}
The quantum metric and Berry curvature obey a trace inequality $\text{tr}[g(\vb{k})] \geq \abs{\Omega(\vb{k})}$ \cite{Roy2014QuantumGeometry}. If the inequality is saturated, then the band is said to have ideal quantum geometry \cite{WangCano, Wang2023OriginFCI, Crepel, MeraOzawa2025GLL}. For bands with uniform ideal quantum geometry, the form factors $\mathcal{F}(\vb{k},\vb{k}')$, which are overlaps of the cell-periodic Bloch states, take the same form as in the lowest Landau level (LLL) \cite{WangCano}
\begin{gather}
    \mathcal{F}_{LLL}(\mathbf{k},\mathbf{k'})=\exp\left(-\frac{\mathcal{B}}{4}\left(\abs{\mathbf{k}-\mathbf{k'}}^2+2i \mathbf{k}\wedge \mathbf{k'} \right)\right)\label{eq:FF_infty}.
\end{gather}
Here we denote the scalar-triple product as $\mathbf{k}\wedge \mathbf{k'}=k_xk'_y-k_yk'_x$. Here $\sqrt{\mathcal{B}}$ is a model parameter which has units of length. For clarity, we emphasize that $\mathcal{B}$ is not an external magnetic field. Here, $\mathcal{B}$ controls the profile of the Berry curvature distribution as explained below.

\begin{figure}
    \centering
    \includegraphics[scale=1.0]{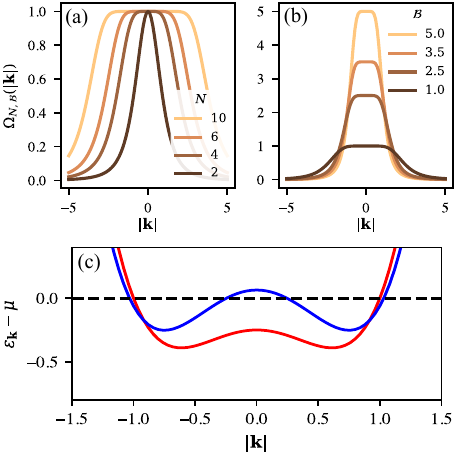}
    \caption{The Berry curvature distribution of the $\lambda_N$-model. $\Omega_{N,\mathcal{B}}(\mathbf{k})$ denotes the Berry curvature: (a) at a constant $\mathcal{B}=1$, increasing $N$ results in broadening of the Berry curvature distribution. The total flux enclosed by the momentum plane increases linearly with N: $\Phi_{N}(\infty)=2\pi (N-1)$. (b) At a constant $N=5$, increasing $\mathcal{B}$ results in the Berry curvature distribution becoming more concentrated near the origin, with the total flux enclosed $\Phi_{N}(\infty)$ remaining constant. The maximum of the Berry curvature distribution has a value of $\mathcal{B}$ occurring at the origin. (c) Two two different dispersions that we used in our study, having to distinct Fermi surface topologies with circular and annular Fermi surfaces respectively.}
    \label{fig:Figure_Lambda_N_BC_dist}
\end{figure}

Recently, it was shown in Ref.~\cite{Jiang2025} that the surface states of RNG satisfy this criterion in a small region around the valley points. In this region, the Berry curvature was found to be constant and nonzero, a property which is absent in the two-orbital RNG models considered in earlier studies \cite{Koshino2009RNG, MacDonaldTrilayer}. The $\lambda_N$-model instead realizes non-uniform ideal quantum geometry, encoded in its lowest-band form factor. Restricting to the lowest band $(n=0)$, the form factor is defined by
\begin{gather}
    \mathcal{F}_{N}(\mathbf{k},\mathbf{k'})=\braket{s_{0\vb{k}}}{s_{0\vb{k}'}}=\mathcal{M}_{N}(\mathbf{k},\mathbf{k'})
    \mathcal{F}_{LLL}(\mathbf{k},\mathbf{k'}).\label{eq:FF_l_N}
\end{gather}
The $N$-dependent prefactor $\mathcal{M}_{N}(\mathbf{k},\mathbf{k'})$ is \cite{desrochers2025}
\begin{gather}
    \mathcal{M}_{N}(\mathbf{k},\mathbf{k'})=\frac{\Gamma(N,\mathcal{B}k_zk_z'^{*})}{\sqrt{\Gamma(N,\frac{\mathcal{B}}{2}|\mathbf{k}|^2)\Gamma(N,\frac{\mathcal{B}}{2}|\mathbf{k'}|^2)}},\label{eq:pre_factor_l_N}
\end{gather}
where $\Gamma(N,x)$ is the upper incomplete Gamma function. In the limit $N\rightarrow \infty$, we find that $\mathcal{M}_{N}(\mathbf{k},\mathbf{k'})\rightarrow 1$, reproducing the LLL limit. In Appendix~\ref{sec_app:LambdaNModelDetails}, we show that the actual form of the eigenstates $\ket{s_{0\vb{k}}}$ as $N\rightarrow \infty$ are identical to those found in the ICB model \cite{Tan2024,TanDevakul2025WavefunctionAHC}. The Berry curvature of the $\lambda_N$ model takes the form
\begin{align}
    \Omega_{N}(\mathbf{k})=\mathcal{B}+\mathcal{B}e^{-\frac{\mathcal{B}|\mathbf{k}|^2}{2}}\frac{E_{1-N}\left(\frac{\mathcal{B}|\mathbf{k}|^2}{2}\right)-E_{-N}\left(\frac{\mathcal{B}|\mathbf{k}|^2}{2}\right)}{\left( E_{1-N}\left(\frac{\mathcal{B}|\mathbf{k}|^2}{2}\right) \right)^2},\label{eq:Berry curvature_l_N}
\end{align}
where the exponential integral function is defined as $E_N(z)=z^{N-1}\Gamma(1-N,z)$. Since the model has ideal quantum geometry, the quantum metric is fixed by the magnitude of the Berry curvature according to
\begin{align}
    g_{\mu\nu}(\vb{k})=\frac{\abs{\Omega_{N}(\vb{k})}}{2}\delta_{\mu\nu}.
\end{align}
The Berry curvature in Eq.~\eqref{eq:Berry curvature_l_N} is plotted in Figs.~\ref{fig:Figure_Lambda_N_BC_dist}(a, b). It forms a plateau of height $\mathcal{B}$ centered at $\mathbf{k}=0$, before decaying to zero as $\abs{\vb{k}}\rightarrow \infty$. Increasing $N$ broadens this plateau while leaving its height approximately fixed. For large $N$, the radius of the plateau is well approximated by $k_{\text{plt}}=\sqrt{2(N-1)/\mathcal{B}}$, so that the model realizes an approximately uniform quantum geometry over the region $|\mathbf{k}| < k_{\text{plt}}$, where $\text{tr}[g(\mathbf{k})] \approx \mathcal{B}$. 
The Berry flux enclosed in a circle of radius $\Lambda$ is given by  
\begin{align}
    \hspace{-0.3cm}\Phi_{N}(\Lambda)=2\pi\int_{0}^{\Lambda} dk\, k\, \Omega_{N}(\vb{k})=\pi \mathcal{B}\Lambda^2-\frac{2\pi e^{-\frac{\mathcal{B}\Lambda^2}{2}}}{E_{1-N}(\frac{\mathcal{B}\Lambda^2}{2})}\label{eq:enc_flux}.
\end{align}
In the limit $\Lambda \rightarrow \infty$, the total flux enclosed is quantized in units of $2\pi$,
\begin{align}
\Phi_{N}(\infty)=2\pi(N-1). \label{flux quantization}
\end{align}
In Appendix~\ref{sec_app:onepointcompactification}, we show that this quantized flux defines the parent-band Chern number 
\begin{align}
    C_{P}=\frac{\Phi_N(\infty)}{2\pi}=N-1.
\end{align}
This identification follows from the one-point compactification of the model's momentum plane, $\mathbb{R}^2 \cup \{\infty\}=\mathbb{S}^2$. As in Fig.~\ref{fig:Figure_Lambda_N_BC_dist}, for fixed $\mathcal{B}$, the plateau widens with increasing $N$, leading to an increase in the total flux enclosed. For fixed $N$, the Berry curvature profile becomes more peaked around the center as we increase $\mathcal{B}$, while the flux enclosed by the entire domain remains constant. In Sec. \ref{sec:results_gap_eqn}, we study the effect of this tunability in $N$ and $\mathcal{B}$ on the formation of momentum-space vortices in the superconducting state.

\subsection{The band dispersion}
In the $\lambda_N$-model, the lowest-band eigenstate is determined independently of the band dispersion $\epsilon_{\mathbf{k}}$. This allows us to investigate pairing using two kinds of dispersions with distinct Fermi surface topologies, while keeping the underlying quantum geometry fixed. In Sec.~\ref{sec_BdG Topology}, we will see that the Fermi surface topology is tied to the topology of the superconducting state.

We use a dispersion with a quadratic and quartic term. The dispersion we use takes the form,
\begin{gather}
\varepsilon(\mathbf{k})=a_2|\mathbf{k}|^2+a_4|\mathbf{k}|^4+a_0.\label{eq:kinetic_term}
\end{gather}
It is parameterized by $\delta$ which controls the curvature of the band. We set $a_2=-(1+\delta)$, $a_4=1$, $a_0=1+\delta$ with $\delta\in [-1,0.5]$. Increasing $\delta$ progressively flattens the bottom of the parabolic band. Beyond a critical value, the dispersion evolves from a simple parabolic form into a Mexican-hat-shaped profile, thereby mimicking the effect of an out-of-plane displacement field \cite{Ghazaryan2021,Ghazaryan2023,Geier2025}. To obtain a circular Fermi surface we chose $\delta=-0.25$ or $(a_2,a_4,a_0,\mu)=(-0.75,1.0,0.75,1.0)$, and for an annular Fermi surface we chose $\delta=0.125$ or $(a_2,a_4,a_0,\mu)=(-1.125,1.0,1.125,1.061)$. The dispersions are shown in Fig.~\ref{fig:Figure_Lambda_N_BC_dist}(c), and we will henceforth call these the parent bands. These parameters are chosen to qualitatively mimic the flat band bottom near a K-valley of RNG \cite{Zhang_2010,Zhang_2019,Jung_2013,Dong_2024}. Additionally, neglecting the effects of trigonal warping, previous studies have shown that the band structure of RNG state scales as $\epsilon \sim |\mathbf{k}|^N$ away from the $K$ points \cite{Koshino2009RNG, MacDonaldTrilayer}. This is well described by a down-folded two-band model \cite{Koshino2009RNG,Slizovskiy2019, Zhang_2019,patri2025}. In the present study we neglect these features — the power-law scaling and trigonal warping — and focus primarily on the qualitative features of the Fermi surface.

\section{Band-Projected Superconductivity}\label{sec:band_proj_SC}
This section introduces the formalism used to analyze superconducting states. We focus on superconducting pairing in the lowest isolated band of the $\lambda_N$ model.  In contrast to conventional band-projected approaches to superconductivity \cite{Guerci2025, Shavit2025, guerci2026, Geier2025, Yang2025, Wang2024, li2025}, where $\mathbf{k}$ is defined over the Brillouin zone of a lattice model, the $\lambda_N$-model is defined on the entire momentum plane.

\subsection{Band projected Hamiltonian} 
We consider an attractive density-density interaction projected onto the lowest band of the $\lambda_N$ model. In real materials, such an interaction can arise from several mechanisms: an effective attraction mediated by phonons \cite{Bardeen1957,VinasBostrom2024}, a Kohn-Luttinger mechanism \cite{Kohn1965,Shavit2025} arising from over-screening of the repulsive Coulomb interaction, or a soft fluctuating mode \cite{Scalapino2012,Chubukov2003,Chubukov2012,Lederer2017} that mediates an attractive channel. The microscopic mechanism giving rise to superconductivity in non-moir\'e RNG remains under investigation. However, for the purpose of this study we make no assumption regarding its microscopic origin. 

The Hamiltonian describing the system takes the form
\begin{align}
H=\sum_{\vb{k}}\epsilon_{\vb{k}}c^{\dag}_{\vb{k}}c_{\vb{k}}+\frac{1}{2A}\sum_{\vb{q}}V(\vb{q}):\rho_{\vb{q}}\rho_{-\vb{q}}:, \label{BandProjectedBCS}
\end{align}
where $:\,\, :$ denotes normal ordering in the interaction term. Here, $V(\vb{q})$ is the interaction, and $A$ is the sample area. Projection onto the lowest band dresses the density operator by form factors
\begin{align}
\rho_{\vb{q}}=\sum_{\vb{k}}\mathcal{F}(\vb{k},\vb{k}+\vb{q})c^{\dag}_{\vb{k}}c_{\vb{k}+\vb{q}}. \label{projected density}
\end{align}
Here, the creation operators create single-particle states of the form
$c^{\dag}_{\vb{k}}\ket{0}=e^{i\vb{k} \cdot \vb{r}}\ket{s_{0\vb{k}}}$. Going forward, we suppress the band index $n=0$ for clarity. The attractive interaction in Eq. \eqref{BandProjectedBCS} is the Gaussian
\begin{align}
    V(\vb{q})=-U e^{-\sigma^2 \abs{\vb{q}}^2},
\end{align}
where $U>0$ sets the interaction strength, and $\sigma^2$ is a parameter controlling the range of the interaction in momentum space. A non-zero $\sigma$ ensures that 
$V(\mathbf{q})\rightarrow0$ in the large $\abs{\mathbf{q}}$ limit. $\sigma^2$ can also be interpreted as the range of the interaction in real space, with $\sigma^2\ll1$ implying a localized interaction in real space.

As discussed in Sec.~\ref{GeometryofLambda}, the form factors $\mathcal{F}(\vb{k},\vb{k}')=\bra{s_{0\vb{k}}}\ket{s_{0\vb{k}'}}$ encode the geometry of the lowest band. Since these form factors enter explicitly in the projected density operators, the resulting interaction Hamiltonian is modulated by the geometry of the band. This can be seen from the interaction vertex in the Cooper channel
\begin{align}
    V_{\mathcal{F}}(\vb{k},\vb{k}')=V(\vb{k}-\vb{k}')\mathcal{F}(\vb{k},\vb{k}')\mathcal{F}(-\vb{k},-\vb{k}').\label{eq:form_factor_dressed_vertex}
\end{align}
Assuming that the pairing occurs between electrons of momentum $\vb{k}$ and $-\vb{k}$ in the entire plane, the interaction in the Cooper channel takes the usual BCS form
\begin{align}
    H_{I}=\frac{1}{2A}\sum_{\vb{k},\vb{k}'}V_{\mathcal{F}}(\vb{k},\vb{k}')c^{\dag}_{\vb{k}}c^{\dag}_{-\vb{k}}c_{-\vb{k}'}c_{\vb{k}'}.
\end{align}
The corresponding mean-field gap is
\begin{align}
    \Delta(\vb{k})=\frac{1}{A}\sum_{\vb{k}'}\mathcal{W}(\vb{k},\vb{k}')\expval{c_{\vb{k}'}c_{-\vb{k}'}}, \label{gapeq}
\end{align}
where the Cooper vertex is anti-symmetrized by the antisymmetric pairing function $\expval{c_{\vb{k}'}c_{-\vb{k}'}}$, i.e.
\begin{gather}
    \mathcal{W}(\vb{k},\vb{k}')=\frac{1}{2}(V_{\mathcal{F}}(\vb{k},\vb{k}')-V_{\mathcal{F}}(\vb{k},-\vb{k}')).
    \label{eq:anti_symm_Cooper_vertex}
\end{gather}
This antisymmetrization ensures the correct symmetry for the gap, $\Delta(\vb{k})=-\Delta(-\vb{k})$, as is required for pairing among spinless (or spin polarized) fermions.

The mean-field order parameter has the same formal structure as in conventional BCS theory. The essential distinction is that the gap is constructed from the band-projected interaction vertex in Eq.~\eqref{eq:form_factor_dressed_vertex}, and annihilation operators defined within the parent band. As a result, the projected order parameter inherits the gauge freedom of the parent-band Bloch states. To see this, consider a local $U(1)$ transformation of the parent-band Bloch state $\ket{s_{\vb{k}}}$, 
\begin{align}
    \ket{s_{\vb{k}}} \rightarrow e^{i\chi(\vb{k})}\ket{s_{\vb{k}}}.
\end{align}
This implies that the projected ladder operators in the parent band transform as
\begin{align}
    c_{\vb{k}}\rightarrow e^{-i\chi(\vb{k})}c_{\vb{k}}.
\end{align}
Furthermore, since the vertex $V_{\mathcal{F}}(\vb{k},\vb{k}')$ is dressed by the form factors $\mathcal{F}(\vb{k},\vb{k}')$, it too transforms
\begin{align}
    V_{\mathcal{F}}(\vb{k},\vb{k}')\rightarrow e^{i(\chi(\vb{k}')+\chi(-\vb{k}')-\chi(\vb{k})-\chi(-\vb{k}))}V_{\mathcal{F}}(\vb{k},\vb{k}').
\end{align}
Therefore, the gap function transforms under gauge transformations of the underlying electronic band
\begin{align}
    \Delta(\vb{k}) \rightarrow e^{-i(\chi(\vb{k})+\chi(-\vb{k}))}\Delta(\vb{k}). \label{eq_gapgaugetransform}
\end{align}
This behavior is unexpected from a gap in momentum space, and it parallels the behavior of the gap in real space minimally coupled to an external magnetic vector potential $\vb{A}_{em}(\vb{r})$. In such cases, the gap behaves as a complex scalar field, transforming under an induced $U(1)$ gauge transformation of the vector potential (see Appendix~\ref{app:differnt_gauge_transformations} for a clear distinction between the electromagnetic and Bloch gauge transformations). Treating Eq.~\eqref{eq_gapgaugetransform} as a momentum space analogy, we infer that the gap is coupled to a pair Berry connection defined as 
\begin{align}
    \mathcal{A}_P(\vb{k})=\mathcal{A}(\vb{k})-\mathcal{A}(-\vb{k}),
\end{align}
where $\mathcal{A}(\vb{k})=i\bra{s_{\vb{k}}}\ket{\nabla_{\vb{k}}s_{\vb{k}}}$ is the Berry connection of the parent band. Since the pair connection transforms as
\begin{align}
    \mathcal{A}_{P}(\vb{k}) \rightarrow \mathcal{A}_{P}(\vb{k})-\nabla_{\vb{k}}(\chi(\vb{k})+\chi(-\vb{k})),
\end{align}
it is natural to define the gauge-invariant quantity
\begin{align}
    \vb{v}_{\Delta}(\vb{k})=\nabla_{\vb{k}}\text{arg}[\Delta(\vb{k})]-\mathcal{A}_{P}(\vb{k}), \label{eq_momentumspacevelocity}
\end{align}
whose form shares the same underlying gauge-theoretic structure as the superfluid velocity in a superconductor. This gauge-invariant quantity was highlighted in Ref.~\cite{HaldaneProjected}, which noted its regularity away from gap nodes. The structure of Eq.~\eqref{eq_momentumspacevelocity} and in particular its singularities underpins the analysis that follows. In Sec.~\ref{sec:results_gap_eqn}, we demonstrate the existence of these nodes with phase singularities by solving the full gap equation at $T=0$. Later in Sec.~\ref{sec_BdG Topology}, we show that the analytic structure of Eq.~\eqref{eq_momentumspacevelocity} is tied to the underlying BdG topology and thermal Hall conductivity. As a preliminary step, we revisit chiral superconductivity and clarify its formulation within the band-projected superconductivity formalism.

\subsection{Chiral superconductivity}
Chiral superconductivity is expected to arise in the lowest band of the $\lambda_N$-model. This is because the form-factors $\mathcal{F}(\mathbf{k},\mathbf{k+q})$ explicitly break time-reversal symmetry, since $\mathcal{F}(\mathbf{k},\mathbf{k+q})\neq \mathcal{F}(\mathbf{-k},-\mathbf{k}-\mathbf{q})^*$. Therefore the Cooper vertex in Eq.~\eqref{eq:form_factor_dressed_vertex} also breaks time-reversal symmetry at the level of the pairing interaction, and the resulting gap function $\Delta(\mathbf{k})$ describes chiral superconducting states. We highlight here that this time-reversal symmetry breaking is not spontaneous but is instead explicitly encoded within the Cooper vertex itself. In conventional chiral superconductors, the weak-coupling gap on a circular Fermi surface can be expressed as 
\begin{gather}
\Delta(\vb{k})\big|_{\vb{k}\in FS}\sim (k_x+ik_y)^{\ell}=\abs{\vb{k}}^{\ell}e^{i\ell \phi(\vb{k})},
\end{gather}
where $\phi(\vb{k})=\text{arg}(k_x+ik_y)$, and $\ell$ is the phase winding of the gap around the Fermi surface. This winding can be understood as arising from a phase singularity at the origin. For spinless pairing, the gap is odd under inversion. Since the point $\vb{k}=0$ is an inversion-invariant point, the gap function necessarily vanishes at the origin, $\Delta(0)=0$. To preserve the antisymmetry of the gap, the phase must wind by an odd integer around any inversion symmetric loop $\gamma$ enclosing the origin, provided that $\Delta(\vb{k}) \neq 0$ on $\gamma$:
\begin{align}
    \ell=\oint_{\gamma} \frac{d\vb{k}}{2\pi} \cdot \nabla_{\vb{k}}\text{arg}[\Delta(\vb{k})] \in 2\mathbb{Z}+1. \label{eq:winding_of_gap}
\end{align}
This defines a momentum-space vortex: an isolated zero of $\Delta(\vb{k})$ around which the phase of the gap winds nontrivially.

\begin{figure*}
    \centering
    \hspace{-0.5cm}
    \includegraphics[scale=0.95]{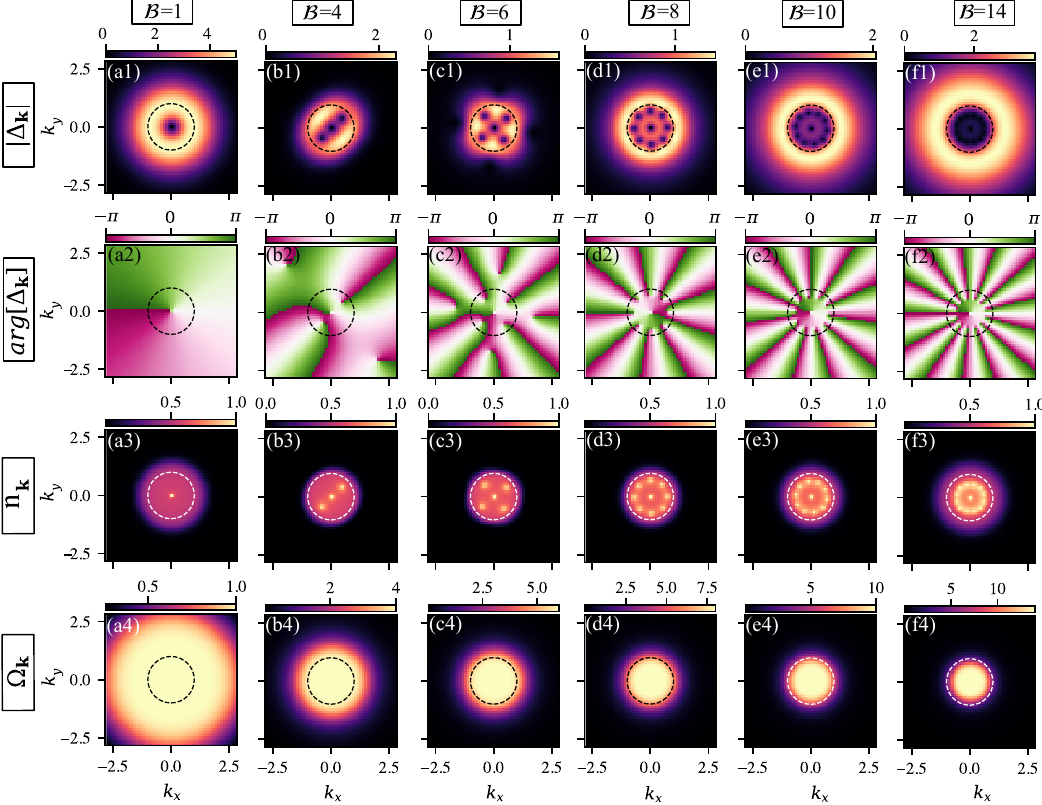}
    \caption{The variation of the magnitude $|\Delta_{\mathbf{k}}|$ and phase $\text{arg}[\Delta_{\mathbf{k}}]$ of the gap function, local occupation $n_{\mathbf{k}}$, and the Berry curvature distribution $\Omega_{\mathbf{k}}$, as the width of the Berry curvature distribution is varied by changing $\mathcal{B}$. These gap functions are calculated for parameters, $N=8$ (for the $\lambda_N$ model), $U=5$, $\sigma^2=1$. The circular Fermi surface is marked by the dashed lines. As $\mathcal{B}$ increases, more vortices are nucleated until a critical value is reached beyond which the number of vortices reach its upper bound. More details shown in Figs. \ref{fig:vortex_nucleation}, \ref{fig:vortex_coalescence}.
    }
    \label{fig:trend_gap_N_8}
\end{figure*}

As we shall see in Sec.~\ref{sec:results_gap_eqn}, the numerical solution of the gap equation away from the weak-coupling regime shows the presence of vortices away from the origin. The gap can therefore be captured using an ansatz
\begin{gather}
    \Delta(\mathbf{k})=\Delta_0 f(\mathbf{k}, \{\mathbf{k}_{i}\})\prod_{i=0}^{N_V-1} e^{i\ell_{i}\phi(\mathbf{k}-\mathbf{k}_{i})},\label{eq:vortex_ansatz}
\end{gather}
where $\mathbf{k}_{i}$ are the location of the vortices in the gap ($i=0,\cdots,N_V-1$). The function $f$ encodes the profile of the gap for all momenta positions, and vanishes at the vortex cores, $f(\mathbf{k}_{i})=0$. Since, the gap is antisymmetric, there is always a central vortex pinned to the origin. Additionally, each vortex at $\mathbf{k}_{i}$ is accompanied by a vortex at $-\mathbf{k}_{i}$ with the same winding. 

For $T>T_c$, the normal state possesses a spatial $SO(2)$ symmetry coming from the form factors of the $\lambda_N$-model and the choice of band structure. However, the formation of momentum-space vortices in the superconducting state, as prescribed by the ansatz in Eq.~\eqref{eq:vortex_ansatz}, can break this rotational symmetry. As shown in Sec.~\ref{sec:results_gap_eqn}, these vortices form a ring enclosing the origin, with $N_V-1$ vortices present on the ring. Therefore, based on the numerical results we find that this reduces the symmetry of the superconducting state to $C_{N_V-1}$.

\section{Solving the gap equation}\label{sec:results_gap_eqn}
In this section, we present numerical solutions to the gap equation. To determine the effect of the underlying Berry curvature on the nature of the gap function, the full BCS gap equation is solved self-consistently across the momentum plane. The gap equation is
\begin{align}
    \Delta(\vb{k})=-\sum_{\vb{k}'}\frac{\Delta(\vb{k}')}{2E_{\vb{k}'}}\mathcal{W}(\vb{k},\vb{k}')\text{tanh}\left(\frac{\beta E_{\vb{k}'}}{2}\right),\label{eq:gap_eqn}
\end{align}
where $E_{\vb{k}}=\sqrt{(\epsilon_{\vb{k}}-\mu)^2+\abs{\Delta(\vb{k})}^2}$ is the Bogoliubov quasiparticle spectrum. We fix a smooth gauge for the parent band state $\ket{s_{\vb{k}}}$ on the momentum patch over which the gap equation is numerically solved, chosen to extend several times larger than the Fermi momentum. The Berry curvature is concentrated within this patch, decaying to a negligible value at its boundary $\abs{\vb{k}_b}=\Lambda$. The patch size is chosen such that the gap function decays to negligible values near its edge, ensuring its qualitative features are fully captured. The numerical procedure used to solve the gap equation is described in Appendix~\ref{sec_app:NLGE}. All results presented here are obtained at $T=0$. We also fix $U = 5$ and $\sigma = 1$ while varying $\mathcal{B}$ and $N$ systematically across a range of values in the $\lambda_N$-models. Additional data with different values of $U$ and $\sigma$ are presented in Appendix~\ref{sec_app:NLGE_additional_data}.

\subsection{Momentum space vortices}\label{sec:mom_vort}
To identify the momentum space vortices in the gap function, we plot both the magnitude $\abs{\Delta({\mathbf{k})}}$ and the phase $\text{arg}[\Delta({\mathbf{k})}]$. In Fig.~\ref{fig:trend_gap_N_8}, we show the results for a fixed $N$ ($N=8$) and with increasing $\mathcal{B}$. For small $\mathcal{B}$ (see Fig. \ref{fig:trend_gap_N_8} (a)), we find the presence of a single momentum space vortex at the origin, having a winding of $+1$. This indicates the formation of a chiral $p_x+ip_y$ state. Furthermore, we find that the gap is concentrated on a wide ring centered around the Fermi surface, which is indicated by the dashed lines in Fig.~\ref{fig:trend_gap_N_8} (a1). The gap is suppressed at large momenta, an effect that is amplified by the non-zero interaction length $\sigma$ (see Appendix~\ref{sec_app:NLGE_additional_data}).

To highlight the role of the parent band's Berry curvature profile in shaping the gap function, we plot the corresponding Berry curvature profile in Fig. \ref{fig:trend_gap_N_8} (a4). We observe that it is almost uniform in the region where the gap is non-zero, and therefore emulates an almost constant momentum space magnetic field of strength $\mathcal{B}$. However, its strength is insufficient to nucleate additional vortices in the gap function within the patch where the gap equation is solved, other than the vortex at the origin which is required by symmetry.

With increasing $\mathcal{B}$, the Berry curvature profile becomes more concentrated and peaked near the origin (see Figs.~\ref{fig:trend_gap_N_8} (b4)-(f4)). This leads to the nucleation of additional momentum space vortices in the gap function as shown in Fig.~\ref{fig:trend_gap_N_8} (b)-(f). All these nucleated vortices have a winding of $+1$, which can be identified from the phase winding of the gap in $\text{arg}[\Delta({\mathbf{k})}]$. As required by the antisymmetry of the gap function, momentum space vortices nucleated away from the origin always appear in pairs. This ensures that the total number of vortices enclosed by a circular Fermi surface remains odd. We also observe in Fig.~\ref{fig:trend_gap_N_8} that the vortices are always nucleated on a ring centered around the origin. This structure was also reported in Ref.~\cite{patri2025}, although for a different Berry curvature distribution. For $SO(2)$ symmetric continuum models, the ring structure is expected as it breaks the least symmetries of the normal state and therefore leads to the lowest free energy. 

We also observe that a strong Berry curvature in the parent band suppresses the magnitude of the gap at the same locations in momentum space. This can be observed in Figs.~\ref{fig:trend_gap_N_8} (d1)-(f1). Since the Berry curvature is concentrated near the origin, the gap function is suppressed in the same region and accumulates largely outside the region of strong Berry curvature. This effect can be traced back to the quantum metric \cite{patri2025}, which for this ideal model is the same as the Berry curvature. To see this analytically, the gap equation can be solved in the limit of $\sigma^2\gg1$ following a similar procedure to Ref. \cite{patri2025}. In this limit the gap can be expressed as (details in Appendix~\ref{sec_app:suppression_of_gap})

\begin{align}
    |\Delta({\mathbf{k}})|^2&\approx \left(\frac{U}{8\pi \sigma^2}\right)^2\left(1-\frac{\text{tr}[g_{\mu\nu}(\mathbf{k})]}{\sigma^2} \right)-\xi_{\mathbf{k}}^2+\cdots\nonumber\\
    &= \left(\frac{U}{8\pi \sigma^2}\right)^2\left(1-\frac{\Omega_{N}(\mathbf{k})}{\sigma^2} \right)-\xi_{\mathbf{k}}^2+\cdots.\label{eq:metric_suppressing_gap}
\end{align}
This demonstrates that the trace of the quantum metric suppresses the gap. In $\lambda_N$-models with ideal quantum geometry, this trace equals the Berry curvature, so the gap is suppressed wherever the Berry curvature is large. However, Eq.~\eqref{eq:metric_suppressing_gap} does not predict the location of the nucleated momentum-space vortices.

At zero temperature in the BCS ground state, the local occupation, $n_{\mathbf{k}}=\langle c_{\mathbf{k}}^{\dagger}c_{\mathbf{k}}\rangle$, of the band is given by
\begin{gather}
    n_{\mathbf{k}}=\frac{1}{2}\left(1-\frac{\epsilon_{\mathbf{k}}-\mu}{E_{\mathbf{k}}}\right).
\end{gather}
In Fig.~\ref{fig:trend_gap_N_8}(a3)-(f3), we find that regions of momentum space hosting the vortices exhibit an enhanced local occupation. This is expected as the condensate is destroyed within the vortex leading to an enhancement of the local occupation $n_{\mathbf{k}}$. 

All the results in Fig.~\ref{fig:trend_gap_N_8} are using a constant $U$ and $\sigma^2$. In Appendix \ref{sec_app:NLGE_additional_data}, we discuss the trend in vortex nucleation as we vary $U$ and $\sigma^2$. We find that increasing $U$ enhances vortex formation. However, increasing $\sigma^2$ suppresses vortex formation and also the magnitude of the gap as was discussed earlier.

\subsection{Vortex number $\mathcal{Q}$ and condensation energy}\label{Sec:IVB}

\begin{figure}
    \centering
    \includegraphics[scale=1]{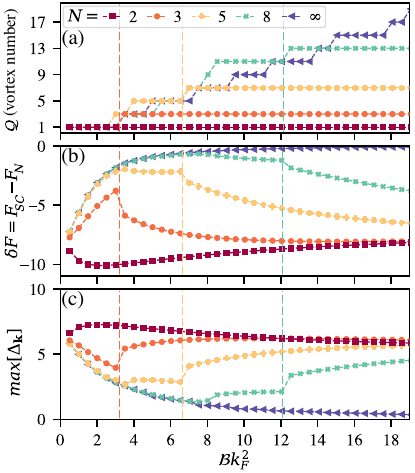}
    \caption{(a) The vortex number $\mathcal{Q}$ with increasing $\mathcal{B}k_F^2$. For finite $N$, more vortices are nucleated with increasing $\mathcal{B}k_F^2$, until vortex number saturation is reached when $\mathcal{Q}=2N-3$. The vertical dashed lines demarcate these two regions of vortex nucleation and vortex saturation. (b) The condensation energy $\delta F=F_{SC}-F_{N}$, shows shows jumps associated with vortex formation in (a), where the formation of vortices leads to a decrease in the condensation energy. Panel (c) shows the maximum value of the gap as these parameters are varied.}
    \label{fig:vortex_nucleation}
\end{figure}

\begin{figure*}
    \centering
    \includegraphics[scale=1.1]{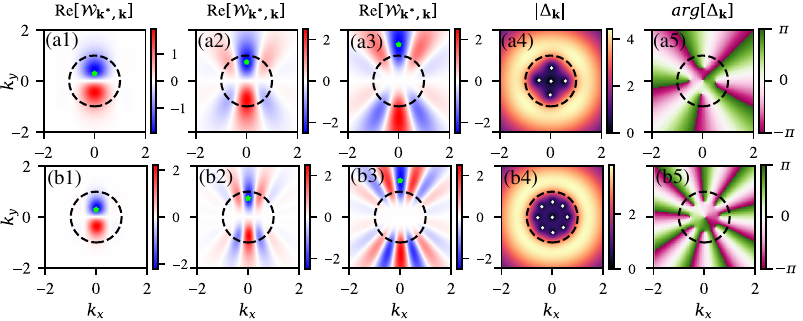}
    \caption{The real part of the Cooper vertex $\mathcal{W}(\mathbf{k}^*,\mathbf{k})$ (Eq. \ref{eq:anti_symm_Cooper_vertex}) is shown in (a1)–(a3) for $N=4$, $\mathcal{B}=5.5$, where $\mathbf{k}^*$ is fixed at the momentum marked by the green star and the dashed line denotes the Fermi surface. The corresponding gap magnitude $|\Delta(\mathbf{k})|$, and argument $\arg[\Delta(\mathbf{k})]$ are shown in (a4), and (a5). Row (b) shows the same for $N=6$, $\mathcal{B}=10$. The vortices in $\text{abs}[\Delta(\mathbf{k})]$ are represented by the white dots for clarity. For small values of $\mathbf{k}^*$ (column 1), one nodal plane is present in $\text{Re}[\mathcal{W}(\mathbf{k}^*,\mathbf{k})]$ implying the formation of a winding $+1$ central vortex. For larger values of $\mathbf{k}^*$ (columns 2, 3), $2N-3$ nodal planes appear implying the formation of additional momentum space vortices on a ring at larger values of $\mathbf{k}$.
    }
    \label{fig:Cooper_vertex_Gap_Omega}
\end{figure*}

In Sec.~\ref{sec:mom_vort}, we observed the formation of momentum-space vortices in the gap function driven by a nontrivial Berry curvature. Now we characterize how the number of momentum-space vortices evolves with $\mathcal{B}$ and $N$. To quantify this behavior, we introduce the occupied vortex charge $\mathcal{Q}$, defined as the sum of total winding of vortices (each with winding $\ell_i$ at location $\mathbf{k}_i$) enclosed within the occupied region of the Fermi surface,
\begin{gather}
    \mathcal{Q}=\sum_{i: \xi_{\vb{k}_i}<0}\ell_i. \label{eq_postulatedvortexnumber}
\end{gather}
From our numerics we always find vortices of winding $\ell_i=+1$. Therefore, in this subsection, $\mathcal{Q}$ can be thought of as the occupied vortex number. We analyze the accompanying change in condensation energy upon vortex formation. Tracking both $\mathcal{Q}$ and the condensation energy as functions of $\mathcal{B}$ and $N$ reveals two distinct regimes: a nucleation regime, where the vortex number increases with increasing $\mathcal{B}$, and a vortex number saturation regime in which $\mathcal{Q}$ has reached its upper bound. The condensation energies are computed within the BCS formulation for spinless fermions. At zero temperature, the free energy in the superconducting state is given by
\begin{gather}
    F_{SC}[\Delta(\vb{k})]=\frac{1}{2}\sum_{\mathbf{k}}\left(\xi_{\mathbf{k}}-E_{\mathbf{k}} +\frac{\abs{\Delta(\vb{k})}^2}{2E_{\mathbf{k}}}\right),\label{eq:free_energy_SC}
\end{gather}
and the zero temperature free energy in the normal state ($\Delta(\mathbf{k})=0$) is 
\begin{gather}
    F_N=\sum_{|\mathbf{k}|<k_F}\xi_{\mathbf{k}}.\label{eq:free_energy_normal}
\end{gather}
We define the condensation energy as $\delta F=F_{SC}-F_{N}$, which in our convention is negative in the superconducting state; a free-energetically stable state will tend to lower its condensation energy. $\delta F$ is calculated using the numerically obtained gap function $\Delta(\mathbf{k})$. Since the model is defined on the continuum, we work on a sufficiently large momentum patch and set $\sigma = 1$, so that the gap function vanishes for $|\mathbf{k}| < \Lambda$ inside the patch, ensuring reliable convergence of the condensation energy.

In Fig.~\ref{fig:vortex_nucleation} (a), we plot the vortex number $\mathcal{Q}$ as a function of $\mathcal{B}k_F^2$ for different values of $N$ ($k_F=1$). A larger $\mathcal{B}$ leads to the nucleation of more momentum-space vortices. $\mathcal{Q}$ increases in steps of two to ensure the correct anti-symmetry of the gap function. For each finite value $N$ (except $N=2$), we observe that the vortex number saturates once we reach some critical value of $\mathcal{B}$, indicated by the dashed lines in Fig.~\ref{fig:vortex_nucleation}. For each $N$, the left side of the dashed line represents the region of vortex nucleation, while the right side denotes the region of vortex number saturation. In the saturation region, $\mathcal{Q}_{\text{sat}}=2N-3$. For $N=2$, the vortex number is bounded at $\mathcal{Q}_{\text{sat}}=1$, so the demarcation between these two regimes is not distinct, and all pairing states are of the form $p+ ip$.

In Fig.~\ref{fig:vortex_nucleation} (b), we plot the condensation energies $\delta F$. We find that the nucleation of vortices tends to lower the condensation energy. This can be identified for $N=3$ in Fig.~\ref{fig:vortex_nucleation} (b). As $\mathcal{B}$ approaches the threshold for nucleating additional vortices, the condensation energy initially rises, reflecting this growing tendency to nucleate. As the threshold is crossed at $\mathcal{B}k_F^2\sim 3.5$, momentum space vortices are nucleated (see jump in $\mathcal{Q}$ in Fig.~\ref{fig:vortex_nucleation}(a)) and the condensation energy decreases as we further increase $\mathcal{B}$. For larger $N$, successive vortex nucleations contribute progressively less to the condensation energy, until vortex number saturation $\mathcal{Q}_{\text{sat}}$ is reached.

Beyond this point, the condensation energy shows a steep drop with further increases in $\mathcal{B}$, and for large values of $\mathcal{B}$ the condensation energies tend towards a common value that is independent of $N$ (see also Fig.~\ref{fig:vortex_nucleation_coalescence_log_scale} in Appendix). This happens because in this vortex number saturation regime, the Berry curvature distribution becomes strongly peaked near the origin, forming a narrow and very tall plateau for all values of $N$. This leads to the suppression of the gap being confined to this same narrow region of strong Berry curvature around the origin (in which the vortices reside). Therefore, the gap away from the origin tends towards an common profile, as shown in Fig.~\ref{fig:vortex_nucleation}(c) where $\text{max}[\Delta({\mathbf{k}})]$ saturates for large $\mathcal{B}k_F^2$. Since the gap function becomes qualitatively identical in this limit for all $N$, the condensation energy in Fig.~\ref{fig:vortex_nucleation}(a) likewise tends to an $N$-independent value.

The condensation energy also increases monotonically with $N$ at fixed $\mathcal{B}$, reflecting the broadening of the Berry curvature plateau: a wider plateau suppresses the gap over a larger region, increasing the condensation energy accordingly. For $N = \infty$ (or the ICB), increasing $\mathcal{B}$ leads to an unbounded growth in the number of nucleated vortices, accompanied by a monotonic increase in the condensation energy, and a suppression of the gap function (see Figs. \ref{fig:vortex_nucleation}(a), (c)). This is expected since the Berry curvature is uniform and the gap cannot avoid regions of strong Berry curvature as it does in the finite $N$-models driving the monotonic growth.

Based on the above results, we find an analogy with type-II superconductors: just as a magnetic field suppresses the gap and nucleates real-space vortices to lower the condensation energy, nucleated vortices in the momentum-space gap function due to a strong Berry curvature exhibit a similar tendency of lowering the condensation energy.

\subsection{Vortex saturation}
We noted that the maximum number of momentum space vortices with winding $+1$ that can be nucleated for a finite $N$, is $\mathcal{Q}_{\text{sat}}=2N-3$. In this configuration, there is one central vortex and a concentric ring of $2N-4$ vortices encircling it. The formation of this structure can be understood by visualizing the Cooper vertex $\mathcal{W}(\mathbf{k},\mathbf{k'})$ (of Eq.~\eqref{eq:anti_symm_Cooper_vertex}) in the regime of large $\mathcal{B}$ where vortex number saturation in $\mathcal{Q}$ has occurred. In Fig.~\ref{fig:Cooper_vertex_Gap_Omega} (a1)-(a3), we plot the real part of the Cooper vertex $\mathcal{W}(\mathbf{k}^*,\mathbf{k})$ (for $N=4$, $\mathcal{B}=5.5$) where one momentum $\mathbf{k}^*$ remains fixed (marked by the green cross). In Fig.~\ref{fig:Cooper_vertex_Gap_Omega} (a1) for small $\mathbf{k}^*$, we observe that the Cooper vertex has one nodal plane with two lobes. Since, the gap function $\Delta({\mathbf{k}})$, obtained from solving the gap equation with the Cooper vertex $\mathcal{W}(\mathbf{k},\mathbf{k}')$, must have a structure commensurate with the Cooper vertex, this implies the presence of a nodal plane in the real part of the gap. This is consistent with the formation of a momentum space vortex of winding $+1$ at small momenta $\mathbf{k}$, in agreement with the vortex observed at the origin. For larger momenta $\mathbf{k}^*$, additional nodal planes emerge in the real part of the Cooper vertex. As seen in Fig.~\ref{fig:Cooper_vertex_Gap_Omega}(a3) five nodal planes appear for $N=4$. This implies that the real part of the gap function also acquires five nodal planes at large $\mathbf{k}$, which can be accommodated by the formation of four additional momentum space vortices with windings $+1$. The corresponding gap function in Figs.~\ref{fig:Cooper_vertex_Gap_Omega} (a4)-(a5) confirms this picture, showing one central vortex encircled by a concentric ring of four vortices.

These nodes in the Cooper vertex can be understood by expanding it in two different limits: $\mathcal{B}k_F^2\ll1$ and $\mathcal{B}k_F^2\gg1$, the details are worked out in Appendix~\ref{sec_app:harmonics of the Cooper vertex}. We also assume $\sigma^2k_F^2\ll1$ to control this expansion. Expanding the Cooper vertex in the dominant winding channel for momenta on the Fermi surface, we find that in the limit
$\mathcal{B}k_F^2\ll1$, $\sigma^2k_F^2\ll1$,
\begin{gather}
\text{Re}[\mathcal{W}_{\mathbf{k}_F,\mathbf{k}'_F}]\approx -\frac{Ue^{-\sigma^2k_F^2}\left(\frac{\sigma^2}{2}+\mathcal{B}\right)k_F^2}{\left( 1+\frac{\mathcal{B}}{2}k_F^2\right)^2}\cos({\phi_{\mathbf{k},\mathbf{k}'}}),
\end{gather}
where $\phi_{\mathbf{k},\mathbf{k}'}$ denotes the angle between the momenta. 
We find that the dominant (most negative) harmonic has winding $\ell=1$. This weak coupling result agrees with the gap equation in the limit $\mathcal{B}k_F^2\ll1$, where the vortex number, which is also winding of the gap around the Fermi surface, are equivalent, i.e., $\mathcal{Q}=\ell=1$. Similarly, in the limit $\mathcal{B}k_F^2\gg1$, $\sigma^2k_F^2\ll1$, the dominant channel of the Cooper vertex has
\begin{align}
    \text{Re}[\mathcal{W}_{\mathbf{k}_F,\mathbf{k}'_F}]\approx-\frac{Ue^{-\sigma^2k_F^2}\left(\frac{4(N-1)}{\mathcal{B}k_F^2}+\frac{\sigma^2 k_F^2}{2}\right)}{\left(1+\frac{2(N-1)}{\mathcal{B}k_F^2}\right)^2} \nonumber\\\ \times \cos{((2N-3)\phi_{\mathbf{k},\mathbf{k}'}}).
\end{align}
Here, the dominant harmonic is $\ell_{\text{sat}}=2N-3$, and is in agreement with the saturation of the vortex number $\mathcal{Q}_{\text{sat}}=\ell_{\text{sat}}=2N-3$.  Therefore, in both limits, the dominant harmonic $\ell$ from a weak-coupling analysis is consistent with the vortex number $\mathcal{Q}$ obtained from the full gap equation. However, a weak coupling analysis is agnostic to the location of these momentum space vortices present within the Fermi surface. Note that although Figs.~\ref{fig:trend_gap_N_8}-\ref{fig:Cooper_vertex_Gap_Omega} present numerical evidence for vortex saturation at $\sigma^2 k_F^2=1$, we also numerically observe the saturation in the regime $\sigma^2 k_F^2\ll 1$, in direct agreement with the weak-coupling analysis presented above.

Expanding the Cooper vertex in weak coupling for $\mathcal{B}k_F^2 \gg 1$ reproduces the numerically obtained value of $\mathcal{Q}_{\text{sat}}$. However, in Sec.~\ref{sec_BdG Topology} we provide a general analytical argument showing that the maximum number of momentum-space vortices that can be nucleated on the momentum-space gap function is related to the parent band Chern number ($C_P$) by the relation $\mathcal{Q}_{\text{sat}} = 2C_P - 1$. For the $\lambda_N$-model, we found in Sec.~\ref{GeometryofLambda} that $C_P = N - 1$, which predicts $\mathcal{Q}_{\text{sat}} = 2N - 3$, in agreement with our previous calculations.

\subsection{Radius of the vortex ring}
\begin{figure}
    \centering
    \includegraphics[scale=0.8]{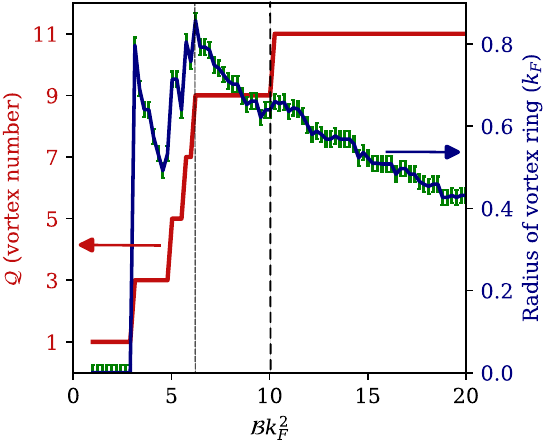}
    \caption{The left axis shows the evolution of $\mathcal{Q}$ with increasing $\mathcal{B}k_F^2$ for $N=7$. The right axis shows the radius of the vortex ring encircling the origin in units of $k_F$ ($k_F=1$). The bold dashed line separates the regions of vortex nucleation and saturation. The thin dashed line marks a discontinuity in $\mathcal{Q}$ and the associated kink in the vortex ring radius. The ring radius carries an uncertainty of $\pm0.018k_F$, arising from the finite grid resolution.}
    \label{fig:ring_radius}
\end{figure}
\begin{figure}[h!]
    \centering
    \hspace{-0.3cm}
    \includegraphics[scale=0.85]{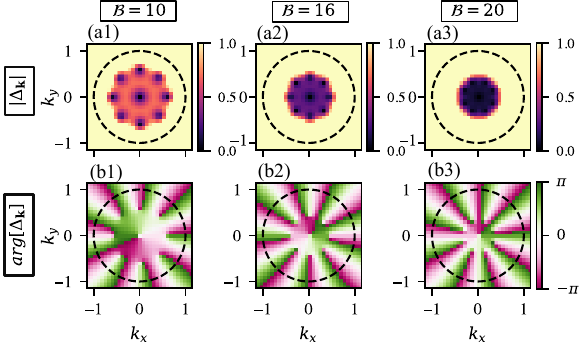}
    \caption{Columns (1)–(3) illustrate vortices migrating toward the origin with increasing $\mathcal{B}$ in the vortex number saturation regime. Row (a) displays the gap magnitude and row (b) its argument. Results are shown for $N=6$, the Fermi surface is denoted by the dashed line.}
    \label{fig:vortex_coalescence}
\end{figure}

We identified that the momentum-space vortices are nucleated on a ring enclosing the origin. The radius of this ring displays a notable trend with increasing $\mathcal{B}$. In Fig.~\ref{fig:ring_radius}, we track the radius of the ring of vortices for $N=7$, as $\mathcal{B}$ increases. We first observe that for a given interval of $\mathcal{B}$ where the vortex number $\mathcal{Q}$ is constant, the radius of the vortex ring shrinks with increasing $\mathcal{B}$, until $\mathcal{Q}$ jumps to a larger value. However, once vortex saturation is reached, the radius of the ring continues to decrease with increasing $\mathcal{B}$ as no further vortices can be nucleated. In this regime, the radius of the ring of $2N-4$ vortices contracts as the vortices migrate towards the origin. This coalescence is also shown in Fig.~\ref{fig:vortex_coalescence}, for $N=6$ as $\mathcal{B}$ increases past the vortex number saturation threshold.

In the vortex number saturation regime, the migration of the vortices towards the origin has a pronounced effect on the condensation energy, which decreases rapidly in this regime and tends to approach a common value in the large $\mathcal{B}$ limit, independent of $N$ (see Figs.~\ref{fig:vortex_nucleation}, \ref{fig:vortex_nucleation_coalescence_log_scale}). The origin of this trend toward an $N$-independent condensation energy for large $\mathcal{B}k_F^2$ was discussed in Sec.~\ref{Sec:IVB}. Additionally, this feature can also be understood using an ansatz for the gap function with the nucleated vortices arranged on a ring, with more details in Appendix~\ref{app_sec:ansatz_for_gap}. Using this ansatz, we find that as the radius of the vortex ring shrinks, the condensation energy decreases in agreement with condensation energy calculated from the numerical solutions.

\subsection{Fermi surface topology}
\begin{figure}
    \centering
    \hspace{-0.5cm}
    \includegraphics[scale=1.0]{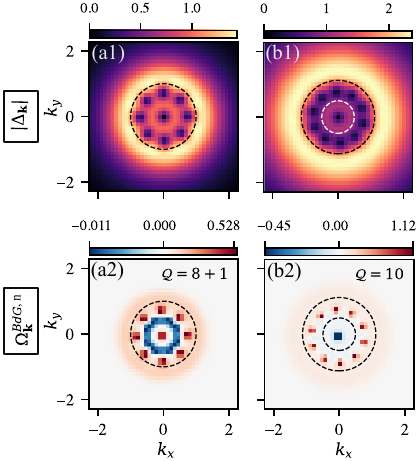}
    \caption{The gap structure for $N=8$ and $\mathcal{B}=8$  shown for two dispersions: (a1) a circular Fermi surface and (b1) an annular Fermi surface with two Fermi sheets. Panels (a2) and (b2) show the numerically calculated BdG Berry curvature $\Omega_{\mathbf{k}}^{BdG,n}$. For the circular Fermi surface in (a1), the vortex number counts the sum of vortex windings of all vortices enclosed by the Fermi surface, leading to $\mathcal{Q}=9$. For the annular Fermi surface, $\mathcal{Q}$ counts the vortex number enclosed between the two Fermi sheets, leading to $\mathcal{Q}=10$.}
    \label{fig:circular_annular_FS_BdG_Berry_curvature}
\end{figure}

Having thus far considered a circular Fermi surface, we now turn to the case of an annular Fermi surface, whose corresponding dispersion is shown in Fig.~\ref{fig:Figure_Lambda_N_BC_dist}(c). For an annular Fermi surface with inversion symmetry, the vortex number $\mathcal{Q}$ counts the sum of vortex windings enclosed within the occupied annular region. Since the anti-symmetry of the gap $\Delta(\mathbf{k})$ guarantees a winding $+1$ vortex at $\mathbf{k}=0$, the vortex number of the annular region must be even valued. We contrast the two cases: a circular, and an annular Fermi surface in Fig. \ref{fig:circular_annular_FS_BdG_Berry_curvature}. Additionally, we evaluate $\mathcal{Q}$ numerically based on the BdG Berry curvature (see details in Appendix~\ref{sec_app:numerical_chern_number}). For a circular Fermi surface, $\mathcal{Q}$ measures the total number of vortices enclosed by this single Fermi surface, while for an annular Fermi surface only the vortices enclosed within the annular region contribute. Therefore, provided the ring of vortices encircling the origin lies within the annular region, $\mathcal{Q}$ is non-zero and even-valued, a feature that would be absent from a weak-coupling analysis. This feature can lead to unique signatures of the momentum space vortices in thermal Hall measurements, as we shall discuss in Sec. \ref{sec_BdG Topology}.

We additionally find that an annular Fermi surface exhibits the same qualitative trends in vortex number $\mathcal{Q}$ with increasing $\mathcal{B}$ and $N$ as a circular Fermi surface (shown in Fig.~\ref{fig:vortex_nucleation}), again admitting a demarcation into regimes of vortex nucleation and vortex number saturation. The condensation energy also has a similar behavior with increasing $\mathcal{B}$. 
%Additional data for the annular Fermi surface can be found in Appendix~\ref{app:annular_FS_results}.

\section{Topology in Band Projected Superconductivity} \label{sec_BdG Topology}
In Sec.~\ref{sec:results_gap_eqn}, we observed the formation of momentum-space vortices in the order parameter $\Delta(\vb{k})$. As noted earlier, a vortex in two dimensions is a topological defect characterized by an integer winding number around its core. This raises the natural question of whether the net winding of vortices in the gap is directly related to the topology of the superconductor. The chiral superconducting states considered here break time-reversal symmetry and, in the absence of additional protecting symmetries, fall in the symmetry class D of superconductivity \cite{Altland_1997,Read_2000, Schnyder_2008,Chiu_2016,Kitaev_2009}. This class is characterized by a $\mathbb{Z}$ topological invariant. For the two-dimensional continuum models studied here, this invariant is the Bogoliubov–de Gennes Chern number \cite{SatoTopology,thermalHallSumiyoshi}, namely the Chern number of the occupied BdG quasiparticle bands. The goal of this section is to derive the relation between momentum-space vortex winding and the BdG Chern number. In band-projected superconductivity, this invariant is often evaluated using an effective $2 \times 2$ BdG Hamiltonian for the active band. Although this representation correctly captures the quasiparticle spectrum, it hides an important gauge structure inherited from the parent Bloch band. As a result, a naive Berry-curvature calculation in the $2 \times 2$ band basis is not manifestly invariant under parent-band gauge transformations. The issue is that the projected gap $\Delta(\vb{k})$ is not itself gauge invariant, but instead transforms like a complex scalar field coupled to the pair Berry connection, as seen in Eq.~\eqref{eq_gapgaugetransform}. Therefore, before relating vortex winding to BdG topology, one must first define the BdG Berry curvature in a form that is invariant under parent-band gauge transformations.

We accomplish this by formulating the problem in the full orbital basis, where parent-band gauge invariance is manifest. We show below that this formulation reveals a gauge-invariant momentum-space phase current whose curl gives the occupied BdG Berry curvature. The current is the momentum-space analogue of the Meissner current, with the pair Berry connection playing the role of the electromagnetic vector potential. Its curl obeys a London-type equation, making precise the analogy between momentum-space vortices and real-space vortices in a type-II superconductor. The topology follows from the boundary structure of this current. Since the current vanishes at large momentum, the outer boundary gives no contribution, so the Chern number reduces to circulation integrals of the current around vortex cores. This gives the central result: the BdG Chern number is the sum of vortex windings in the occupied region. This occupied vortex charge generalizes the weak-coupling angular-momentum formula for chiral topological superconductors \cite{Read_2000} and controls the thermal Hall response. For multiple Fermi surfaces bounding an annular occupied region, this gives a result absent from the weak-coupling angular momentum picture: the BdG Chern number can take even integer values when a ring of vortices lies inside the annulus. Details of these constructions are explained in the following subsections.

\subsection{Gauge Ambiguity of the Low Energy Model} 
Before constructing the gauge-invariant orbital formulation, we first identify the gauge ambiguity in the usual projected $2 \times 2$ description. The projected low-energy BdG Hamiltonian is
\begin{align}
    H_{\text{BdG}}(\vb{k})=\begin{pmatrix}
        \xi_{\vb{k}} & \Delta(\vb{k}) \\
        \Delta^{*}(\vb{k}) & -\xi_{-\vb{k}} 
    \end{pmatrix}. \label{BdG2by2}
\end{align}
This is the standard band-basis description of quasiparticles in a superconductor projected into an isolated parent band. The order parameter $\Delta(\vb{k})$ is the same band-projected order parameter introduced in Eq.~\eqref{gapeq}. A representative occupied BdG eigenstate is
\begin{align}
    \ket{\Psi^{-}_{\vb{k}}}=\begin{pmatrix}
        -v^{*}_{\vb{k}} \\
        u^{*}_{-\vb{k}} 
    \end{pmatrix}. \label{eq_2x2eigenstate}
\end{align}
Here 
\begin{align}
    E_{\vb{k}}=\sqrt{\xi_{\vb{k}}^2+\abs{\Delta(\vb{k})}^2}\label{eq_quasiparticleenergy}
\end{align}
is the positive quasiparticle energy, and the coherence factors satisfy
\begin{align}
    u_{\vb{k}}&=e^{i\theta_{u}(\vb{k})}\sqrt{\frac{1}{2}\left(1+\frac{\xi_{\vb{k}}}{E_{\vb{k}}}\right)}, \\ v_{\vb{k}}&=e^{i\theta_{v}(\vb{k})}\sqrt{\frac{1}{2}\left(1-\frac{\xi_{\vb{k}}}{E_{\vb{k}}}\right)}.
\end{align}
The phases $\theta_{u}(\vb{k})$ and $\theta_{v}(\vb{k})$ are left arbitrary, since their transformation properties are central to the gauge issue below. The corresponding occupied energy is
\begin{align}
    E_{\vb{k}}^{-}=\bra{\Psi^{-}_{\vb{k}}}H_{\text{BdG}}(\vb{k})\ket{\Psi^{-}_{\vb{k}}}=-E_{\vb{k}}. \label{eq_spectrum}
\end{align}
Thus the spectrum depends only on the gauge-invariant magnitude $\abs{\Delta(\vb{k})}$. The phase of $\Delta(\vb{k})$, however, is not itself gauge invariant. Under a gauge transformation of the parent band
\begin{align}
    \ket{s_{\vb{k}}} \rightarrow e^{i\chi(\vb{k})}\ket{s_{\vb{k}}},
\end{align}
the projected gap transforms as
\begin{align}
    \Delta(\vb{k}) \rightarrow e^{-i(\chi(\vb{k})+\chi(-\vb{k}))}\Delta(\vb{k}).
\end{align}
Accordingly, the $2 \times 2$ BdG Hamiltonian is not invariant as a fixed matrix; it transforms covariantly under a momentum-dependent change of the projected Nambu basis. The occupied eigenstate in Eq.~\eqref{eq_2x2eigenstate}, therefore transforms as
\begin{align}
    \ket{\Psi^{-}_{\vb{k}}} \rightarrow \begin{pmatrix}
        e^{i\chi(\vb{k})} & 0 \\
        0 & e^{-i\chi(-\vb{k})}
    \end{pmatrix} \ket{\Psi^{-}_{\vb{k}}}. \label{eq_2x2Gauge}
\end{align}
This leaves the expectation value in Eq.~\eqref{eq_spectrum} invariant. However, it affects any Berry curvature computed directly from $\ket{\Psi^{-}_{\vb{k}}}$. For an isolated physical BdG band, the only intrinsic eigenstate freedom should be the local $U(1)$ phase. Instead, $\ket{\Psi^{-}_{\vb{k}}}$ transforms under the diagonal parent-band frame rotation in Eq.~\eqref{eq_2x2Gauge}, which acts differently on its particle and hole components. Therefore, the Berry curvature derived from Eq.~\eqref{BdG2by2}, inherits this frame dependence. This Berry curvature is defined as $\Tilde{\mathcal{F}}(\vb{k})=i\epsilon^{\mu\nu}\braket{\partial_{\mu}\Psi^{-}_{\vb{k}}}{\partial_{\nu}\Psi^{-}_{\vb{k}}}$. Under Eq.~\eqref{eq_2x2Gauge}, it transforms as 
\begin{align}
    \tilde{\mathcal{F}}(\vb{k})\rightarrow  \tilde{\mathcal{F}}(\vb{k})+\nabla_{\vb{k}}\wedge \vb{\omega}(\vb{k})\label{eq_2x2shift},
\end{align}

where $\vb{\omega}(\vb{k})$ is a vector field generated by the parent-band gauge transformation:
\begin{align}
    \vb{\omega}(\vb{k})=\abs{u_{-\vb{k}}}^2 \nabla_{\vb{k}}\chi(-\vb{k})-\abs{v_{\vb{k}}}^2 \nabla_{\vb{k}}\chi(\vb{k}). \label{eq_gaugefunctionbad}
\end{align}
If $\omega(\vb{k})$ were globally smooth and single valued on the integration domain, the additional curl term in Eq.~\eqref{eq_2x2shift} would integrate to a vanishing boundary contribution on a closed compactified manifold. In that case, the integrated Chern number $\tilde{C}$ would be unchanged. The subtlety is that, in the presence of gap vortices, the phase of $\Delta(\vb{k})$ is not differentiable at the vortex cores. The Berry curvature must therefore be evaluated on a punctured domain, and singular parent-band frame transformations can produce finite boundary contributions around the punctures. We show below that these boundary terms shift the Chern number $\tilde{C}$

The Chern number $\tilde{C}$ is obtained by integrating the Berry curvature over a disk $D_{\Lambda}$ of radius $\Lambda$, followed by the limit $\Lambda \rightarrow \infty$. When this limit exists, we denote the limiting domain by $D_{\infty}$, so that

\begin{align}
    \tilde{C}=\int_{D_{\infty}} \frac{d^2\vb{k}}{2\pi}\tilde{\mathcal{F}}(\vb{k}).
\end{align}
The existence of this limit is tied to a one-point compactification of the model's domain to the sphere 
\begin{align}
    \mathbb{S}^2=\mathbb{R}^2 \cup \{\infty\},
\end{align}
as discussed in Appendix~\ref{sec_app:onepointcompactification} and \ref{sec_app:BdGCompactification}. For the continuum models considered here, this compactification is valid provided the occupied BdG projector ($\ket{\Psi^{-}_{\vb{k}}}\bra{\Psi^{-}_{\vb{k}}}$) approaches a unique limit at large momentum. Equivalently, the outer boundary contribution from $\partial D_{\Lambda}$ vanishes as $\Lambda \rightarrow \infty$. 

The outer boundary, however, is not the only boundary relevant to the problem. Since the Berry curvature $\tilde{\mathcal{F}}$ depends on the momentum-space phase of $\Delta(\vb{k})$, it is singular at vortex cores, where the phase is not differentiable. We therefore remove small disks $D_{i}(\epsilon)$ of radius $\epsilon$ centered at each vortex position $\vb{k}_{i}$ ($i=1,\cdots , N_{V}$). The Berry curvature is then evaluated on the punctured domain
\begin{align}
    D_{\Lambda,\epsilon}=D_{\Lambda} \setminus \bigcup_{i=0}^{N_V-1}D_{i}(\epsilon), \label{eq_punctureddomain}
\end{align}
whose boundary is
\begin{align}
    \partial{D_{\Lambda, \epsilon}}=\partial D_{\Lambda} \cup \bigcup_{i=0}^{N_V-1}\left(-\partial D_{i}(\epsilon)\right). \label{eq_punctureboundary}
\end{align}
On this punctured domain, singular parent-band gauge transformations can wind around removed vortex cores. This is the same mechanism as the Wu--Yang description of a monopole bundle \cite{WuYang}, adapted here to a compactified continuum Chern band. As reviewed in Appendix~\ref{sec_app:singulargauge}, when $C_P \neq 0$, the parent-band state $\ket{s_{\vb{k}}}$ cannot be chosen smoothly everywhere on the compactified momentum sphere. In any single gauge patch, this obstruction appears as phase winding of $\ket{s_{\vb{k}}}$ on a boundary. In one gauge, this winding may be placed at infinity; after a singular change of gauge, the same winding may instead be placed at a finite puncture. Because the vortex core has been removed from the domain, such a singular gauge transformation is well defined on the punctured domain. In the $\lambda_N$ model used here, the parent-band state is smooth on the finite momentum plane, but winds by $-C_P$ at infinity (as shown in Appendix~\ref{sec_app:lambdaN}). We can remove this winding by the following gauge transformation on the punctured domain
\begin{align}
    \ket{s_{\vb{k}}} \rightarrow e^{iC_P \phi(\vb{k})}\ket{s_{\vb{k}}}, \label{eq_windinggauge}
\end{align}
where $\phi(\vb{k})=\text{arg}(k_x+ik_y)$. As shown in Appendix~\ref{sec_app:WuYang}, this transformation does not change the parent-band Chern number. It only transfers the phase winding of the chosen parent-band state from infinity to the origin. At the level of the BdG Berry curvature $\tilde{\mathcal{F}}$, defined in Eq.~\eqref{eq_2x2shift}, this transformation makes the vector field $\vb{\omega}(\vb{k})$ in Eq.~\eqref{eq_gaugefunctionbad} wind nontrivially around the origin, such that the Chern number changes. Integrating Eq.~\eqref{eq_2x2shift}, we have
\begin{align}
    \tilde{C} \rightarrow  \tilde{C}+\delta \tilde{C},
\end{align}
where the shift $\delta \tilde{C}$ is given by
\begin{align}
    \delta \tilde{C}=\lim_{\epsilon \rightarrow 0}\int_{D_{\infty, \epsilon}}\frac{d^2\vb{k}}{2\pi}\nabla_{\vb{k}} \wedge \omega(\vb{k}).
\end{align}
To explicitly evaluate $\delta \tilde{C}$, we use Green's theorem on the punctured domain to obtain
\begin{align}
    \delta \tilde{C}=\oint_{\partial D_{\infty}} \frac{d\vb{k}}{2\pi}\cdot \vb{\omega}(\vb{k})-\lim_{\epsilon\rightarrow 0}\oint_{\partial D_0(\epsilon)}\frac{d\vb{k}}{2\pi}\cdot \vb{\omega}(\vb{k}). \label{eq_deltaC2x2}
\end{align}
Here $\partial D_{0}(\epsilon)$ denotes a small circle around the origin enclosing the winding $C_P$ introduced in Eq.~\eqref{eq_windinggauge}. The shift in $\tilde{C}$ is therefore controlled by the boundary circulation of $\vb{\omega}$. We explicitly evaluate $\delta \tilde{C}$ for this gauge transformation. The field $\omega(\vb{k})$ takes the form
\begin{align}
    \omega(\vb{k})=C_P(\abs{u_{-\vb{k}}}^2-\abs{v_{\vb{k}}}^2)\nabla_{\vb{k}}\phi(\vb{k}).
\end{align}
At infinity, the coherence factors satisfy $\abs{u_{-\vb{k}}}^2 \rightarrow 1$ and  $\abs{v_{\vb{k}}}^2 \rightarrow 0$, so the outer boundary integral contributes
\begin{align}
    \oint_{\partial D_{\infty}} \frac{d\vb{k}}{2\pi} \cdot \omega(\vb{k})=C_P.
\end{align}
Near the origin, assuming the central vortex lies inside the occupied region $\xi_{\vb{k}} <0$, the coherence factors obey the opposite limit: $\abs{u_{-\vb{k}}} \rightarrow 0$ and $\abs{v_{\vb{k}}}^2 \rightarrow 1$. Therefore, the inner boundary integral contributes
\begin{align}
    \oint_{\partial D_{0}(\epsilon)} \frac{d\vb{k}}{2\pi} \cdot \omega(\vb{k})=-C_P,
\end{align}
or equivalently, the circulation around $-\partial D_{0}(\epsilon)$ is $+C_P$. Combining the outer and inner boundary terms using Eq.~\eqref{eq_deltaC2x2} gives
\begin{align}
    \delta \tilde{C}=C_P-(-C_P)=2C_P.
\end{align}
This shows that the Chern number obtained using the eigenstate of the $2\times 2$ Hamiltonian (in Eq.~\eqref{BdG2by2}) is not invariant under parent-band gauge transformations on the punctured domain. The physical parent band and the quasiparticle spectrum have not changed; what has changed is the distribution of phase winding between the projected gap and the parent-band state. The corresponding Berry curvature is sensitive to this distribution, and can therefore convert a change of parent-band frame into an apparent change of Chern number.

This fact also explains why previous works using the effective band projected Hamiltonian similar to Eq.~\eqref{BdG2by2} need not be incorrect. If the parent-band state is chosen smoothly on a disk enclosing the occupied region and pairing window, using Eq.~\eqref{BdG2by2} can reproduce the correct Chern number. However, that agreement relies on an implicit gauge choice. A related gauge-covariance issue was identified in Ref.~\cite{sedov2025probingsuperconductivitytunnelingspectroscopy}, where an orbital-basis with a restricted pairing was proposed. Our purpose below is different: starting from the orbital-basis formulation, we derive a generic expression for the gauge-invariant BdG Berry curvature in band projected superconductivity, and reduce the full Chern number to a vortex-counting rule.

\subsection{The Orbital Basis}
Motivated by the gauge ambiguity of the previous section, we now formulate the same projected superconductor in the orbital basis. Here orbitals refer to the internal degrees of freedom of the $\lambda_N$-model as was identified earlier in Sec.~\ref{sec:recap}. The advantage of this basis is twofold: parent-band gauge invariance is manifest, and the inert higher bands are retained, and have consequences on the resulting geometry and topology of the state. The generic form of the BdG Hamiltonian in the orbital basis is the $2N \times 2N$ matrix 
\begin{align}
    H_{\mathcal{O}}(\vb{k})=\begin{pmatrix}
        h(\vb{k}) & \Delta_{\mathcal{O}}(\vb{k}) \\
        \Delta_{\mathcal{O}}^{\dag}(\vb{k}) & -h^{T}(-\vb{k})
    \end{pmatrix}.
\end{align} 
The orbital basis, is denoted by $\mathcal{O}$. In what follows, $h(\vb{k})$ is an $N \times N$ normal-state Hamiltonian with an isolated parent band. The order parameter in the orbital basis is the $N \times N$ matrix $\Delta_{\mathcal{O}}(\vb{k})$. If we restrict pairing to the parent band $\ket{s_{\vb{k}}}$, then the order parameter takes the form
\begin{gather}
    \Delta_{\mathcal{O}}(\mathbf{k})=\Delta(\vb{k})\ket{s_{\vb{k}}}\bra{s^{*}_{-\vb{k}}},
\end{gather}
as shown in Appendix~\ref{sec_app:BdGHamiltonian}. The Bloch state pair $\ket{s_{\vb{k}}}\bra{s^{*}_{-\vb{k}}}$ carries exactly the opposite parent-band gauge phase to the scalar projected gap $\Delta(\vb{k})$. Indeed, under $\ket{s_{\vb{k}}} \rightarrow e^{i\chi(\vb{k})}\ket{s_{\vb{k}}}$, we have
\begin{align}
    \ket{s_{\vb{k}}}\bra{s^{*}_{-\vb{k}}} \rightarrow e^{i(\chi(\vb{k})+\chi(-\vb{k}))}\ket{s_{\vb{k}}}\bra{s^{*}_{-\vb{k}}},
\end{align}
so that their product, $\Delta_{\mathcal{O}}(\vb{k})$, is gauge invariant.

The $2N \times 2N$ Hamiltonian has $N$ negative-energy occupied states. With pairing restricted to the parent band, only one of these is an active superconducting BdG state
\begin{align}
    \ket{\Psi_{0\vb{k}}^{-}}=-v^{*}_{\vb{k}}\ket{s_{\vb{k}}} \oplus u^{*}_{-\vb{k}}\ket{s^{*}_{-\vb{k}}},\label{orbitalBdGEigenstate}
\end{align}
with quasiparticle energy $-E_{\vb{k}}$. This active state is invariant under parent-band gauge transformations, up to the usual overall BdG $U(1)$ phase, because the coherence factors transform oppositely to the Bloch states. The remaining $N-1$ occupied inert hole states take the form
\begin{align}
    \ket{\Psi^{-}_{n\vb{k}}}=0\oplus \ket{s^{*}_{n,-\vb{k}}}, \label{inertstate}
\end{align}
where $n=1,\cdots , N-1$ labels the normal-state bands other than the active parent band. In what follows, we assume that the active superconducting BdG band remains isolated from the inert states. A sufficient condition is
\begin{align}
    J \gg \text{max}_{\vb{k}}\abs{\Delta(\vb{k})},
\end{align}
where $J$ is the minimum normal-state band gap. 

In summary, the orbital-basis construction gives a gauge-invariant representation of band projected superconductivity as a $2N \times 2N$ BdG problem with pairing restricted to the parent band. The active sector has the same quasiparticle dispersion as the projected $2 \times 2$ model and contains the same gap-phase information. The difference is that the orbital-basis formulation is manifestly invariant under parent-band gauge transformations and retains the inert occupied hole states. In the next subsection, we show that these inert contributions play a key role in determining the BdG Berry curvature. 

\subsection{BdG Berry Curvature: Momentum Space Phase Current}
We now compute the full occupied BdG Berry curvature and make the role of the inert-bands explicit. Since the occupied subspace contains both the active superconducting state and the inert hole states, it is convenient to use the projector formula \cite{MitscherlingQG}
\begin{align}
    \mathcal{F}_{\text{BdG}}(\vb{k})=i\epsilon^{\mu\nu}\text{Tr}[P(\vb{k})\partial_{\mu}P(\vb{k})\partial_{\nu}P(\vb{k})]. \label{eq_ProjectorBerrycurvature}
\end{align}
Here $\epsilon^{\mu\nu}$ is the antisymmetric symbol, $\partial_{\mu} \equiv \partial/\partial k^{\mu}$, and $P(\vb{k})$ denotes the occupied BdG projector,
\begin{align}
    P(\vb{k})=\sum_{E_{n\vb{k}}<0}\ket{\Psi_{n\vb{k}}}\bra{\Psi_{n\vb{k}}}.
\end{align}
Under the condition $J \gg \text{max}_{\vb{k}}\abs{\Delta(\vb{k})}$, the active superconducting BdG band remains separated from the inert hole bands. The projector decomposes as
\begin{align}
    P(\vb{k})=P_{\text{SC}}(\vb{k})+P_{h}(\vb{k}),
\end{align}
where $P_{\text{SC}}(\vb{k})=\ket{\Psi^{-}_{0\vb{k}}}\bra{\Psi^{-}_{0\vb{k}}}$, and $P_h(\vb{k})$ projects onto the $N-1$ inert occupied hole states $0 \oplus \ket{s_{n,-\vb{k}}^{*}}$. The projector formula (Eq.~\eqref{eq_ProjectorBerrycurvature}) separates into an active superconducting contribution and an inert-hole contribution,
\begin{align}
    \mathcal{F}_{\text{BdG}}(\vb{k})=\mathcal{F}_{\text{SC}}(\vb{k})+\mathcal{F}_{h}(\vb{k}).
\end{align}
The active contribution comes from $P_{\text{SC}}(\vb{k})$. As derived in Appendix~\ref{sec_app:BdGBerry}, the Berry curvature of the active sector is
\begin{align}
    \mathcal{F}_{\text{SC}}(\vb{k})=-\nabla_{\vb{k}}\wedge \left(\abs{v_{\vb{k}}}^2 \vb{v}_{\Delta}(\vb{k})\right)-\Omega(-\vb{k}). \label{SCcontribution}
\end{align}
Here $\vb{v}_{\Delta}(\vb{k})$ is the gauge-invariant phase velocity, defined in Eq.~\eqref{eq_momentumspacevelocity}, and $-\Omega(-\vb{k})$ is the residual curvature of the active parent band hole components. The full occupied BdG subspace, however, is not exhausted by this active state. It also contains the $N-1$ inert hole states, encoded in the projector $P_h(\vb{k})$. As shown in Appendix~\ref{sec_app:inertholecancellation}, the inert contribution to the BdG Berry curvature, $\mathcal{F}_h(\vb{k})$, cancels the residual parent-band Berry curvature in Eq.~\eqref{SCcontribution}. The cancellation follows from completeness of the parent-band hole sector: the Berry curvature of the complete set of hole bands vanishes because their projector is identity. Hence the cancellation occurs between the active parent band hole components which contributes $-\Omega(-\vb{k})$, and the remaining inert hole bands which contribute $\mathcal{F}_h(\vb{k})=\Omega(-\vb{k})$. The full occupied BdG Berry curvature is therefore
\begin{align}
    \mathcal{F}_{\text{BdG}}(\vb{k})=-\nabla_{\vb{k}} \wedge \vb{J}_{\Delta}(\vb{k}),
\end{align}
where the gauge-invariant vector field $\vb{J}_{\Delta}(\vb{k})$ is given by
\begin{align}
    \vb{J}_{\Delta}(\vb{k})=\abs{v_{\vb{k}}}^2\left(\nabla_{\vb{k}}\text{arg}[\Delta(\vb{k})]-\mathcal{A}_{P}(\vb{k})\right). \label{eq_phasecurrent}
\end{align}
Eq.~\eqref{eq_phasecurrent} should not be confused with a $U(1)$ Berry connection. It is a gauge-invariant phase current: the phase gradient of the projected gap measured relative to the pair Berry connection of the parent band, and weighted by the BdG occuptation factor $\abs{v_{\vb{k}}}^2$. 

The terminology is motivated by its close analogy with the Meissner current. Under a parent-band gauge transformation, the phase gradient $\nabla_{\vb{k}}\text{arg}[\Delta(\vb{k})]$ and the pair Berry connection $\mathcal{A}_{P}(\vb{k})$ shift by the same amount, namely $-\nabla_{\vb{k}}\left(\chi(\vb{k})+\chi(-\vb{k})\right)$. Their difference is therefore invariant. Thus phase windings can be redistributed between the gap and the parent-band connection, while $\vb{J}_{\Delta}(\vb{k})$ itself remains unchanged. This is the same gauge-invariant structure that appears in the Meissner current of a real-space superconductor.

The factor $\abs{v_{\vb{k}}}^2$ also controls the large-momentum behaviour of the phase current. In the continuum models considered here, $\abs{v_{\vb{k}}}^2 \rightarrow 0$ as $\abs{\vb{k}}\rightarrow \infty$. Thus, even if there is a vortex at infinity, its winding is suppressed in $\vb{J}_{\Delta}(\vb{k})$. In this sense, the phase current has no outer-boundary circulation in the compactification limit. In the next section, we define the total vorticity of the gap and seperate this quantity from the BdG-weighted phase current.

\subsection{Vorticity and the Singular Connection}
The phase current introduced above contains two conceptually distinct ingredients: the coherence factor $\abs{v_{\vb{k}}}^2$, which is determined by the BdG spectrum and gives a weight to the phase response, and the gauge-invariant phase velocity $\vb{v}_{\Delta}(\vb{k})$ which contains the vorticity of the gap $\Delta(\vb{k})$ and its coupling to the parent band's topology. Here, we isolate the second ingredient, and show that the total vorticity of the gap on the compactified momentum sphere is fixed by the parent-band Chern number. 

The appropriate formalism to understand the source of vorticity in the gap is the singular-connection approach \cite{MeraSingularConnection, paiva2024shiftpolarizationexcitonsquantum}. Away from zeros of $\Delta(\vb{k})$, we define the unit-normalized gap $\hat{\Delta}(\vb{k})=\Delta(\vb{k})/\abs{\Delta(\vb{k})}$. This isolates the phase of the gap while discarding its amplitude. Since $\Delta(\vb{k})$ transforms like a charged scalar field under parent-band gauge transformations, $\hat{\Delta}(\vb{k})$ transforms in the same way. We therefore define the singular connection associated with the gap by
\begin{align}
    \mathcal{A}_{\Delta}(\vb{k})=-i\hat{\Delta}^{*}(\vb{k})\nabla_{\vb{k}}\hat{\Delta}(\vb{k})=\nabla_{\vb{k}}\text{arg}[\Delta(\vb{k})].
\end{align}
Under parent-band gauge transformations, this connection shifts in the same way as the pair Berry connection $\mathcal{A}_{P}(\vb{k})$. Unlike $\mathcal{A}_{P}(\vb{k})$, the singular connection is globally defined, but with singularities at vortex cores. The singularities of $\mathcal{A}_{\Delta}(\vb{k})$ act as sources of the singular curvature $\mathcal{F}_{\Delta}(\vb{k})=\nabla_{\vb{k}}\wedge \mathcal{A}_{\Delta}(\vb{k})$. Away from vortex cores, $\mathcal{A}_{\Delta}(\vb{k})$ is locally a pure gradient, so $\mathcal{F}_{\Delta}(\vb{k})=0$ here. At a vortex core, however, the phase is not single-valued. The curvature is therefore concentrated at the zeros of $\Delta(\vb{k})$ and should be understood distributionally:
\begin{align}
    \mathcal{F}_{\Delta}(\vb{k})=\sum_{i=0}^{N_V-1}2\pi \ell_i \delta^{(2)}(\vb{k}-\vb{k}_i).
\end{align}
Here, $\ell_i$ is the winding of a vortex located at position $\vb{k}_i$. The total vorticity is obtained by integrating $\mathcal{F}_{\Delta}(\vb{k})$ over the compactified sphere $\mathbb{S}^2$,
\begin{align}
    V_{\Delta}=\int_{\mathbb{S}^2} \frac{d^2\vb{k}}{2\pi}\mathcal{F}_{\Delta}(\vb{k})=\sum_{i=0}^{N_V-1}\ell_i. \label{eq_vorticity}
\end{align}
To relate the vorticity of the gap $V_{\Delta}$ to the parent band Chern number $C_P$, we compute the curl of the phase velocity $\vb{v}_{\Delta}(\vb{k})$:
\begin{align}
    \nabla_{\vb{k}} \wedge \vb{v}_{\Delta}(\vb{k})= \mathcal{F}_{\Delta}(\vb{k})-\mathcal{F}_{P}(\vb{k}), \label{eq_Londonssecondeq}
\end{align}
where $\mathcal{F}_P(\vb{k})$ is the pair Berry curvature of the parent band. Eq.~\eqref{eq_Londonssecondeq} is the momentum-space analogue of London's second equation for a type-II superconductor. It states that the vorticity is set by a competition between the pair Berry curvature and the singular vortex charge encoded in the singular curvature. To determine the net vorticity, we integrate Eq.~\eqref{eq_Londonssecondeq} over the sphere. Since $\vb{v}_{\Delta}(\vb{k})$ is the gauge-invariant difference between the singular connection and the pair Berry connection, its total curl integrates to zero on the closed sphere. Because $\vb{v}_{\Delta}(\vb{k})$ is singular at the zeros of $\Delta(\vb{k})$, this statement must be understood by excising small disks around the vortex cores and taking the shrinking-puncture limit; a rigorous derivation is given in Appendix~\ref{sec_app:vorticityproof}. The result is
\begin{align}
    0=\int_{\mathbb{S}^2} \frac{d^2 \vb{k}}{2\pi} \nabla_{\vb{k}} \wedge \vb{v}_{\Delta}(\vb{k})=V_{\Delta}-\int_{\mathbb{S}^2} \frac{d^2\vb{k}}{2\pi}\mathcal{F}_{P}(\vb{k}). 
\end{align}
Since the pair Berry curvature has flux $2C_P$ through the sphere, we obtain
\begin{align}
    V_{\Delta}=2C_P. \label{eq_netvorticity}
\end{align}
Eq.~\eqref{eq_netvorticity} gives the precise sense in which the parent-band topology sources vorticity in the projected gap. The pair Berry curvature carries total flux $2C_P$, and the singular connection of $\Delta(\vb{k})$ must supply the same total vorticity on the compactified sphere. In this way, the Chern number of the parent band is converted into momentum-space vortices of the gap. This global constraint, however, does not fix how the vorticity is distributed. A further constraint is imposed by the fermionic antisymmetry of the gap, $\Delta(-\vb{k})=-\Delta(\vb{k})$. At inversion fixed points along the compactified sphere, the gap must then vanish. The first point is the origin $\vb{k}=\vb{0}$. Since the gap is an odd function $\vb{k}$, the integral of its singular connection along an inversion symmetric loop enclosing the origin must give an odd integer. Thus, there must always be a vortex of odd winding at the origin. Similarly, since the point at infinity is included as a single point in the domain, $\vb{k}=\infty$ is an inversion fixed point. Therefore, a vortex of odd winding must be found at infinity as well. This explains why the finite-plane vortex count observed in Sec.~\ref{sec:mom_vort} can saturate at  $\mathcal{Q}_{\text{sat}}=2N-3$. The $\lambda_N$ model has Chern number $C_P=N-1$, so that the net vorticity of the gap is $V_{\Delta}=2N-2$. This would imply that the vortex at infinity carries unit winding, leaving $2C_P-1=2N-3$ vortices visible in the finite plane. The fact that $2C_P-1$ is an odd integer is consistent with the fact that around a sufficiently large but finite loop enclosing the origin, the net vorticity must always be odd. 

While Eq.~\eqref{eq_netvorticity} fixes the total vorticity of the gap, it does not yet determine BdG Chern number. The BdG invariant is sensitive only to vortices weighted by the coherence factors, or equivalently to the phase current $\vb{J}_{\Delta}(\vb{k})$. We now show how this weighting selects vortices in the occupied region and gives the BdG Chern number.

\subsection{BdG Chern Number: Flux Quantization} \label{sec_thermalhall}
Having shown that the total gap vorticity is $2C_P$, we now determine which vortices contribute to the BdG Chern number, and infer the associated thermal Hall conductivity. While the total vorticity is controlled by $\vb{v}_{\Delta}(\vb{k})$, the BdG Chern number is controlled by $\vb{J}_{\Delta}(\vb{k})=\abs{v_{\vb{k}}}^2 \vb{v}_{\Delta}(\vb{k})$. The coherence factor $\abs{v_{\vb{k}}}^2$ acts as an occupation weight, selecting which vortex circulations contribute to $C_{\text{BdG}}$. At a vortex core $\vb{k}_i$, where $\abs{\Delta(\vb{k}_i)}=0$, and assuming $\xi_{\vb{k}_i} \neq 0$, it reduces to
\begin{align}
    \abs{v_{\vb{k}_i}}^2
    =\frac{1}{2}\left(1-\frac{\xi_{\vb{k}_i}}{\abs{\xi_{\vb{k}_i}}}\right)
    =\begin{cases}
        1, & \xi_{\vb{k}_i}<0,\\[0.1cm]
        0, & \xi_{\vb{k}_i}>0.
    \end{cases} \label{eq_coherencecore}
\end{align}
Therefore, vortices within the occupied region of the Fermi sea, defined by
\begin{align}
    D_{\text{occ}}=\{\vb{k} \in \mathbb{R}^2 \text{ }| \text{ } \xi_{\vb{k}}<0 \},
\end{align}
contribute to the circulation of $\vb{J}_{\Delta}(\vb{k})$, while vortices outside this region are suppressed due to the coherence factor. Here we assume that no vortex lies exactly on the Fermi surface (or $\partial D_{\text{occ}}$). If a vortex crosses the Fermi surface, then the quasi-particle gap closes as both $\xi_{\vb{k}_i}=0$ and $\Delta(\vb{k}_i)=0$, this allows the BdG Chern number to jump.

The BdG Chern number is obtained by integrating the BdG Berry curvature over the punctured compactified domain $D_{\infty,\epsilon}$:
\begin{align}
    C_{\text{BdG}}=\lim_{\epsilon \rightarrow 0}\int_{D_{\infty,\epsilon}} \frac{d^2\vb{k}}{2\pi}\mathcal{F}_{\text{BdG}}(\vb{k}).
\end{align}
Since $\mathcal{F}_{\text{BdG}}(\vb{k})=-\nabla_{\vb{k}} \wedge \vb{J}_{\Delta}(\vb{k})$, Green's theorem reduces the Chern number to the circulation of $\vb{J}_{\Delta}(\vb{k})$ along the boundary of the punctured domain $D_{\infty,\epsilon}$. Using the boundary orientation in Eq.~\eqref{eq_punctureboundary}, and writing $\partial D_i(0)$ as shorthand for the $\epsilon \rightarrow 0$ limit of $\partial D_i(\epsilon)$, we obtain
\begin{align}
    C_{\text{BdG}}=-\oint_{\partial D_{\infty}}\frac{d\vb{k}}{2\pi}\cdot \vb{J}_{\Delta}(\vb{k})+\sum_{i=0}^{N_V-1}\oint_{\partial D_{i}(0)} \frac{d\vb{k}}{2\pi}\cdot \vb{J}_{\Delta}(\vb{k}). \label{eq_BdGChernExpression}
\end{align}
The outer boundary contribution vanishes in the $\Lambda \rightarrow \infty$ limit. In the continuum models considered here, this is ensured by $\abs{v_{\vb{k}}}^2 \rightarrow 0$ as $\abs{\vb{k}}\rightarrow \infty$, which suppresses $\vb{J}_{\Delta}(\vb{k})$ even if the phase of $\Delta(\vb{k})$ carries winding at infinity. This is the boundary form of the one-point compactification condition: after compactification, the sphere has no outer boundary, so the phase-current circulation at $\partial D_{\infty}$ must vanish. This condition is analogous to flux quantization in a real-space superconductor: far from any real-space vortices, the vanishing of the super-current allows one to relate the quantized phase winding of the order parameter to the magnetic flux. Here, the vanishing of the phase-current circulation at the outer boundary leaves only the circulations around vortex-core punctures, giving a quantized Chern number.

\begin{figure}
    \centering
    \hspace{-0.3cm}\includegraphics[scale=1.35]{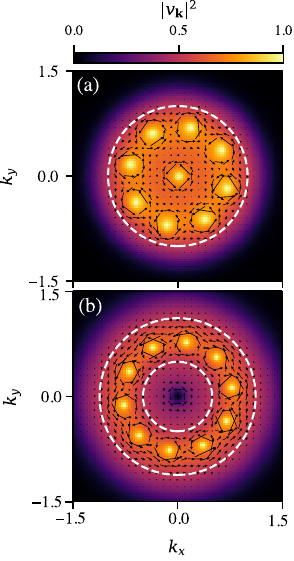}
    \caption{The coherence factor $|v_{\mathbf{k}}|^2$ (color scale) and momentum space current $\mathbf{J}_{\Delta}(\mathbf{k})$ (arrows) for $N=8$ and $\mathcal{B}=8$, shown for two dispersions: (a)  circular, and (b) annular Fermi surface (marked with white dashed lines). The vortices in the occupied region have $|v_{\mathbf{k}}|^2=1$, highlighted by bright yellow. Vortices outside this region are suppressed, highlighted by dark purple. The momentum space current $\mathbf{J}_{\Delta}(\mathbf{k})$ winds around the cores of all vortices, but is highly suppressed for vortices outside the occupied region. Arrows close to the vortex cores are not shown for clarity.}
    \label{fig:cohence_factor_and_currents}
\end{figure}

This boundary-circulation picture is visualized in Fig.~\ref{fig:cohence_factor_and_currents}. The arrows show the phase current $\vb{J}_{\Delta}(\vb{k})$, while the background color shows the coherence factor $\abs{v_{\vb{k}}}^2$. Thus the figure displays both ingredients entering the BdG Chern number: the circulation of the phase current around the vortex cores, and the coherence factor weight determining whether that circulation contributes. In Fig.~\ref{fig:cohence_factor_and_currents}(a), the occupied region of the Fermi sea $D_{\text{occ}}$ is the disk enclosed by the white dashed Fermi surface. The nine visible vortex cores lie inside this region, where $\abs{v_{\vb{k}_i}}^2 = 1$, and the phase current circulates around each core. These nine boundary circulations therefore contribute to $C_{\text{BdG}}$. In Fig.~\ref{fig:cohence_factor_and_currents}(b), the occupied region is an annulus bounded by the two white dashed Fermi surfaces. In this case, eight vortices lie in the occupied annulus and contribute, while the central vortex lies outside the occupied Fermi sea. Since at $\mathbf{k}=0$, $\abs{v_{0}}^2=0$, the phase-current circulation around the central vortex is suppressed and does not contribute to the BdG Chern number.  

We may now simplify Eq.~\eqref{eq_BdGChernExpression} in the smooth parent-band gauge used to compute the self-consistent gap. In this gauge, the Bloch states $\ket{s_{\vb{k}}}$ are smooth on any finite disk $D_{\Lambda}$ enclosing the pairing window. Therefore the pair Berry connection $\mathcal{A}_{P}(\vb{k})$ is smooth at each vortex core. Its circulation around a shrinking loop $\partial D_i(\epsilon)$ then vanishes as $\epsilon \rightarrow 0$. The only surviving contribution to the puncture integral comes from the singular connection $\mathcal{A}_{\Delta}(\vb{k})=\nabla_{\vb{k}}\text{arg}[\Delta(\vb{k})]$ present in $\vb{J}_{\Delta}(\vb{k})$. Thus Eq.~\eqref{eq_BdGChernExpression} reduces to
\begin{align}
    C_{\text{BdG}}=\lim_{\epsilon \rightarrow 0}\sum_{i=0}^{N_V-1}\oint_{\partial D_i(\epsilon)}\frac{d\vb{k}}{2\pi}\cdot \left(\abs{v_{\vb{k}}^2}\mathcal{A}_{\Delta}(\vb{k})\right).
\end{align}
In the same limit, $\abs{v_{\vb{k}}}^2$ can be replaced by its value at the vortex core. From Eq.~\eqref{eq_coherencecore}, this value is equal to $1$ if the vortex lies within the occupied, $\vb{k}_i \in D_{\text{occ}}$, and is equal to zero otherwise. Therefore only vortices inside the occupied region contribute to the BdG Chern number, giving
\begin{align}
    C_{\text{BdG}}=\mathcal{Q}=\sum_{i:\vb{k}_i \in D_{\text{occ}}}\ell_i.
\end{align}
This provides the topological derivation of the occupied vortex charge $\mathcal{Q}$ introduced phenomenologically in Eq.~(\ref{eq_postulatedvortexnumber}). Changing to a singular parent-band gauge on the patch would redistribute phase winding between $\mathcal{A}_{\Delta}(\vb{k})$ and $\mathcal{A}_P(\vb{k})$. In such a gauge, the combination $\mathcal{A}_{\Delta}(\vb{k})-\mathcal{A}_P(\vb{k})$ remains unchanged, and so does the corresponding BdG Chern number. However, because the vortex at infinity is suppressed by the coherence factor $\abs{v_{\vb{k}}}^2$ whenever the one-point compactification condition holds, only finite-momentum vortices can contribute to $C_{\text{BdG}}$. We may therefore choose a finite disk $D_{\Lambda} \supset D_{\text{occ}}$ containing all vortices that contribute to the BdG Chern number. Since $D_{\Lambda}$ is contractible, there is no topological obstruction to choosing a smooth parent-band gauge on the disk. In this gauge, $\mathcal{A}_P(\vb{k})$ carries no singular circulation around vortex cores, so $C_{\text{BdG}}$ is represented entirely by the vortex charge $\mathcal{Q}$, determined by the singularity structure of the singular connection $\mathcal{A}_{\Delta}(\vb{k})$. Thus, the occupied vortex charge $\mathcal{Q}$ is guaranteed to exist as a smooth-gauge representative of the BdG Chern number. In the examples shown in Fig.~\ref{fig:cohence_factor_and_currents}, counting the unit vortices enclosed by the occupied region gives $\mathcal{Q}=9$ in panel (a), and $\mathcal{Q}=8$ in panel (b), up to the overall sign set by our orientation convention.

The occupied-vortex formula reduces to the familiar weak-coupling result in the rotationally symmetric limit.
For a trivial parent band, the only symmetry-enforced vortices of a spinless chiral gap are located at the inversion-invariant points $\vb{k}=0$ and $\vb{k}=\infty$. If the occupied region is a disk enclosing the origin, then only the vortex at $\vb{k}=\vb{0}$ contributes to the BdG Chern number. In a rotationally symmetric weak-coupling state, this winding is precisely the angular momentum of the Cooper pair, reproducing the Read--Green result \cite{Read_2000}. If the occupied region is instead an annulus that encloses neither inversion-invariant point, this simple weak-coupling picture gives $C_{\text{BdG}}=0$.

The formalism developed here extends this picture in two ways. First, it does not rely on weak coupling or rotational symmetry. The relevant quantity is the occupied vortex charge $\mathcal{Q}$, not an angular-momentum quantum number. Second, in a Chern parent band, the projected gap can contain additional vortices away from the inversion-invariant points. If these vortices lie inside an annular occupied region, then $\mathcal{Q}$ can be nonzero even though the annulus excludes both $\vb{k}=0$ and $\vb{k}=\infty$. This is the mechanism illustrated in Fig.~\ref{fig:cohence_factor_and_currents}(b), and it leads directly to a nonzero thermal Hall response.

\subsection{The Thermal Hall Effect}
We now apply the occupied vortex charge formula to the thermal Hall conductivity of the $\lambda_N$ model. In a smooth parent-band gauge on a disk $D_{\Lambda}$ containing the occupied region, the BdG Chern number is represented by the occupied vortex charge, $C_{\text{BdG}}=\mathcal{Q}$. In natural units, the corresponding thermal Hall conductivity is 
\begin{align}
    \kappa_{xy}=\frac{\pi T}{12}\mathcal{Q}=\frac{\pi T}{12}\sum_{i: \vb{k}_i \in D_{\text{occ}}}\ell_i.
\end{align}
Thus, the thermal Hall response directly measures the occupied vortex charge.

Because $\kappa_{xy}$ is tied to the BdG Chern number, it is invariant under smooth parameter changes and can jump only when the BdG quasiparticle gap closes. This occurs when a momentum-space vortex crosses a Fermi surface. At such a point, $\Delta(\vb{k}_i)=0$ and $\xi_{\vb{k}_i}=0$, so the quasiparticle energy vanishes. Equivalently, changes in the thermal Hall response occur when the occupied region gains or loses vortex charge.

\begin{figure}
    \centering
    \includegraphics[scale=1.0]{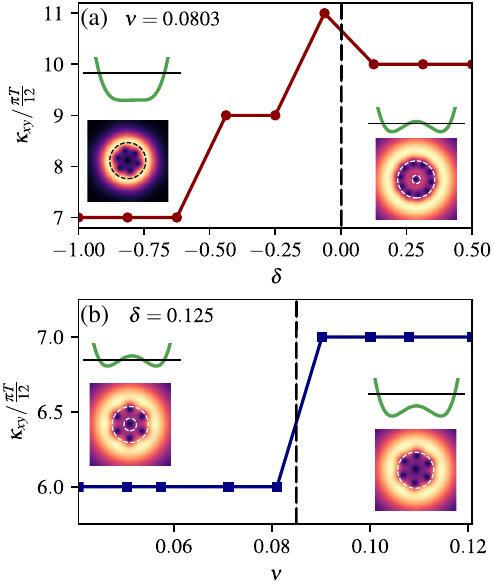}
    \caption{The thermal Hall conductance $\kappa_{xy}=\frac{\pi T}{12}C_{\text{BdG}}$ with increasing (a) band curvature $\delta$ that emulates the effect of a out-of-plane displacement field (for $N=8$, $\mathcal{B}=8$), and (b) filling of the band $\nu$ (for $N=6$, $\mathcal{B}=6$). The Lifshitz transition is marked by the dashed line. For a circular Fermi surface the thermal Hall contribution from the BdG Chern number is odd quantized, and for an annular Fermi surface it is even quantized. If the ring of vortices is present within the annular region, then the thermal Hall response provides an unique signature of the momentum space vortices. The insets show the dispersion and the chemical potential, and the gap function with the vortices in either regime.}
    \label{fig:thermal_hall}
\end{figure}
Fig.~\ref{fig:thermal_hall} illustrates this mechanism in the $\lambda_N$ model by tuning the topology of the Fermi surface across a Lifshitz transition. In Fig.~\ref{fig:thermal_hall}(a), we tune the band-curvature parameter $\delta$ at fixed electron filling $\nu$, where
\begin{gather}
    \nu=\frac{1}{A}\sum_{\mathbf{k}}n_{F}(\xi_{\mathbf{k}}).\label{eq:filling}
\end{gather}
For the values of $\mathcal{B}$ and $N$ shown, $\kappa_{xy}/\frac{\pi T}{12}$ increases in steps of two as additional vortices enter the occupied region. This reflects the fact that the occupied region is bounded by a circular Fermi surface, so the resulting response is odd-valued. After the Lifshitz transition (at $\delta=0$), the Fermi surface becomes annular and the occupied region excludes the central vortex. The thermal Hall response is then determined only by vortices inside the annulus, giving an even-valued response when a ring of off-center vortices lies inside that region. 

Similarly, Fig.~\ref{fig:thermal_hall}(b) shows the evolution of $\kappa_{xy}$ as the band filling $\nu$ is varied at fixed band curvature. As the filling is tuned through a Lifshitz transition, the occupied region changes from annular to disk-like, and $\kappa_{xy}/\frac{\pi T}{12}$ jumps from even to odd values. The odd--even change in the quantized thermal Hall response across the Lifshitz transition is therefore a distinctive signature of momentum-space vortex structure in the superconducting state.

In dual-gated samples of rhombohedral RNG, the band filling $\nu$ and out-of-plane displacement field $D$ can be controlled independently \cite{Han2025}. The displacement field tunes the curvature of the lowest spin-valley polarized band, progressively flattening the dispersion and driving it toward a Mexican-hat profile, an effect captured in our model by the band curvature parameter $\delta$. Therefore, both the filling and the displacement field can independently tune the topology of the occupied region, and hence the occupied vortex charge. This makes the thermal Hall signatures of momentum-space vortices proposed above experimentally relevant for rhombohedral graphene multilayers. We note however, that realistic rhombohedral graphene has a Berry-curvature profile different from the $\lambda_N$ model, which can modify the detailed vortex structure of the gap~\cite{patri2025}.

\section{Discussions}\label{sec:discussions}
In this work, we elucidated the role of non-uniform quantum geometry on superconducting pairing in an isolated spin-polarized band. To this end, we used the continuum $\lambda_N$-model \cite{desrochers2025} and solved the full BCS gap equation using an attractive interaction. We observed the nucleation of momentum-space vortices in the gap function in regions of strong and concentrated Berry curvature, while a strong quantum metric suppresses the gap. In these chiral superconducting states, the nucleation of these momentum-space vortices tends to lower the condensation energy ($\delta F=F_{SC}-F_N$). This behavior is analogous to type-II superconductors, where real-space vortices are nucleated in the gap function in the presence of a magnetic field. Beyond the numerical analysis, we also introduced a gauge-invariant framework for describing the BdG topology of the superconducting state, generalizing similar calculations in Ref.~\cite{Read_2000,sedov2025probingsuperconductivitytunnelingspectroscopy,christos2025}. Using our formalism, we established a general result applicable to any continuum model with an isolated Chern band: the sum of all vortex windings in the gap is fixed by the parent band Chern number, but the BdG Chern number is determined by only those vortices which lie within the occupied Fermi sea. 

The numerical solution to the gap equation, revealed a distinct feature: the vortex configuration always consists of a central vortex of winding $+1$ at the origin, encircled by a ring of additional winding $+1$ vortices whose number is set by both $\mathcal{B}$ and $N$. A similar ring of vortices was observed in Ref.~\cite{patri2025} using a two-orbital model of RNG. There the radius of this ring coincided with the the ring of concentrated Berry curvature. For the $\lambda_N$-model however, the ring radius is set by how sharply the Berry curvature is concentrated — a narrow, strongly peaked profile pulls the vortex ring in toward the origin. 

Even though we always observed a single ring of vortices, it would be interesting, from a free-energy perspective, to understand in what regimes multiple vortex rings may form. Furthermore, the numerical solutions revealed that in general enhancing the Berry flux enclosed by the Fermi surface, drives the nucleation of additional vortices within the gap. A similar trend was reported in Ref.~\cite{patri2025}, and proposed a flux quantization rule, associating the number of vortices to the enclosed Berry flux by the Fermi surface, in the limit of a localized interaction in momentum space. Our mean-field calculations, on the other hand, show no such of flux quantization rule. We also observed that for fixed $N$, the maximum number of vortices present within the gap saturates at $\mathcal{Q}_{\text{sat}} = 2N - 3$ in the regime of large $\mathcal{B}$. Analytically, we found that this value is fixed by the parent-band Chern number alone, $\mathcal{Q}_{\text{sat}}=2C_P-1$, with no dependence on the details of the $\lambda_N$-model.

To classify the topology of the superconducting state, previous studies have often used the eigenstates of a $2\times2$ band-projected BdG Hamiltonian to determine the BdG Chern number \cite{sedov2025,christos2025}. We showed that this $2\times 2$ formulation depends on the parent-band gauge chosen to represent the Bloch states. To restore gauge invariance, we developed a general gauge-invariant prescription to compute the BdG Chern number of band-projected superconductors for any continuum model with an isolated parent band. In the course of this calculation, we identified the BdG Berry curvature as arising from the curl of a momentum-space phase current $\vb{J}_{\Delta}(\mathbf{k})=|v_{\mathbf{k}}^2|\left(\nabla_{\vb{k}}\text{arg}[\Delta(\vb{k})]-\mathcal{A}_{P}(\vb{k})\right)$. This current is analogous to the conventional real-space super-current in superconductors, a similar correspondence was previously identified in \cite{HaldaneProjected,chen2026quantumgeometricquadrupolecooper}. Since the current is weighted by the coherence factor $|v_{\mathbf{k}}^2|$, it is confined to regions within the occupied part of the Fermi surface. The BdG Chern number turns out to be equal to the sum of the windings of all vortices within the occupied region of the Fermi sea. In the weak-coupling limit, our general formula reproduces the known result that the BdG Chern number is the net winding of the gap around the Fermi surface \cite{Read_2000,patri2025,maymann2025,Geier2025}.

The relation between the vortex windings and the BdG Chern number presents a scenario not captured by a weak-coupling ansatz. In systems with an annular Fermi surface, if the ring of vortices lies within the annular occupied region, the BdG Chern number becomes non-zero and even-quantized. This can yield a distinctive signature of momentum-space vortices in thermal Hall measurements. Motivated by the filling and band-curvature tunability of RNG, we propose that driving the system across a Lifshitz transition into a circular Fermi surface — either by tuning the filling or by deforming the band structure — should produce a transition from an even- to an odd-quantized thermal Hall conductivity, providing a unique experimental signature for momentum-space vortices in the gap function.

Although thermal Hall measurements can serve as an indirect means of inferring the presence of momentum-space vortices, a dual-gated quantum twisting microscope (QTM) \cite{Waschitz2026,Wei2026,Sukhachov2023} may offer a more direct route by enabling momentum-resolved probing of the gap structure. By biasing and rotating an extended graphene tip, the QTM facilitates tunneling between the metallic Dirac cone of the tip and the Bogoliubov excitations of the superconducting sample. These tunneling processes are strongest when the Fermi surfaces of the tip and sample are mutually aligned, allowing the tunneling conductance to encode the coherence factors $|u_{\mathbf{k}}|^2$ and $|v_{\mathbf{k}}|^2$, from which the momentum-space resolved gap function can be reconstructed.

Looking forward, several directions emerge. First, demonstrating the presence of momentum-space vortices in real materials requires a more microscopic calculation with the details of the band-structure and Berry curvature profiles. Although the $\lambda_N$-model adopted in this study draws its motivation from a model of spin-valley polarized RNG in a displacement field, a more realistic treatment requires the inclusion of effects of trigonal warping \cite{Koshino2009RNG,MacDonaldTrilayer,Yang2025,christos2025}, leading to the breaking of $SO(2)$ symmetry. This effect would be important for spin-polarized triplet pairing, as trigonal warping breaks inversion symmetry, removing the protection that enforces the appearance of momentum-space vortices in pairs at positions $\vb{k}$ and $-\vb{k}$. Secondly, when the Fermi surface is only partially spin-polarized, it would be interesting to investigate how momentum-space vortices modify the singlet and triplet gaps, and identify the leading pairing channels. Finally, in systems that preserve time reversal symmetry, it would be interesting to investigate band-projected superconductivity using the framework developed here. In this case, the association between the occupied momentum-space vortex windings and the $\mathbb{Z}_2$ invariant remains an open question.

\begin{acknowledgements}
The authors acknowledge Valentin Crépel, Adarsh Patri and Carolina Paiva for useful discussions. This work was supported by the Natural Sciences and Engineering Research
Council of Canada (NSERC) and the Centre for Quantum Materials at the University of Toronto. Computations were performed on the Fir cluster, which is hosted by the Digital Research Alliance of Canada.
\end{acknowledgements}

\section*{Data Availability}
The data supporting the findings of this work are available from the authors upon reasonable request.

\appendix

\onecolumngrid

\section{Continuum parent-band models and the $\lambda_N$ model} \label{sec_app:LambdaNModelDetails}
We begin by specifying the class of continuum parent-band models for which the topological arguments of the main text are well defined. We then verify that the $\lambda_N$ model used satisfies these requirements.

\subsection{General continuum model setup}\label{sec_app:setup}
Let $H(\vb{k})$ be a family of $N \times N$ normal-state Hamiltonians acting on a fixed $\vb{k}$-independent orbital space $\mathbb{C}^N$. We denote a fixed orthonormal orbital basis by $\{\ket{\alpha}\}_{\alpha=0}^{N-1}$. In microscopic valley models, $\alpha$ may label sublattice, layer, or other internal orbital degrees of freedom. The Hamiltonian satisfies the eigenproblem
\begin{align}
    H(\vb{k})\ket{s_{n\vb{k}}}=\epsilon_{n\vb{k}}\ket{s_{n\vb{k}}}, \quad n=0,\cdots, N-1.
\end{align}
Here $\ket{s_{n\vb{k}}}$ denotes the internal, cell-periodic part of the Bloch eigenstate in the chosen orbital basis. Since $H(\vb{k})$ is Hermitian, its eigenstates may be chosen to form a complete orthonormal basis of $\mathbb{C}^N$:
\begin{align}
    \braket{s_{m\vb{k}}}{s_{n\vb{k}}}=\delta_{mn}, \quad \sum_{n=0}^{N-1}\ket{s_{n\vb{k}}}\bra{s_{n\vb{k}}}=\mathds{1}_N.
\end{align}
The momentum coordinate $\vb{k} \in \mathbb{R}^2$ is measured relative to a fixed valley point $K$ of an underlying microscopic band structure, so that the full crystal momentum near the valley is $\vb{q}=K+\vb{k}$. In this appendix, we treat the continuum Hamiltonian $H(\vb{k})$ as a model defined on the full momentum plane. 

On any patch where the eigenstates are chosen smoothly, we collect them into an $N \times N$ unitary matrix $U(\vb{k})$ whose columns are the Bloch eigenstates,
\begin{align}
    U(\vb{k})=\begin{pmatrix}
        \ket{s_{0\vb{k}}},\cdots , \ket{s_{N-1,\vb{k}}}
    \end{pmatrix}.
\end{align}
This transformation diagonalizes the normal state Hamiltonian
 \begin{align}
     H_{D}(\vb{k})=U^{\dag}(\vb{k})H(\vb{k})U(\vb{k})=\text{diag}(\epsilon_{0\vb{k}},\cdots,\epsilon_{N-1,\vb{k}}).
 \end{align}
The Bloch states can then be expanded as
\begin{align}
    \ket{s_{n\vb{k}}}=\sum_{\alpha=0}^{N-1}U_{\alpha n }(\vb{k})\ket{\alpha},
\end{align}
where $\ket{\alpha}$ is the fixed $\vb{k}-$independent orbital basis vector.

\subsection{Active-band isolation}\label{sec_app:bandisolation}
For the band-projected superconductivity problem considered in this work, we choose a single isolated active band. In the $\lambda_N$ model this is the lowest band, $n=0$. We assume that the active band is separated from the remaining bands by a normal-state gap $J$,
\begin{align}
    \underset{\vb{k},n\neq 0}{\text{min}}\abs{\epsilon_{n\vb{k}}-\epsilon_{0\vb{k}}} \geq J.
\end{align}
We also assume that the chemical potential $\mu$ crosses only the active band. A sufficient condition for neglecting superconducting mixing with the inert bands is
\begin{align}
    J \gg \underset{\vb{k}}{\text{max}}\abs{\Delta(\vb{k})},
\end{align}
where $\Delta(\vb{k})$ is the projected superconducting gap introduced in the main text.

For simplicity, we denote the isolated eigenstate of band $n=0$ as $\ket{s_{0\vb{k}}}\equiv \ket{s_{\vb{k}}}$. Because the active band is isolated, it defines a complex line bundle $L$ over momentum space. At each $\vb{k} \in \mathbb{R}^2$, the fiber is the one-dimensional eigenspace
\begin{align}
    L_{\vb{k}}=\text{span}\{\ket{s_{\vb{k}}}\} \subset \mathbb{C}^N.
\end{align}
Equivalently, $L_{\vb{k}} \cong \mathbb{C}$, and a choice of normalized eigenstate $\ket{s_{\vb{k}}}$ gives a local frame identifying the fiber with the standard complex line $\mathbb{C}$. This frame is not unique, it has the local $U(1)$ gauge freedom
\begin{align}
    \ket{s_{\vb{k}}} \rightarrow e^{i\chi(\vb{k})}\ket{s_{\vb{k}}}.
\end{align}
Since $\ket{\alpha}$ is $\vb{k}-$independent, this gauge transformation occurs at the level of its components, i.e. 
\begin{align}
    U_{\alpha 0}(\vb{k}) \rightarrow e^{i\chi(\vb{k})}U_{\alpha 0}(\vb{k}). \label{componentGaugeTransformation}
\end{align}
These are the local assumptions underlying the active-band. We now state the global condition needed to assign an integer Chern number to the continuum models considered here.

\subsection{One-point compactification of the parent projector}\label{sec_app:onepointcompactification}
Unlike a lattice band, whose momentum space is the compact Brillouin-zone torus, a continuum model is defined on the noncompact plane $\mathbb{R}^2$. As a complex line bundle over $\mathbb{R}^2$, the active band is topologically trivial. This is because the base space $\mathbb{R}^2$ is contractible, implying that an isolated active band admits a smooth global frame $\ket{s_{\vb{k}}}$ over the plane. Consequently, the Berry-flux integral over $\mathbb{R}^2$ is not automatically an integer-valued Chern number. To define an integer Chern number, we must replace the noncompact base by a compact oriented two-dimensional manifold. In the continuum setting, this is achieved by a one-point compactification of the momentum plane, provided the active band has a well-defined limit at infinity. Recall that the active band assigns to each $\vb{k} \in \mathbb{R}^2$ the one dimensional fiber $L_{\vb{k}}=\text{span}\{\ket{s_{\vb{k}}}\}$. Equivalently, this fiber is the image of the gauge-invariant active-band projector
\begin{align}
    P(\vb{k})=\ket{s_{\vb{k}}}\bra{s_{\vb{k}}}, \quad L_{\vb{k}}=\text{Im}P_{0}(\vb{k}).
\end{align}
We say that the parent band admits a one-point compactification when
\begin{align}
    \lim_{\abs{\vb{k}} \rightarrow \infty}P_{0}(\vb{k})=P_{\infty}
\end{align}
exists and is independent of the direction of approach. In that case, all directions at $\abs{\vb{k}}=\infty$ correspond to the same physical line
\begin{align}
    L_{\infty}=\text{Im}P_{\infty},
\end{align}
so the momentum plane may be compactified as
\begin{align}
    \mathbb{S}^2=\mathbb{R}^2 \cup \{\infty\}.
\end{align}
The projector then extends to a continuous map
\begin{align}
    P_{0}:\mathbb{S}^2 \rightarrow \mathbb{CP}^{N-1},
\end{align}
and the active band defines a complex line bundle over the compactified sphere. As a complex line bundle over the sphere, there is no guarantee that $\ket{s_{\vb{k}}}$ extends smoothly throughout all of $\mathbb{S}^2$. This obstruction is encoded in the first Chern number of the parent band $C_P$, defined by
\begin{align}
    C_P=\int_{\mathbb{S}^2} \frac{\Omega}{2\pi} \in \mathbb{Z}.
\end{align}
Here, $\Omega$ is the two-form Berry curvature, which can be defined locally
\begin{align}
    \Omega=\Omega(\vb{k})d^2 \vb{k}=i\epsilon^{\mu\nu}\text{Tr}[P_0(\vb{k})\partial_{\mu}P_{0}(\vb{k})\partial_{\nu}P_{0}(\vb{k})]d^2 \vb{k}.
\end{align}
In practice, the Chern number is obtained by integrating the scalar form of the Berry curvature $\Omega(\vb{k})$ over a disk $D_{\Lambda}$ of radius $\Lambda$ and taking the limit as $\Lambda \rightarrow \infty$:
\begin{align}
    C_P=\lim_{\Lambda \rightarrow \infty}\int_{D_{\Lambda}} \frac{d^2\vb{k}}{2\pi}\Omega(\vb{k})=\int_{\mathbb{S}^2}\frac{\Omega}{2\pi}.
\end{align}

\subsection{The $\lambda_N$ model as a compactified Chern Band}\label{sec_app:lambdaN}
We now verify that the $\lambda_N$ model used in the main text satisfies the criteria introduced above. Let the Hilbert space of the $\lambda_N$ model be
\begin{align}
    \mathcal{H}_{N}=\text{span}\{\ket{\alpha}\}_{\alpha=0}^{N-1} \cong \mathbb{C}^N.
\end{align}
We define the raising and lowering operators $b^{\dagger}/b$ by their action on the basis
\begin{align}
    b\ket{\alpha}=\sqrt{\alpha}\ket{\alpha-1}, \quad b^{\dag}\ket{\alpha}=\sqrt{\alpha+1}\ket{\alpha+1}, \quad \comm{b}{b^{\dag}}=1.
\end{align}
We also define the orthogonal projector
\begin{align}
    P^{\perp}_{N-1}=\mathds{1}_{N}-\ket{N-1}\bra{N-1},
\end{align}
and the operator
\begin{align}
    D_{N}(\vb{k})=(b^{\dag}-\sqrt{\mathcal{B}}k_z P^{\perp}_{N-1})(b-\sqrt{\mathcal{B}}k_{\overline{z}}P^{\perp}_{N-1}).
\end{align}
Here $\mathcal{B}$ is the real Berry curvature parameter used in the main text. We use the convention $k_z=(k_x+ik_y)/\sqrt{2}$. The $\lambda_N$ model is constructed by adding $D_N(\vb{k})$ to an electron gas with energy $\epsilon_{\vb{k}}$:
\begin{align}
    H_{\lambda_N}(\vb{k})=\epsilon_{0\vb{k}}\mathds{1}_N+JD_N(\vb{k}).
\end{align}
In this case, the operator $D_N(\vb{k})$ gives rise to topology in the eigenstates of the $\lambda_N$ model. Since $D_{N}(\vb{k})$ is positive semidefinite, its zero modes satisfy
\begin{align}
    D_{N}(\vb{k})\ket{s_{0\vb{k}}}=0. \label{eq_app_zeromode}
\end{align}
Thus $\ket{s_{0\vb{k}}}$ is an eigenstate of $H_{\lambda_N}(\vb{k})$ with energy $\epsilon_{0\vb{k}}$, which we are free to choose. Solving Eq.~\eqref{eq_app_zeromode} gives the normalized eigenstate
\begin{align}
    \ket{s_{0\vb{k}}}=\sqrt{\frac{\text{exp}\left(-\frac{\mathcal{B}\abs{\vb{k}}^2}{2}\right)(N-1)!}{\Gamma\left(N,\frac{\mathcal{B}\abs{\vb{k}}^2}{2}\right)}}\sum_{\alpha=0}^{N-1}\frac{\left(\sqrt{\mathcal{B}}k_{\overline{z}}\right)^\alpha}{\sqrt{\alpha!}}\ket{\alpha},
\end{align}
where $\Gamma(N,\mathcal{B}\abs{\vb{k}}^2/2)$ is the upper incomplete gamma function. At large $\abs{\vb{k}}$, the highest power $\alpha=N-1$ dominates the normalized spinor. The state itself approaches
\begin{align}
    \ket{s_{0\vb{k}}} \sim e^{-i(N-1)\phi(\vb{k})}\ket{N-1}, \quad \text{as } \abs{\vb{k}} \rightarrow \infty,
\end{align}
with $\phi(\vb{k})=\text{arg}(k_x+ik_y)$. This reflects the fact that the state cannot be extended smoothly to the point at infinity; it winds by $-(N-1)$ at $\abs{\vb{k}}=\infty$. Its projector however, reaches the fixed point 
\begin{align}
    \lim_{\abs{\vb{k}} \rightarrow \infty}P_{0}(\vb{k})=\ket{N-1}\bra{N-1},
\end{align}
independent of $\phi(\vb{k})$. Thus every finite-N $\lambda_N$ model defines a compactified continuum Chern band. It may be viewed as a finite-dimensional regularization of the infinite Chern band in Ref.~\cite{Tan2024,TanDevakul2025WavefunctionAHC}, whose $N\rightarrow \infty$ limit carries unbounded Berry flux, rather than a finite Chern number.

Because the projector compactifies, the quantized Berry flux $\Phi_N(\infty)$ coincides with the parent band Chern number:
\begin{align}
    C_P=\lim_{\Lambda \rightarrow \infty}\int_{D_{\Lambda}} \frac{d^2 \vb{k}}{2\pi}\Omega_N(\vb{k})=N-1.
\end{align}

\section{Wu-Yang Description of a Continuum Model}
We now describe the compactified parent band using the Wu--Yang two-patch construction \cite{WuYang}. This makes makes explicit how a nonzero parent-band Chern number appears as a winding of local Bloch frames, even though the projector is globally well defined. The same boundary-winding logic underlies the singular connection description of momentum-space vortices used below.

\subsection{Singular Gauge Winding}\label{sec_app:singulargauge}
In the finite-plane gauge used for the $\lambda_N$ model, the normalized Bloch frame $\ket{s_{\vb{k}}}$ winds by $-(N-1)$ as $\abs{\vb{k}} \rightarrow \infty$, while the projector approaches a direction-independent limit. The fact that $C_P=N-1$ is the same value as this winding is no coincidence. Let the point at infinity be referred to as the "north pole". The origin is then the "south pole". The finite-plane gauge used for the $\lambda_N$ models parent band state $\ket{s_{0\vb{k}}}$ is therefore naturally identified with the south-patch gauge, which we denote as $\ket{u_S}$. More generally, for a compactified continuum Chern band, choose a south-patch frame $\ket{u_S}$ that is smooth on $U_S=\mathbb{S}^2 \setminus \{\infty\} \cong \mathbb{R}^2$. The asymptotic behaviour of the frame is then
\begin{align}
     \ket{u_S} \sim e^{-iC_P \phi(\vb{k})}\ket{c}, \quad \text{as } \abs{\vb{k}} \rightarrow \infty.
\end{align}
On $U_S$, the Berry connection takes the form $\mathcal{A}_{S}=i\bra{u_S}\ket{du_S}$. Asymptotically, it behaves like
\begin{align}
    \mathcal{A}_{S} \sim C_P d\phi, \quad \text{as } \abs{\vb{k}} \rightarrow \infty.
\end{align}
The one-form $d\phi$ is not smooth at the added point $\infty$, so $\mathcal{A}_{S}$ does not extend as a smooth connection one-form through the north pole. However, its circulation around a large circle is finite. On the disk $D_{\Lambda}$, the Chern number can be picked up as a boundary winding of the Berry connection in the south gauge
\begin{align}
    \lim_{\Lambda \rightarrow \infty}\int_{D_{\Lambda}}\frac{d^2\vb{k}}{2\pi}\Omega(\vb{k})=\lim_{\Lambda \rightarrow \infty}\oint_{\partial D_{\Lambda}} \frac{\mathcal{A}_S}{2\pi}=\oint \frac{C_P d\phi}{2\pi}=C_P
\end{align}

\subsection{North and South Gauges}\label{sec_app:WuYang}
The complementary Wu--Yang description uses a north-patch frame on $U_N=\mathbb{S}^2 \setminus \{0\}$, which is smooth at infinity but singular at the origin. The corresponding frame is
\begin{align}
    \ket{u_{N}}=e^{iC_P \phi(\vb{k})}\ket{u_{S}}.
\end{align}
At the origin, the frame behaves like
\begin{align}
    \ket{u_{N}} \sim e^{iC_P\phi(\vb{k})}\ket{c'}, \quad \text{as } \abs{\vb{k}} \rightarrow 0.
\end{align}
This is the Wu--Yang description of a nontrivial complex line bundle over $\mathbb{S}^2$: no single frame is smooth on both patches when $C_P \neq 0$, but the projector and curvature are globally well defined. On the overlap $U_{S} \cap U_N$, and in particular on any circle separating the two poles, the two frames are related by the transition function
\begin{align}
    \ket{u_N}=g_{NS}\ket{u_S}, \quad g_{NS}(\phi)=e^{iC_P \phi}.
\end{align}
The Chern number is computed by choosing $M_{S} \subset U_S$ and $M_N \subset U_N$ such that $\mathbb{S}^2=M_S \cup M_N$ with $\partial M_S=-\partial M_N=\gamma$, and integrating the Berry curvature as
\begin{align}
    \int_{\mathbb{S}^2}\frac{\Omega}{2\pi}=\int_{M_N}\frac{\Omega}{2\pi}+\int_{M_S} \frac{\Omega}{2\pi}=\frac{1}{2\pi}\oint_{\gamma} (\mathcal{A}_S-\mathcal{A}_N)=C_P.
\end{align}
In the last line we used Stoke's theorem to convert $\Omega=d\mathcal{A}$ and then used the transition function $\mathcal{A}_S-\mathcal{A}_N=C_P d\phi$.

\section{Orbital-basis formulation of projected superconductivity}
Here we give a detailed overview of the superconducting formalism, including the orbital and band bases, restricting the interaction channels, and the corresponding eigenstates of the BdG Hamiltonian.

\subsection{Orbital and band bases}\label{sec_app:BdGHamiltonian}
We start by defining the creation and annihilation operators in the orbital basis in momentum space
\begin{align}
    a_{\vb{q}\alpha}=\frac{1}{\sqrt{N_c}}\sum_{\vb{R}}e^{-i\vb{q}\cdot (\vb{R}+\vb{r}_{\alpha})}a_{\vb{R}\alpha}, \quad 
    a_{\vb{q}\alpha}^{\dag}=\frac{1}{\sqrt{N_c}}\sum_{\vb{R}}e^{i\vb{q}\cdot (\vb{R}+\vb{r}_{\alpha})}a_{\vb{R}\alpha}^{\dag}.
\end{align}
Here $N_c$ is the number of unit cells in the crystal, $\vb{R}$ labels the unit cells, $\alpha$ labels orbitals/sublattices inside the unit cell, and $\vb{r}_{\alpha}$ is the position of orbital $\alpha$ inside the unit cell. Any resulting topological quantities are generically independent of $\vb{r}_{\alpha}$ \cite{SimonGeometry}, therefore we may safely set $\vb{r}_{\alpha}=0$. Since we are interested in superconductivity emerging within a valley, we define our crystal momentum
\begin{align}
    \vb{q}=K+\vb{k}.
\end{align}
The momentum $\vb{k}$ is a small continuum momentum measured relative to the valley center. We then label our creation and anhilation operators by the continuum momentum $\vb{k}$, assumed to be valid within some cutoff $\abs{\vb{k}}<\Lambda$. The many-body Hamiltonian with a two-body interaction is
\begin{align}
    H=\sum_{\vb{k}}\sum_{\alpha \beta}a^{\dag}_{\vb{k}\alpha}(H_{\alpha \beta}(\vb{k})-\mu \delta_{\alpha \beta}) a_{\vb{k}\beta}+\frac{1}{2A}\sum_{\vb{k},\vb{k}',\vb{q}}\sum_{\alpha \beta \delta \gamma}V_{\alpha \beta;\gamma \delta}(\vb{k},\vb{k}';\vb{q})a^{\dag}_{\vb{k}+\vb{q}\alpha}a^{\dag}_{\vb{k}'-\vb{q}\beta}a_{\vb{k}'\delta}a_{\vb{k}\gamma}.
\end{align}
Here $H_{\alpha \beta}(\vb{k})$ is the continuum normal state Hamiltonian written in the orbital basis. In the main text, this is the $\lambda_N$ model, however the results here hold for any continuum model satisfying the criteria in Appendix~\ref{sec_app:bandisolation}.

We assume there exists an instability between pairs of opposite momentum, e.g. $\vb{k}$ and $-\vb{k}$, so that scattering is dominated in the Cooper channel. The resulting Hamiltonian takes the conventional BCS form
\begin{align}
    H_{\text{BCS}}=\sum_{\vb{k}}\sum_{\alpha \beta}a^{\dag}_{\vb{k}\alpha}(H_{\alpha \beta}(\vb{k})-\mu \delta_{\alpha \beta})a_{\vb{k}\beta}+\frac{1}{2A}\sum_{\vb{k},\vb{k}'}\sum_{\alpha \beta \delta \gamma}V_{\alpha \beta;\gamma \delta}(\vb{k},\vb{k}')a^{\dag}_{\vb{k}\alpha}a^{\dag}_{-\vb{k}\beta}a_{-\vb{k}'\delta}a_{\vb{k}' \gamma}.
\end{align}
At this point, we assume the existence of off diagonal long range order, giving rise to the BCS ground state $\ket{\text{BCS}}$, from which we compute the anomalous average $\bra{\text{BCS}}a^{\dag}_{\vb{k}\alpha}a^{\dag}_{-\vb{k}\beta}\ket{\text{BCS}}\equiv \expval{a^{\dag}_{\vb{k}\alpha}a^{\dag}_{-\vb{k}\beta}}$. Expanding the four-point operator in powers of fluctuations about this average yields the mean-field approximation
\begin{align}
    a^{\dag}_{\vb{k}\alpha}a^{\dag}_{-\vb{k}\beta}a_{-\vb{k}'\delta}a_{\vb{k}' \gamma} \approx a^{\dag}_{\vb{k}\alpha}a^{\dag}_{-\vb{k}\beta}\expval{a_{-\vb{k}'\delta}a_{\vb{k}'\gamma}}+\expval{a^{\dag}_{\vb{k}\alpha}a^{\dag}_{-\vb{k}\beta}}a_{-\vb{k}'\delta}a_{\vb{k}'\gamma}-\expval{a^{\dag}_{\vb{k}\alpha}a^{\dag}_{-\vb{k}\beta}}\expval{a_{-\vb{k}'\gamma}a_{\vb{k}'\delta}}.
\end{align}
We further define the order parameter in the orbital basis to be
\begin{align}
    \Delta_{\alpha \beta}(\vb{k})=\frac{1}{A}\sum_{\vb{k}'}\sum_{\gamma \delta}V_{\alpha \beta;\gamma \delta}(\vb{k},\vb{k}')\expval{a_{-\vb{k}'\delta} a_{\vb{k}'\gamma}},
\end{align}
so that the mean-field BCS Hamiltonian takes the form
\begin{align}
    H_{\text{MF}}=E_{0}+\frac{1}{2}\sum_{\vb{k}}\text{Tr}h(\vb{k})+\frac{1}{2}\sum_{\vb{k}}\Psi_{\mathcal{O}}^{\dag}(\vb{k})H_{\mathcal{O}}(\vb{k})\Psi_{\mathcal{O}}(\vb{k}).
\end{align}
The subscript $\mathcal{O}$ indicates that this representation is written in the orbital basis. Here, the Nambu spinor $\Psi_{\mathcal{O}}(\vb{k})=(a_{\vb{k}},(a^{\dag}_{-\vb{k}})^{T})^T$ is the full $2N$-dimensional vector of particle and hole ladder operators, defined by $a_{\vb{k}}=(a_{\vb{k},\alpha=0},\cdots, a_{\vb{k},\alpha=N-1})^T$, and $a^{\dag}_{-\vb{k}}=(a_{-\vb{k},\alpha=0}^{\dag},\cdots, a^{\dag}_{-\vb{k},\alpha=N-1} )$. Here Tr denotes trace over the orbital indices of the normal state Hamiltonian $h=H-\mu$. The BdG Hamiltonian in the orbital basis takes the form
\begin{align}
    H_{\mathcal{O}}(\vb{k})=\begin{pmatrix}
        H(\vb{k})-\mu & \Delta_{\mathcal{O}}(\vb{k}) \\
        \Delta^{\dag}_{\mathcal{O}}(\vb{k}) & -H^{T}(-\vb{k})+\mu
    \end{pmatrix},
\end{align}
where $\mu$ is the chemical potential. At this point, we would like to restrict to pairing channels only within the lowest isolated band. To do so, it is advantageous to work in the band basis. This is not the quasi-particle band basis, but rather the basis which diagonalizes the normal state Hamiltonian into its constituent bands. Since the BdG Hamiltonian includes both a particle and hole sector, we define the $2N \times 2N$ unitary matrix
\begin{align}
    V(\vb{k})=\begin{pmatrix}
        U(\vb{k}) & 0 \\
        0 & U^{*}(-\vb{k})
    \end{pmatrix}.
\end{align}
The BdG Hamiltonian in the band basis takes the form
\begin{align}
    H_{\mathcal{B}}(\vb{k})=V^{\dag}(\vb{k})H_{\mathcal{O}}(\vb{k})V(\vb{k})=\begin{pmatrix}
        H_{D}(\vb{k})-\mu & \Delta_{\mathcal{B}}(\vb{k}) \\
        \Delta^{\dag}_{\mathcal{B}}(\vb{k}) & -H_{D}^{T}(-\vb{k})+\mu
    \end{pmatrix}. \label{BandBasisBdG}
\end{align}
Here the subscript $\mathcal{B}$ stands for the band basis. The components of the band basis order parameter, $\Delta_{\mathcal{B}}(\vb{k})$, are written using the same indices as the normal state bands, namely $n$ and $m$. They take the form
\begin{align}
    \Delta_{mn}(\vb{k})=\frac{1}{A}\sum_{\vb{k}'}\sum_{m'n'}V_{mn;m'n'}(\vb{k},\vb{k}')\expval{c_{-\vb{k}'n'}c_{\vb{k}'m'}}.
\end{align}
In this basis, the Cooper vertex takes the form
\begin{align}
    V_{mn;m'n'}(\vb{k},\vb{k}')=\sum_{\alpha \beta \gamma \delta}\overline{U}_{\alpha m}(\vb{k})\overline{U}_{\beta n}(-\vb{k})V_{\alpha\beta;\gamma \delta}(\vb{k},\vb{k}')U_{\gamma m'}(\vb{k}')U_{\delta n'}(-\vb{k}').
\end{align}
The ladder operator $c^{\dag}_{\vb{k}n}$ creates a local representation of a Bloch state in band $n$, i.e. $c^{\dag}_{\vb{k}n}\ket{0}=e^{i\vb{k}\cdot \vb{r}}\ket{s_{n\vb{k}}}$. Since these states are related to the orbital basis through the unitary $U(\vb{k})$, we have the change of basis for ladder operators
\begin{align}
    a_{\vb{k}\alpha}=\sum_{n=0}^{N-1}U_{\alpha n}(\vb{k})c_{n\vb{k}}, \qquad a^{\dag}_{\vb{k}\alpha}=\sum_{n=0}^{N-1}c^{\dag}_{n\vb{k}}U^{*}_{n\alpha}(\vb{k}). \label{eq_ladderorbital}
\end{align}
Band projected superconductivity amounts to selecting the non-zero components of the tensor $V_{mn;m'n'}(\vb{k},\vb{k}')$ represented in the band basis. If we assume pairing is to only occur within the lowest band, then $m=n=0$. Furthermore, if intraband pairing is restricted, then $m'=n'=0$. Even with this restriction, the operators in Eq.~\eqref{eq_ladderorbital} which contribute to the superconducting sector themselves contain $\alpha=0,\cdots, N-1$ orbital degrees of freedom, encoded in the components of the parent band state $U_{\alpha 0}(\vb{k})$. The key distinction is that higher bands are not allowed to participate in the formation of the superconducting condensate, which is determined solely by the interaction $V_{mn;m'n'}(\vb{k},\vb{k}')$. With this restriction, the order parameter takes the form
\begin{align}
    \Delta(\vb{k})=\frac{1}{A}\sum_{\vb{k}'}V_{00;00}(\vb{k},\vb{k}')\expval{c_{-\vb{k}'}c_{\vb{k}}}, \label{ProjectedOrder}
\end{align}
where we once again have suppressed the band index $n=0$ for cleanliness. In the main text, we have assumed that the interaction vertex is given by the following Gaussian $V_{\alpha \beta;\gamma \delta}(\vb{k},\vb{k}')=-U_0 e^{-\sigma^2 \abs{\vb{k}-\vb{k}'}^2}\delta_{\alpha \gamma}\delta_{\beta\delta}$, leading to Eq.~(\ref{eq:form_factor_dressed_vertex}). The pairing combination in the density-density channel $\delta_{\alpha \gamma}\delta_{\beta\delta}$ is the conventional one used in most studies on band projected superconductivity. In what follows, we do not restrict ourselves to this particular form of the pairing interaction; instead, we treat the interaction as completely general. Applying a local gauge transformation to the components of the Bloch states (as in Eq.~(\ref{componentGaugeTransformation})), we find that the restricted band-projected vertex transforms like
\begin{align}
    V_{00;00}(\vb{k},\vb{k}')\rightarrow e^{-i(\chi(\vb{k})+\chi(-\vb{k}))}e^{i(\chi(\vb{k}')+\chi(-\vb{k}'))}V_{00;00}(\vb{k},\vb{k}').
\end{align}
The projected ladder operators will also transform, since
\begin{align}
    c^{\dag}_{\vb{k}}\ket{0}=e^{i\vb{k}\cdot \vb{r}}\ket{s_{\vb{k}}} \rightarrow e^{i\chi(\vb{k})}e^{i\vb{k}\cdot \vb{r}}\ket{s_{\vb{k}}} \implies c_{\vb{k}} \rightarrow e^{-i\chi(\vb{k})}c_{\vb{k}}.
\end{align}
Thus, counterintuitively, the projected order parameter transforms under gauge transformations of the parent band
\begin{align}
    \Delta(\vb{k}) \rightarrow e^{-i(\chi(\vb{k})+\chi(-\vb{k}))}\Delta(\vb{k}).
\end{align}
This is quite unlike ordinary superconductivity in a trivial band, where the order parameter takes the form of an ordinary function on momentum space. Instead, $\Delta(\vb{k})$ transforms like the following pair state $(\ket{s_{\vb{k}}}\bra{s^{*}_{-\vb{k}}})^{*}$. This property was remarked in Ref.~\cite{DaidoGaugeSC,sedov2025probingsuperconductivitytunnelingspectroscopy}. Most works currently fix a gauge for the Bloch states. While this approach can work for certain goals, one loses the underlying interpretation of band projected superconductivity.

Inserting the projected order parameter of Eq.~(\ref{ProjectedOrder}) into the band basis BdG Hamiltonian of Eq.~(\ref{BandBasisBdG}), and changing the basis, we find
\begin{align}
    H_{\mathcal{O}}(\vb{k})=V(\vb{k})H_{\mathcal{B}}(\vb{k})V^{\dag}(\vb{k})=\begin{pmatrix}
        H(\vb{k})-\mu & \Delta_{\mathcal{O}}(\vb{k}) \\
        \Delta^{\dag}_{\mathcal{O}}(\vb{k}) & -H^{T}(-\vb{k})+\mu
    \end{pmatrix}. \label{ProjectedOrbitalBdG}
\end{align}
In this case, the order parameter takes the form
\begin{align}
    \Delta_{\mathcal{O}}(\vb{k})=\Delta(\vb{k})U(\vb{k})\ket{0}\bra{0}U^{T}(-\vb{k})=\Delta(\vb{k})\ket{s_{\vb{k}}}\bra{s^{*}_{-\vb{k}}} \label{orderparameterOrbital}
\end{align}
This order parameter is clearly gauge invariant. In the mathematical language, the order parameter $\Delta(\vb{k})$ is a section of a vector bundle. The matrix $\ket{s_{\vb{k}}}\bra{s^{*}_{-\vb{k}}}$ is known as a $\vb{k}-$dependent frame, and transforms covariantly (oppositely to $\Delta(\vb{k})$), i.e. $\ket{s_{\vb{k}}}\bra{s^{*}_{-\vb{k}}} \rightarrow e^{i(\chi(\vb{k})+\chi(-\vb{k}))}\ket{s_{\vb{k}}}\bra{s^{*}_{-\vb{k}}}$, implying that $\Delta_{\mathcal{O}}(\vb{k}) \rightarrow \Delta_{\mathcal{O}}(\vb{k})$. While $\Delta_{\mathcal{O}}(\vb{k})$ is gauge invariant, its exact form in Eq.~(\ref{orderparameterOrbital}) is only a local representation in the gauge chosen for $\ket{s_{\vb{k}}}$. For a valley polarized superconductor, where the base is compactified to a sphere, we can generically choose a local gauge on the disk $D_{\Lambda}$ surrounding the valley. For this reason, the order parameters component in the frame, $\Delta(\vb{k})$, is still meaningful.

\subsection{Contrasting the electromagnetic and Bloch gauge transformations}\label{app:differnt_gauge_transformations}
The electromagnetic $U(1)$ gauge redundancy of charged matter is implemented by minimally coupling the electron fields, defined in the orbital basis, to the external electromagnetic gauge potential $\mathbf{A}_{em}(\vb{r})$. This structure is encoded in the real-space BCS Hamiltonian
\begin{align}
    H_{\text{BCS}}=\int d^2\vb{r}\sum_{\alpha\beta}a_{\alpha}^{\dag}(\vb{r})\left(\frac{1}{2m}(-i \nabla_{\vb{r}}+e\vb{A}(\vb{r}))^2\delta_{\alpha\beta}+t_{\alpha\beta}(\vb{r})\right)a_{\beta}(\vb{r})\nonumber\\+\sum_{\alpha\beta}\frac{1}{2}\int d^2\vb{r}_1 d^2\vb{r}_2\left(\Delta_{\alpha\beta}(\vb{r}_1,\vb{r}_2)a_{\alpha}^{\dag}(\vb{r}_1)a_{\beta}^{\dag}(\vb{r}_2)+h.c.\right).
\end{align}
where $\alpha,\beta$ are orbital indices. Under a local change of phase we have
$a_{\alpha}(\vb{r}) \rightarrow e^{i\theta(\vb{r})}a_{\alpha}(\vb{r}) $, and $ \vb{A}(\vb{r}) \rightarrow \vb{A}(\vb{r})-\frac{1}{e}\nabla_{\vb{r}}\theta(\vb{r})$. The real space gap defined in the orbital basis must then transform as
\begin{align}
    \Delta_{\alpha\beta}(\vb{r}_1,\vb{r}_2) \rightarrow e^{i(\theta(\vb{r}_1)+\theta(\vb{r}_2))}\Delta_{\alpha\beta}(\vb{r}_1,\vb{r}_2).
\end{align}
This local $U(1)$ electromagnetic gauge redundancy is always present in the formalism.

The superconducting state breaks the global charge-$U(1)$ symmetry corresponding to particle number conservation by picking a definite phase $\phi$. Under this transformation $U(\theta_0)=e^{i\theta_0 \hat{N}}$, the BCS ground state obeys
\begin{align}
    U(\theta_0)\ket{\text{BCS}(\phi)}=\ket{\text{BCS}(\phi+2\theta_0)}.
\end{align}
Equivalently, the operators transform as
\begin{align}
    a_{\alpha}(\vb{r})\rightarrow U^{\dagger}(\theta_0)a_{\alpha}(\vb{r}) U(\theta_0)=e^{i\theta_0}a_{\alpha}(\vb{r}),
\end{align}
and the the gap function transforms as
\begin{gather}
    \Delta_{\alpha\beta}(\vb{r}_1,\vb{r}_2)\rightarrow e^{2i\theta_0}\Delta_{\alpha\beta}(\vb{r}_1,\vb{r}_2).
\end{gather}
A superconducting mean-field solution with $\Delta_{\alpha\beta}\neq 0$ therefore selects a definite overall condensate phase and is not invariant under this transformation.

By contrast, the Bloch-gauge redundancy discussed in the main text originates from the arbitrary momentum-dependent phase convention used to define each Bloch eigenstate. This local $U(1)$ gauge redundancy can be expressed as
$\ket{u_{\vb{k}}} \sim e^{i\chi(\vb{k})}\ket{u_{\vb{k}}}$, with the associated $U(1)$ Berry connection
$\mathcal{A}_{\mu}(\vb{k})=i\bra{u_{\vb{k}}}\ket{\partial_{\mu}u_{\vb{k}}}$, which transforms as $\mathcal{A}_{\mu}(\vb{k})\rightarrow \mathcal{A}_{\mu}(\vb{k})-\partial_{\mu}\chi(\mathbf{k})$. As shown in the main text, upon projecting the pairing interaction onto the active band, the corresponding band-projected gap function transforms as
\begin{align}
    \Delta(\vb{k})\rightarrow e^{-i(\chi(\vb{k})+\chi(-\vb{k}))}\Delta(\vb{k}).
\end{align}
Therefore, note that this Bloch-gauge redundancy is distinct from both the local electromagnetic $U(1)$ gauge redundancy, and the spontaneous breaking of global $U(1)$ symmetry associated with the superconducting order parameter in the orbital basis. The Bloch gauge theory structure in momentum space is what underlies the formation
of momentum-space vortices of the projected gap and the corresponding relation to the BdG Chern number. 

\subsection{Many-body gap and eigenstates of the BdG Hamiltonian} \label{sec_app:BdGeigenstatesandFreeEnergy}
Here we diagonalize the BdG Hamiltonian in the orbital basis after band projection.

The projected Hamiltonian (Eq.~(\ref{ProjectedOrbitalBdG})) has size $2N \times 2N$. Therefore, we expect $2N$ eigenstates. These can be split up by sectors. The first sector is the superconductor sector. This corresponds to the Nambu spinor eigenstates that form from the pairing of electrons in lowest band ($n=0$). It consists of two eigenstates; a particle eigenstate, and a hole eigenstate labeled with a $+$ and $-$ sign respectively. They take the form
\begin{align}
    \left.
\begin{aligned}
\ket{\Psi_{0\vb{k}}^{-}} &= -v^{*}_{\vb{k}}\ket{s_{\vb{k}}} \oplus u^{*}_{-\vb{k}}\ket{s^{*}_{-\vb{k}}} \\
\ket{\Psi_{0\vb{k}}^{+}} &= u_{\vb{k}}\ket{s_{\vb{k}}} \oplus v_{-\vb{k}}\ket{s^{*}_{-\vb{k}}}\label{eq:define_Psi-}
\end{aligned}
\right\}
\quad
\text{Superconducting sector.}
\end{align}
Here, $u_{\mathbf{k}}$ and $v_{\mathbf{k}}$ are the coherence factors, defined in the main text. The second sector is the inert sector. This consists of inert normal state bands, not affected by any superconducting interaction. There are $2N-2$ of these eigenstates, $N-1$ of them belonging to the electron branch ($+$), and $N-1$ of them belonging to the hole branch ($-$). They take the form
\begin{align}
    \left.
\begin{aligned}
\ket{\Psi_{n\vb{k}}^{-}} &= 0 \oplus \ket{s^{*}_{n(-\vb{k})}} \\
\ket{\Psi_{n\vb{k}}^{+}} &= \ket{s_{n\vb{k}}} \oplus 0
\end{aligned}
\right\}
\quad
\text{Inert sector.}
\end{align}
The quasi-particle energies are
\begin{align}
    \underbrace{E_{0\vb{k}}^{\pm}=\pm \sqrt{\xi_{0\vb{k}}^2+\abs{\Delta(\vb{k})}^2}}_{\text{Superconducting sector}}, \qquad \underbrace{E_{ n\vb{k}}^{\pm}=\pm (\epsilon_{n,\pm \vb{k}}-\mu)}_{\text{Inert sector}}, \quad n=1,\cdots, N-1.
\end{align}
The coherence factors take the usual form
\begin{align}
    u_{\vb{k}}=e^{i\theta_{u}(\vb{k})}\sqrt{\frac{1}{2}\left(1+\frac{\xi_{\vb{k}}}{E_{\vb{k}}^{+}}\right)}, \qquad v_{-\vb{k}}=e^{i\theta_{v}(\vb{k})}\sqrt{\frac{1}{2}\left(1-\frac{\xi_{-\vb{k}}}{E^{+}_{-\vb{k}}}\right)}.
\end{align}
While the coherence factors have the expected form, they also behave as sections of a vector bundle. This can be seen by considering the following expectation value in the superconducting sector
\begin{align}
    E^{\pm}_{0\vb{k}}=\bra{\Psi^{\pm}_{0\vb{k}}}H_{\mathcal{O}}(\vb{k})\ket{\Psi^{\pm}_{0\vb{k}}}. \label{EnergyExpVal}
\end{align}
Since the quasi-particle energy is gauge invariant by definition, and the Hamiltonian $H_{\mathcal{O}}(\vb{k})$ is gauge invariant, then we require that $\ket{\Psi_{0\vb{k}}^{\pm}}$ be gauge invariant, up to a local $U(1)$ phase. Indeed, the Nambu states themselves are permitted to transform as
\begin{align}
    \ket{\Psi^{\pm}_{0\vb{k}}} \rightarrow e^{i\theta({\vb{k}})}\ket{\Psi^{\pm}_{0\vb{k}}},
\end{align}
leaving Eq.~(\ref{EnergyExpVal}) invariant. This is a gauge transformation of the quasi-particle states of the effective BdG Hamiltonian. If we perform gauge transformations of the parent band however, the particle and hole sectors will generically mix (i.e. they do not simultaneously transform under a total $U(1)$ transformation). The only way that Eq.~(\ref{EnergyExpVal}) remains gauge invariant under parent band gauge transformations is if the coherence factors themselves transform: 
\begin{align}
    \ket{s_{\vb{k}}} &\rightarrow e^{i\chi(\vb{k})}\ket{s_{\vb{k}}} \implies u_{\vb{k}} \rightarrow e^{-i\chi(\vb{k})}u_{\vb{k}} \\
    \ket{s^{*}_{-\vb{k}}} &\rightarrow e^{-i\chi(-\vb{k})}\ket{s^{*}_{-\vb{k}}} \implies \quad v_{-\vb{k}} \rightarrow e^{i\chi(-\vb{k})}v_{-\vb{k}}.
\end{align}
Thus, much like $\Delta(\vb{k})$ is a section of a vector bundle which transforms, the coherence factors are sections of the superconducting sector that also transform oppositely to their frame, here consisting of the parent band electron and hole states. We now proceed to diagonalize the BdG Hamiltonian into its constituent quasiparticle bands. We begin by defining the unitary which diagonalizes the orbital basis Hamiltonian as
\begin{align}
    U_{\mathcal{O}}(\vb{k})&=\begin{pmatrix}
        \ket{\Psi^{+}_{0\vb{k}}} & \ket{\Psi^{+}_{1\vb{k}}} & \cdots & \ket{\Psi^{+}_{N-1,\vb{k}}}; & \ket{\Psi^{-}_{0\vb{k}}} & \ket{\Psi^{-}_{1\vb{\vb{k}}
        }} & \cdots & \ket{\Psi^{-}_{N-1,\vb{k}}}
    \end{pmatrix} \\
    &=\begin{pmatrix}
u_{\vb{k}}\ket{s_{0\vb{k}}}
& \ket{s_{1\vb{k}}}
& \cdots
& \ket{s_{N-1,\vb{k}}}
& -v^{*}_{\vb{k}}\ket{s_{0\vb{k}}}
& 0
& \cdots
& 0
\\[0.15cm]
v_{-\vb{k}}\ket{s^{*}_{0,-\vb{k}}}
& 0
& \cdots
& 0
& u^{*}_{-\vb{k}}\ket{s^{*}_{0,-\vb{k}}}
& \ket{s^{*}_{1,-\vb{k}}}
& \cdots
& \ket{s^{*}_{N-1,-\vb{k}}}
\end{pmatrix} \\
&=\begin{pmatrix}
    U(\vb{k})\mathcal{U}_{\vb{k}} & -U(\vb{k})\mathcal{V}_{\vb{k}}^{*} \\
    U^{*}(-\vb{k})\mathcal{V}_{-\vb{k}} & U^{*}(-\vb{k})\mathcal{U}^{*}_{-\vb{k}}
\end{pmatrix}=\begin{pmatrix}
    U(\vb{k}) & 0 \\
    0 & U^{*}(-\vb{k})
\end{pmatrix} \begin{pmatrix}
        \mathcal{U}_{\vb{k}} & -\mathcal{V}^{*}_{\vb{k}} \\
        \mathcal{V}_{-\vb{k}} & \mathcal{U}^{*}_{-\vb{k}}.
    \end{pmatrix}
\end{align}
The frame written in the band basis is then
\begin{align}
    U_{\mathcal{B}}(\vb{k})=V^{\dag}(\vb{k})U_{\mathcal{O}}(\vb{k})=\begin{pmatrix}
        \mathcal{U}_{\vb{k}} & -\mathcal{V}^{*}_{\vb{k}} \\
        \mathcal{V}_{-\vb{k}} & \mathcal{U}^{*}_{-\vb{k}},
    \end{pmatrix}
\end{align}
where the coherence factor matrices are defined as
\begin{align}
    \mathcal{U}_{\vb{k}}
    =
    \begin{pmatrix}
        u_{\vb{k}} & 0 & \cdots & 0 \\
        0 & 1 & \cdots & 0 \\
        \vdots & \vdots & \ddots & \vdots \\
        0 & 0 & \cdots & 1
    \end{pmatrix}, \quad \text{and } \quad \mathcal{V}_{\vb{k}}
    =
    \begin{pmatrix}
        v_{\vb{k}} & 0 & \cdots & 0 \\
        0 & 0 & \cdots & 0 \\
        \vdots & \vdots & \ddots & \vdots \\
        0 & 0 & \cdots & 0
    \end{pmatrix}.
\end{align}
Accordingly, the quasiparticle operators are obtained from
\begin{align}
    \Gamma_{\mathcal{B}}(\vb{k})=\begin{pmatrix}
        \gamma_{0\vb{k}} \\
        \gamma_{1\vb{k}} \\
        \vdots \\
        \gamma_{N-1,\vb{k}} \\
        \gamma^{\dag}_{0,-\vb{k}} \\
        \gamma^{\dag}_{1,-\vb{k}} \\
        \vdots \\
        \gamma^{\dag}_{N-1,-\vb{k}}
    \end{pmatrix}
    =\begin{pmatrix}
        \mathcal{U}_{\vb{k}} & -\mathcal{V}^{*}_{\vb{k}} \\
        \mathcal{V}_{-\vb{k}} & \mathcal{U}^{*}_{-\vb{k}},
    \end{pmatrix} \begin{pmatrix}
        c_{0\vb{k}} \\
        c_{1\vb{k}} \\
        \vdots \\
        c_{N-1,\vb{k}} \\
        c^{\dag}_{0,-\vb{k}} \\
        c^{\dag}_{1,-\vb{k}} \\
        \vdots \\
        c^{\dag}_{N-1,-\vb{k}}
    \end{pmatrix}=\begin{pmatrix}
        u^{*}_{\vb{k}}c_{0\vb{k}}+v^{*}_{-\vb{k}}c^{\dag}_{0,-\vb{k}} \\
        c_{1\vb{k}} \\
        \vdots \\
        c_{N-1 ,\vb{k}} \\
        -v_{\vb{k}}c_{0\vb{k}}+u_{-\vb{k}}c^{\dag}_{0,-\vb{k}} \\
        c^{\dag}_{1,-\vb{k}} \\
        \vdots \\
        c^{\dag}_{N-1,-\vb{k}}
    \end{pmatrix}.
\end{align}
Using these quasiparticle operators, the mean-field Hamiltonian is diagonalized as
\begin{align}
    H_{\text{MF}}&=E_{\text{SC}}[\Delta]+\frac{1}{2}\sum_{\vb{k}}\left(E_{0\vb{k}}\gamma^{\dag}_{0\vb{k}}\gamma_{0\vb{k}}+E_{0,-\vb{k}}\gamma^{\dag}_{0,-\vb{k}}\gamma_{0,-\vb{k}}\right)+\frac{1}{2}\sum_{\vb{k}}\sum_{n=1}^{N-1}\left(\xi_{n\vb{k}}c^{\dag}_{n\vb{k}}c_{n\vb{k}}+\xi_{n,-\vb{k}}c^{\dag}_{n,-\vb{k}}c_{n,-\vb{k}}\right).
\end{align}
The free energy of the superconducting ground state at $T=0$ is
\begin{align}
    F_{\text{SC}}[\Delta,T=0]=E_{\text{SC}}[\Delta]=\frac{1}{2}\sum_{\vb{k}}\left(\xi_{0\vb{k}}-E_{0\vb{k}}+\frac{\abs{\Delta(\vb{k})}^2}{2E_{0\vb{k}}}\right).
\end{align}

\section{The Singular Connection and Vorticity}
In this appendix we define the singular connection associated with the phase of the projected gap and use a punctured-domain Stokes argument to prove the vorticity constraint $V_{\Delta}=2C_P$. The proof avoids treating the vortex curvature as a delta function until the final distributional reformulation.

\subsection{The singular connection}\label{sec_app:singularconnectiondef}
As shown in the orbital basis, the full order parameter $\Delta_{\mathcal{O}}(\vb{k})$ is the gauge invariant combination
\begin{align}
    \Delta_{\mathcal{O}}(\vb{k})=\Delta(\vb{k})\ket{s_{\vb{k}}}\bra{s^{*}_{-\vb{k}}},
\end{align}
where $\Delta(\vb{k})$ is the band projected gap which transforms under gauge transformations of the parent band. The form of this order parameter is mathematically equivalent to the transition dipole for a pair of quantum states \cite{Ahn_2021,MeraSingularConnection,chen2026quantumgeometricquadrupolecooper}. The correct mathematical structure is known as homomorphism bundle, a vector bundle describing a pair of quantum states. In our case, the fiber of the relevant bundle is spanned by the pair frame $e_{P}(\vb{k})=\ket{s_{\vb{k}}}\bra{s^{*}_{-\vb{k}}}$. We may define a Berry connection using the pair frame $e_{P}(\vb{k})$ by
\begin{align}
    \mathcal{A}_{P}(\vb{k})=i\text{Tr}[e^{\dag}_P(\vb{k})de_{P}(\vb{k})]=\mathcal{A}(\vb{k})-\mathcal{A}(-\vb{k}). \label{eq_app_pairconnection}
\end{align}
However, since we are already given a choice of field $\Delta_{\mathcal{O}}(\vb{k})$ on the homomorphism bundle, we may additionally define a connection using its component $\Delta(\vb{k})$, which transforms opposite to the frame. Suppressing the argument $\vb{k}$ for clarity, we define the unit normalized band projected gap $\hat{\Delta}=\Delta / \abs{\Delta}$ away from zeros. We define the singular connection on the homomorphism bundle to be
\begin{align}
    \mathcal{A}_{\Delta}=-i\hat{\Delta}^{*}d\hat{\Delta}.
\end{align}
The connection $\mathcal{A}_{\Delta}$ is the phase only part of the singular connection. As explained in Ref.~\cite{MeraSingularConnection}, it is the phase only part of the singular connection which generically contributes to the Chern number, and it is this object which appears naturally in band projected superconductivity. Specifically, away from the zeros of $\Delta$, the singular connection is simply
\begin{align}
    \mathcal{A}_{\Delta}=d\text{arg}\Delta,
\end{align}
which transforms in the same way as $\mathcal{A}_{P}$. Thus the phase velocity $v_{\Delta}=\mathcal{A}_{\Delta}-\mathcal{A}_P$ defined in the main text is globally well defined away from zeros. If $v_{\Delta}$ were a smooth global one-form on $\mathbb{S}^2$, then $\int_{\mathbb{S}^2}dv_{\Delta}=0$ would follow immediately from Stokes theorem. Here, however, $\mathcal{A}_{\Delta}$ is singular at the zeros of $\Delta$. We therefore prove the corresponding statement by removing small disks around the zeros, applying Stokes theorem on the punctured sphere, and only then taking the limit in which the disks shrink to points.

\subsection{Proof of the vorticity theorem}\label{sec_app:vorticityproof}
In the subsection above, we argued that on the complement of the zeros of $\Delta$, the one-form $v_{\Delta}=\mathcal{A}_{\Delta}-\mathcal{A}_P$ is globally defined and smooth. We now apply Stokes theorem on the punctured sphere $\mathbb{S}^2_{\epsilon}$,
\begin{align}
    \mathbb{S}^2_{\epsilon}=\mathbb{S}^2 \setminus \bigcup_{i=0}^{N_V-1}D_{i}(\epsilon), \quad \text{with boundary   } \quad \partial \mathbb{S}^2_{\epsilon}=\bigcup_{i=0}^{N_V-1}(-\partial D_i(\epsilon)). \label{eq_app_puncturedsphere}
\end{align}
The curves $\partial D_{i}(\epsilon)$ are oriented counterclockwise as boundaries of the removed disks. Since the punctured sphere has a boundary component, we use Stoke's theorem to integrate $dv_{\Delta}$:
\begin{align}
    \int_{\mathbb{S}^2_{\epsilon}}dv_{\Delta}=\int_{\partial \mathbb{S}^2_{\epsilon}}v_{\Delta}=-\sum_{i=0}^{N_V-1}\oint_{\partial D_{i}(\epsilon)}(\mathcal{A}_{\Delta}-\mathcal{A}_{P}).
\end{align}
The left hand side of the equation can be simplified to
\begin{align}
    \int_{\mathbb{S}^2_{\epsilon}}dv_{\Delta}=\int_{\mathbb{S}^2_{\epsilon}}(d\mathcal{A}_{\Delta}-d\mathcal{A}_P)=-\int_{\mathbb{S}^2_{\epsilon}}\mathcal{F}_P,
\end{align}
where we used the fact that away from the singular points, $\mathcal{A}_{\Delta}=d\text{arg}\Delta$, so that $d\mathcal{A}_{\Delta}=d^2\text{arg}\Delta=0$ on $\mathbb{S}^2_{\epsilon}$. In the limit that $\epsilon \rightarrow 0$, the integral over $\mathcal{A}_P$ vanishes since the Berry connection can be chosen smooth on those points. Thus
\begin{align}
    \frac{1}{2\pi}\int_{\mathbb{S}^2}\mathcal{F}_P=\lim_{\epsilon \rightarrow 0}\frac{1}{2\pi}\sum_{i=0}^{N_V-1}\oint_{\partial D_i(\epsilon)}d\text{arg}\Delta. \label{eq_app_equivalencehomomorphismbundle}
\end{align}
The left hand side gives the Chern number of the homomorphism bundle, which is just $2C_P$. The right-hand side is a sum of terms of the form
\begin{align}
    \ell_i=\frac{1}{2\pi}\lim_{\epsilon \rightarrow 0}\oint_{\partial D_i(\epsilon)}d \text{arg}\Delta.
\end{align}

These correspond to winding numbers, and are independent of the choice of loops which enclose the vortices. This can be proven as follows. Let $\gamma_0$ and $\gamma_1$ be two oriented loops enclosing the same vortex and no other vortices. These are defined as the following maps
\begin{align}
    \gamma_0 : \mathbb{S}^1 \rightarrow \mathbb{S}^2 \setminus Z_{\Delta}, \quad \gamma_1 : \mathbb{S}^1 \rightarrow \mathbb{S}^2 \setminus Z_{\Delta},
\end{align}
where $\gamma(s) \in \mathbb{S}^2 \setminus Z_{\Delta}$, and $Z_{\Delta}=\{\vb{k}_i\}$ denotes the set of gap vortices. Suppose that $\gamma_1$ and $\gamma_2$ are homotopic in $\mathbb{S}^2 \setminus Z_{\Delta}$, meaning that there exists a smooth family of loops
\begin{align}
    H: \mathbb{S}^1 \times [0,1] \rightarrow \mathbb{S}^2 \setminus Z_{\Delta}
\end{align}
such that
\begin{align}
    \gamma_0(s)=H(s,0), \quad \gamma_1(s)=H(s,1).
\end{align}
Since the homotopy never crosses a zero $\Delta(\vb{k})$, the phase map $\Delta/ \abs{\Delta}:\mathbb{S}^2 \setminus Z_{\Delta} \rightarrow U(1)$ remains well-defined throughout the deformation. Therefore the maps
\begin{align}
    \frac{\Delta(\gamma_0(s))}{\abs{\Delta(\gamma_0(s))}}, \quad \frac{\Delta(\gamma_1(s))}{\abs{\Delta(\gamma_1(s))}}
\end{align}
are homotopic and hence have the same degree,
\begin{align}
    \frac{1}{2\pi}\oint_{\gamma_0}d\text{arg}\Delta=\frac{1}{2\pi}\oint_{\gamma_0}d\text{arg}\Delta.
\end{align}
Thus the winding number depends only on the homotopy class of the loop in the punctured momentum space. The total vorticity is then
\begin{align}
    V_{\Delta}=\sum_{i=0}^{N_V-1}\ell_i,
\end{align}
independent of how we choose $\gamma_i=\partial D_i(\epsilon)$. Equating this result to the left-hand side of Eq.~\eqref{eq_app_equivalencehomomorphismbundle}, we obtain the desired result
\begin{align}
    V_{\Delta}=2C_P.
\end{align}
Thus the singular curvature of $\mathcal{A}_{\Delta}$ may equivalently be represented by the distributional identity
\begin{align}
    d\mathcal{A}_{\Delta}=\sum_{i=0}^{N_V-1}2\pi \ell_i \delta^{(2)}(\vb{k}-\vb{k}_i)dk^x \wedge dk^y.
\end{align}
Consequently, we can integrate $dv_{\Delta}$ over $\mathbb{S}^2$ to find
\begin{align}
    0=\frac{1}{2\pi}\int_{\mathbb{S}^2}dv_{\Delta}=V_{\Delta}-2C_P,
\end{align}
which proves the result in the main text.

\section{BdG Topology}
Here we show the requirements for the validity of our Chern number calculation in the main text. Namely, we explain how the one-point compactification of the BdG projector is tied to the one-point compactification of the parent bands projector. We then derive the Berry curvature of the active superconducting sector, and show that the full BdG Berry curvature is obtained using particle hole symmetry of the BdG Hamiltonian. Finally, we reduce the BdG Chern number to boundary integrals on the punctures around vortices, highlighting that the topology is encoded on the boundary.

\subsection{Occupied BdG projector}\label{sec_app:BdGCompactification}
In order to have a well defined Chern number, we must perform a one-point compactification of the base space by proving that the BdG projector of the superconducting band, $P_{\text{SC}}(\vb{k})=\ket{\Psi^{-}_{0\vb{k}}}\bra{\Psi^{-}_{0\vb{k}}}$, reaches a fixed direction independent limit as $\abs{\vb{k}} \rightarrow \infty$. The projector onto the active sector takes the form
\begin{align}
    P_{\text{SC}}(\vb{k})=\begin{pmatrix}
        \abs{v_{\vb{k}}}^2 \ket{s_{\vb{k}}}\bra{s_{\vb{k}}} & -v^{*}_{\vb{k}}u_{-\vb{k}}\ket{s_{\vb{k}}}\bra{s^{*}_{-\vb{k}}} \\
        -u^{*}_{-\vb{k}}v_{\vb{k}}\ket{s^{*}_{-\vb{k}}}\bra{s_{\vb{k}}} & \abs{u_{-\vb{k}}}^2\ket{s^{*}_{-\vb{k}}}\bra{s^{*}_{-\vb{k}}}
    \end{pmatrix}. \label{ProjectorBdG}
\end{align}
As $\abs{\vb{k}}\rightarrow \infty$, at least one of the coherence factors must go to zero. This is satisfied if $\Delta(\vb{k}) \rightarrow 0$ at infinity. Numerically, we find that this is indeed the case. Therefore, we assume that $\abs{v_{\vb{k}}}^2 \rightarrow 0$ as $\abs{\vb{k}}\rightarrow 0$. In this limit, the projector approaches the fixed point
\begin{align}
    \lim_{\abs{\vb{k}} \rightarrow \infty}P_{\text{SC}}(\vb{k})=\lim_{\abs{\vb{k}} \rightarrow \infty}\begin{pmatrix}
        0 & 0 \\
        0 & P^{*}_0(\vb{k})
    \end{pmatrix}=\begin{pmatrix}
        0 & 0 \\
        0 & P^{*}_{\infty}
    \end{pmatrix}.
\end{align}
Where in the last line, we have used the fact that the parent band projector $P_{0}(\vb{k})$ reaches a fixed direction independent limit, and therefore so too does the corresponding hole projector $P_{0}^{*}(\vb{k})$. Therefore, we may include the point at infinity such that the domain is $\mathbb{S}^2=\mathbb{R}^2 \cup \{\infty\}$.

\subsection{BdG Berry Curvature}\label{sec_app:BdGBerry}
In order to define the BdG Berry curvature, we will need to define a punctured disk, since the phase of the coherence factors isn't well defined at the vortex cores. As explained in the main text, we define the punctured disk $D_{\Lambda,\epsilon}$, where we remove circles $D_{\alpha}(\epsilon)$ around each vortex location. The resulting domain is 
\begin{align}
D_{\Lambda,\epsilon}=D_{\Lambda} \setminus \bigcup_{\alpha=0}^{N_V-1}D_{\alpha}(\epsilon), \quad \text{with boundary} \quad \partial D_{\Lambda,\epsilon}=\partial D_{\Lambda}\cup \bigcup_{\alpha=0}^{N_V-1}(-\partial D_{\alpha}(\epsilon)).
\end{align}

The BdG Berry curvature is computed on this punctured domain by assuming that the superconducting band is isolated from the inert bands of the BdG Hamiltonian. This leads to the decomposition of the occupied projector $P(\vb{k})$ into
\begin{align}
    P(\vb{k})=P_{\text{SC}}(\vb{k})+P_h(\vb{k}),
\end{align}
where 
\begin{align}
    P_{\text{SC}}(\vb{k})=\ket{\Psi_{0\vb{k}}}\bra{\Psi_{0\vb{k}}} \quad \text{and} \quad P_h(\vb{k})=\sum_{n=1}^{N-1}\ket{\Psi^{-}_{n\vb{k}}}\bra{\Psi^{-}_{n\vb{k}}}
\end{align}
are the relevant projectors onto the active superconducting sector, and the inert hole sector. Since the superconducting band is isolated from the hole contributions, the BdG Berry curvature decomposes as
\begin{align}
    \mathcal{F}_{\text{BdG}}(\vb{k})=\mathcal{F}_{\text{SC}}(\vb{k})+\mathcal{F}_{h}(\vb{k}).
\end{align}
Since the superconducting band is isolated from the remaining inert hole bands, the Berry curvature contributions are computed from the formulas
\begin{align}
    \mathcal{F}_{\text{SC}}(\vb{k})=\nabla_{\vb{k}} \wedge \mathcal{A}_{\text{SC}}(\vb{k}), \quad \mathcal{F}_{h}(\vb{k})=i\epsilon^{\mu\nu}\text{Tr}[P_{h}(\vb{k})\partial_{\mu}P_h(\vb{k})\partial_{\nu}P_h(\vb{k})].
\end{align}
Here $\mathcal{A}_{\text{SC}}(\vb{k})=i\bra{\Psi^{-}_{0\vb{k}}}\ket{\nabla_{\vb{k}}\Psi^{-}_{0\vb{k}}}$ is the Berry connection of the superconducting sector. Using $\ket{\Psi^{-}_{0\vb{k}}}$ (defined in Eq.\eqref{eq:define_Psi-}), the Berry connection is
\begin{align}
    \mathcal{A}_{\text{SC}}(\vb{k})=i\left(v_{\vb{k}}\nabla_{\vb{k}}v^{*}_{-\vb{k}}+u_{-\vb{k}}\nabla_{\vb{k}}u^{*}_{-\vb{k}}\right)+\abs{v_{\vb{k}}}^2 \mathcal{A}(\vb{k})+\abs{u_{-\vb{k}}}^2\mathcal{A}(-\vb{k}),
\end{align}
where $\mathcal{A}(\vb{k})=i\bra{s_{\vb{k}}}\ket{ds_{\vb{k}}}$ is the parent band Berry connection. The Berry curvature is obtained by taking the exterior derivative:
\begin{align}
    \mathcal{F}_{\text{SC}}(\vb{k})=i\left(\nabla_{\vb{k}}v_{\vb{k}}\wedge \nabla_{\vb{k}}v^{*}_{-\vb{k}}+\nabla_{\vb{k}}u_{-\vb{k}}\wedge \nabla_{\vb{k}}u^{*}_{-\vb{k}}\right)+\nabla_{\vb{k}}\abs{v_{\vb{k}}}^2 \wedge \mathcal{A}_P(\vb{k})+\abs{v_{\vb{k}}}^2\mathcal{F}(\vb{k})-\abs{u_{-\vb{k}}}^2\mathcal{F}(-\vb{k}).
\end{align}
Here $\mathcal{A}_P(\vb{k})$ is the pair connection defined in Eq.~\eqref{eq_app_pairconnection}. This equation is simplified by inserting the local polar form of the coherence factors, $v_{\vb{k}}=\abs{v_{\vb{k}}}e^{i\theta_{v}(\vb{k})}$, and $u_{-\vb{k}}=\abs{u_{-\vb{k}}}e^{i\theta_{u}(\vb{k})}$, into the Berry curvature. We obtain
\begin{align}
    \mathcal{F}_{\text{SC}}(\vb{k})&=i\left(-i\nabla_{\vb{k}}\abs{v_{\vb{k}}}^2 \wedge \nabla_{\vb{k}}\theta_v(\vb{k})-i\nabla_{\vb{k}}\abs{u_{-\vb{k}}}^2 \wedge \nabla_{\vb{k}}\theta_u(\vb{k})\right)+\nabla_{\vb{k}}\abs{v_{\vb{k}}}^2 \wedge \mathcal{A}_P(\vb{k})+\abs{v_{\vb{k}}}^2\mathcal{F}(\vb{k})-\abs{u_{-\vb{k}}}^2\mathcal{F}(-\vb{k}) \\
    &=-\nabla_{\vb{k}}\abs{v_{\vb{k}}}^2 \wedge \nabla_{\vb{k}}\text{arg}\Delta(\vb{k})+\nabla_{\vb{k}}\abs{v_{\vb{k}}}^2 \wedge \mathcal{A}_P(\vb{k})+\abs{v_{\vb{k}}}^2 (\mathcal{F}(\vb{k})+\mathcal{F}(-\vb{k}))-\mathcal{F}(-\vb{k}) \\
    &=-\nabla_{\vb{k}}\wedge \left(\abs{v_{\vb{k}}}^2\left(\nabla_{\vb{k}}\text{arg}\Delta(\vb{k})-\mathcal{A}_P(\vb{k})\right)\right)-\mathcal{F}(-\vb{k}) \\
    &=-\nabla_{\vb{k}} \wedge \vb{J}_{\Delta}(\vb{k})-\mathcal{F}(-\vb{k}). \label{eq_app_SCBerry}
\end{align}
In going from the first line to the second, we used the fact that $\abs{v_{\vb{k}}}^2+\abs{u_{-\vb{k}}}^2=1$ to simplify the last two parent band Berry curvature terms. We also used this identity to obtain $\nabla_{\vb{k}}\abs{u_{-\vb{k}}}^2=-\nabla_{\vb{k}}\abs{v_{\vb{k}}}^2$, allowing the first two phase terms involving derivatives of $\theta_u(\vb{k})$ and $\theta_v(\vb{k})$ to be combined. The phases are combined by noting that $\theta_{u}(\vb{k})-\theta_v(\vb{k})=\text{arg}\Delta(\vb{k})$. In going from the second line to the third line, we used the fact that $\mathcal{F}_P(\vb{k})=\nabla_{\vb{k}} \wedge \mathcal{A}_P(\vb{k})=\mathcal{F}(\vb{k})+\mathcal{F}(-\vb{k})$, allowing us to rewrite the first four terms of the second line in terms of a total derivative. In doing so we have used the fact that on the punctured domain, $\nabla_{\vb{k}} \wedge \nabla_{\vb{k}}\text{arg}\Delta(\vb{k})=0$. Finally, in the last line, we have defined the phase current
\begin{align}
    \vb{J}_{\Delta}(\vb{k})=\abs{v_{\vb{k}}}^2\left(\mathcal{A}_{\Delta}(\vb{k})-\mathcal{A}_{P}(\vb{k})\right), 
\end{align}
where $\mathcal{A}_{\Delta}=\nabla_{\vb{k}}\text{arg}\Delta(\vb{k})$ is the singular connection. We now proceed to show that the residual parent band curvature term $\mathcal{F}(-\vb{k})$ does not contribute to the full BdG Berry curvature.

\subsection{Inert-hole cancellation}\label{sec_app:inertholecancellation}
The normal state Hamiltonian has projector
\begin{align}
    \mathds{1}=P_{0}(\vb{k})+Q(\vb{k}).
\end{align}
where $P_0(\vb{k})$ is the projector onto the parent band, and $Q_{\vb{k}}$ is the projector along the remaining $N-1$ bands which do not participate in superconductivity. This fact will be important in what follows. We perform a particle-hole transformation on the normal state Hamiltonian $H(\vb{k})$ to find
\begin{align}
    H(\vb{k}) \rightarrow -H^{T}(-\vb{k}).
\end{align}
The corresponding hole-band projectors are
\begin{align}
    P_0(\vb{k}) \rightarrow P_{0}^{*}(-\vb{k}), \quad \text{and} \quad \mathcal{Q}(\vb{k}) \rightarrow Q^{*}(-\vb{k}),
\end{align}
We note that the hole projector $Q^{*}(-\vb{k})$ corresponds to the same hole projector $P_h(\vb{k})$ of the inert occupied BdG bands. Therefore, we can relate the parent band hole Berry curvature to the Berry curvature of the inert sector by computing
\begin{align}
    \mathcal{F}_{h}(\vb{k})=\Omega_{Q^{*}}(\vb{k})=i\epsilon^{\mu\nu}\text{Tr}[Q^{*}(\vb{k})\partial_{\mu}Q^{*}(\vb{k})\partial_{\nu}Q^{*}(\vb{k})]=-i\epsilon^{\mu\nu}\text{Tr}[P_0^{*}(\vb{k})\partial_{\mu}P_0^{*}(\vb{k})\partial_{\nu}P_0^{*}(\vb{k})]=-\Omega(-\vb{k})^{*}=\Omega(-\vb{k}).
\end{align}
Thus $\mathcal{F}_{h}(\vb{k})=\Omega(-\vb{k})$ cancels the residual parent band Berry curvature Eq.~\eqref{eq_app_SCBerry} to give (on the punctured disk)
\begin{align}
    \mathcal{F}_{\text{BdG}}(\vb{k})=-\nabla_{\vb{k}}\wedge \vb{J}_{\Delta}(\vb{k}).
\end{align}

\section{Numerical solution of the gap equation}\label{sec_app:NLGE}
To solve the full gap equation we use the anti-symmertized Cooper vertex $\mathcal{W}_{\mathbf{k},\mathbf{k'}}$ for all pairs of points $\mathbf{k}$ and $\mathbf{k'}$ within some momentum cutoff $\Lambda$ of the origin $\mathbf{k=0}$. This cut-off is chosen to be $\sim 2.5k_{F}-3k_{F}$. We solve the full-gap equation Eq. \eqref{eq:gap_eqn}, self-consistently starting from different initial choices of $\Delta({\mathbf{k}})$. The gap is solved for on either a $81\times 81$ or a $121\times 121$ grid. We find that the results obtained are robust to changes in grid discretization for the grid sizes used. The presence of vortices in the gap can be identified from the phase winding of the gap in $\arg[\Delta({\mathbf{k}})]$, coinciding with zeros of the gap in $|\Delta({\mathbf{k}})|$. Additionally, for certain parameter values, we also self-consistently solved for the chemical potential $\mu$, keeping the electron filling $\nu$ a constant. With this constant electron filling, the primary results — including the identification of the nucleation and vortex number saturation regimes for vortices, and the behavior of the condensation energy with increasing $N$ and $\mathcal{B}$ – remain qualitatively unchanged from those obtained at constant chemical potential $\mu$.

\section{Numerical computation of a single band BdG Chern number}\label{sec_app:numerical_chern_number}

As seen earlier, the BdG Chern number of the $\lambda_N$-model model can be written as $C_{\text{BdG}}=\mathcal{Q}$ in an appropriate choice of gauge. For the $2\times 2$ BdG Hamiltonian in Eq.~\eqref{BdG2by2} (projected onto the lowest band), a BdG Chern number can be calculated as well using the corresponding BdG eigenstates, where the pair connection is defined in a smooth gauge. We calculate this single band BdG Chern number numerically using the algorithm by Fukui et. al \cite{Fukui2005}. Using the gap function $\Delta_{\mathbf{k}}$ from solving the gap equation in Eq.~\eqref{BdG2by2}, we find that this numerically computed Chern number is exactly equal to the sum of windings of the vortices enclosed by the Fermi surface, i.e. $C_{\text{BdG,numerical}}=\mathcal{Q}$. This relation also holds for annular Fermi surfaces, where $C_{\text{BdG,numerical}}$ only counts the sum of the windings of the vortices enclosed within the annulus region of the Fermi surface. 

\onecolumngrid
\section{Identifying leading harmonics by expanding the Cooper vertex}\label{sec_app:harmonics of the Cooper vertex}
We can identify the winding of the leading superconducting instability from expanding the Cooper vertex $\mathcal{W}_{\mathbf{k},\mathbf{k}'}$ in various limits. The anti-symmetrized Cooper vertex is expressed in terms of the form factors as in Eq. \eqref{eq:form_factor_dressed_vertex}. The form factors can be expanded in the limit of $\mathcal{B}k_F^2\gg1$, or $\mathcal{B}k_F^2\ll1$, and $\sigma^2k_F^2\ll1$ as shown below.
\subsection{$\mathcal{B}k_F^2\gg1$ and $\sigma^2k_F^2\ll1$}
In the limit of $\mathcal{B}|\vb{k}|^2\gg1$ and $\mathcal{B}|\vb{k}'|^2\gg1$, the form factor can be expanded as follows
\begin{gather}
\mathcal{F}(\mathbf{k},\mathbf{k}')\approx\frac{e^{i\phi_{\mathbf{k},\mathbf{k}'}(N-1)}+\frac{2(N-1)}{\mathcal{B}|\mathbf{k}||\mathbf{k}'|} e^{i\phi_{\mathbf{k},\mathbf{k}'}(N-2)} }{\sqrt{\left(1+\frac{2(N-1)}{\mathcal{B}|\mathbf{k}|^2}\right)\left(1+\frac{2(N-1)}{\mathcal{B}|\mathbf{k}'|^2}\right)}}\\
    \implies\mathcal{F}(\mathbf{k},\mathbf{k}')^2\approx\frac{e^{i\phi_{\mathbf{k},\mathbf{k}'}(2N-2)}+\frac{4(N-1)}{\mathcal{B}|\mathbf{k}||\mathbf{k}'|} e^{i\phi_{\mathbf{k},\mathbf{k}'}(2N-3)}}{\left(1+\frac{2(N-1)}{\mathcal{B}|\mathbf{k}|^2}\right)\left(1+\frac{2(N-1)}{\mathcal{B}|\mathbf{k}'|^2}\right)}.
\end{gather}
Using Eq. \eqref{eq:anti_symm_Cooper_vertex}, we can express the Cooper vertex in this limit as
\begin{gather}
\mathcal{W}_{\mathbf{k},\mathbf{k}'}
\approx-\frac{U_0e^{-\frac{\sigma^2}{2}(|\mathbf{k}|^2+|\mathbf{k}'|^2)}}{\left(1+\frac{2(N-1)}{\mathcal{B}|\mathbf{k}|^2}\right)\left(1+\frac{2(N-1)}{\mathcal{B}|\mathbf{k}'|^2}\right)}\left(e^{i\phi_{\mathbf{k},\mathbf{k}'}(2N-2)}\sinh{\left(\sigma^2\mathbf{k}\cdot \mathbf{k'}\right)}+\frac{4(N-1)}{\mathcal{B}|\mathbf{k}||\mathbf{k}'|}e^{i\phi_{\mathbf{k},\mathbf{k}'}(2N-3)} \cosh{\left(\sigma^2\mathbf{k}\cdot \mathbf{k'}\right)}\right).
\end{gather}
Now assuming the momenta are on the Fermi surface, i.e., $|\mathbf{k}|=|\mathbf{k}'|=k_F$, we further expand $\mathcal{W}_{\mathbf{k},\mathbf{k}'}$ in the limit $\sigma^2k_F^2\ll1$ in addition to $\mathcal{B}k_F^2\gg1$. Finally we obtain
\begin{gather}
\mathcal{W}_{\mathbf{k}_F,\mathbf{k}_F'}\approx-\frac{U_0e^{-\sigma^2k_F^2}}{\left(1+\frac{2(N-1)}{\mathcal{B}k_F^2}\right)^2}\left( \frac{\sigma^2 k_F^2}{2}e^{i\phi_{\mathbf{k},\mathbf{k}'}(2N-1)}+\left(\frac{4(N-1)}{\mathcal{B}k_F^2}+\frac{\sigma^2 k_F^2}{2}\right)e^{i\phi_{\mathbf{k},\mathbf{k}'}(2N-3)}\right)\label{eq:exp_vertex_1}
\end{gather}
where we have expanded to linear order in $\sigma^2k_F^2$. From Eq. \eqref{eq:exp_vertex_1}, we find that the most negative harmonic, with the highest $T_c$, has a winding of 
$\ell=2N-3$ in this limit of $\mathcal{B}k_F^2\gg1$ and $\sigma^2k_F^2\ll1$. Thus in this limit the leading superconducting instability has its gap winds around the Fermi surface $\ell=2N-3$ times. This weak coupling result matches with the full-gap equation in the limit of large $\mathcal{B}k_F^2$, as shown in Fig. \ref{fig:vortex_nucleation}. In particular in this limit, we find that increasing $\mathcal{B}$ leads to saturation of vortices, with $\mathcal{Q}_{\text{sat}}=2N-3$. $\mathcal{Q}_{\text{sat}}$ counts the sum the windings of all the vortices enclosed by the Fermi surface, and is also equal to the winding of the gap around the Fermi surface, similar to $\ell$. This upper bound on the maximum winding of the gap, is not present in the ICB model, where the winding of the gap function keeps on increasing with $\mathcal{B}$, as also shown in Fig. \ref{fig:vortex_nucleation}.
\subsection{The limit $\mathcal{B}k_F^2\ll1$ and $\sigma^2k_F^2\ll 1$}
We now expand the form factors in the limit $\mathcal{B}|\mathbf{k}|^2\ll1$ and $\mathcal{B}|\mathbf{k}'|^2\ll1$,
\begin{gather}
\mathcal{F}(\mathbf{k},\mathbf{k}')\approx \frac{1+\frac{\mathcal{B}}{2}|\mathbf{k}||\mathbf{k}'|e^{i\phi_{\mathbf{k},\mathbf{k}'}}}{\sqrt{\left(1+\frac{\mathcal{B}}{2}|\mathbf{k}|^2\right)\left(1+\frac{\mathcal{B}}{2}|\mathbf{k}'|^2\right)}}\\
\implies\mathcal{F}(\mathbf{k},\mathbf{k}')^2\approx\frac{1+\mathcal{B}|\mathbf{k}||\mathbf{k}'|e^{i\phi_{\mathbf{k},\mathbf{k}'}}}{\left(1+\frac{\mathcal{B}}{2}|\mathbf{k}|^2\right)\left(1+\frac{\mathcal{B}}{2}|\mathbf{k}'|^2\right)}.
\end{gather}
Assuming the momenta are on the Fermi surface, i.e., $|\mathbf{k}|=|\mathbf{k}'|=k_F$, we further expand $\mathcal{W}_{\mathbf{k},\mathbf{k}'}$ in the limit $\sigma^2k_F^2\ll1$ to linear order
\begin{gather}
\mathcal{W}_{\mathbf{k}_F,\mathbf{k}_F'}\approx -\frac{U_0e^{-\sigma^2k_F^2}}{\left( 1+\frac{\mathcal{B}}{2}k_F^2\right)^2}\left( \frac{\sigma^2k_F^2}{2}e^{-i\phi_{\mathbf{k},\mathbf{k}'}}+\left(\frac{\sigma^2}{2}+\mathcal{B}\right)k_F^2e^{i\phi_{\mathbf{k},\mathbf{k}'}}\right).\label{eq:exp_vertex_2}
\end{gather}
We see from Eq. \eqref{eq:exp_vertex_2} that the leading harmonic in the limit of $\mathcal{B}k_F^2\ll1$ and $\sigma^2k_F^2\ll1$ has winding $\ell=1$, independent of $N$. We find agreement in Fig. \ref{fig:vortex_nucleation}, where in this limit we find $\mathcal{Q}=1$.
\\\\

\twocolumngrid

\section{Ansatz for the gap function}\label{app_sec:ansatz_for_gap}
In this section we use expand on the ansatz for the gap function in the presence of momentum space vortices. As in Eq.~\eqref{eq:vortex_ansatz}, the gap can be written as 
\begin{gather}
    \Delta(\mathbf{k})=\Delta_0 f(\mathbf{k}, \{\mathbf{k}_{i}\})\prod_{i=0}^{N_V-1} e^{i\ell_{i}\phi(\mathbf{k}-\mathbf{k}_{i})},\label{eq:vortex_ansatz_app}
\end{gather}
where $\mathbf{k}_{i}$ are the location of the vortices in the gap ($i=0,\cdots,N_V-1$), and $\phi(\mathbf{k}-\mathbf{k}_{i})=\arctan(\frac{k_y-k_{i,y}}{k_x-k_{i,x}})$. At the vortex core $f(\mathbf{k}_{i})=0$ ensuring that the gap goes vanishes at those points. 

Since from our numerics we find that the vortices always occur on a ring enclosing the origin, the position of the vortices can be written as:
\begin{gather}
    \mathbf{k}_0=(0,0):\quad \text{central vortex}\\
    \mathbf{k}_i=\rho\left(\cos{\phi_i},\sin{\phi_i}\right):\quad \text{ring of vortices}
\end{gather}
where $\phi_i=\frac{2\pi i}{(N_V-1)}+\Delta\phi$ and $\rho$ is the radius of the vortex ring. In subsequent calculations we choose $\Delta \phi=0$. Furthermore, motivated by the numerics, we also take $\ell_i=+1$ for all vortices. In our ansatz we express
\begin{gather}
    f(\mathbf{k},\{\mathbf{k}_i\})=e^{-\frac{(|\mathbf{k}|-k_R)^2}{2\beta^2}}\prod_{i=0}^{N_V-1} h(|\mathbf{k}_i|)
\end{gather}
where $h(\abs{\vb{k}})=\tanh{(\abs{\vb{k}}/\xi_l)}$. For weak interactions, one may set $k_R=k_F$, since the gap remains localized near the Fermi surface (see Appendix~\ref{sec_app:NLGE_additional_data}). In all subsequent calculations using this ansatz we set $k_R=k_F$. The momentum scale $\xi_l$ is a phenomenological parameter that controls the extent of the halo surrounding a vortex core.

\begin{figure}
    \centering
    \includegraphics[scale=1.1]{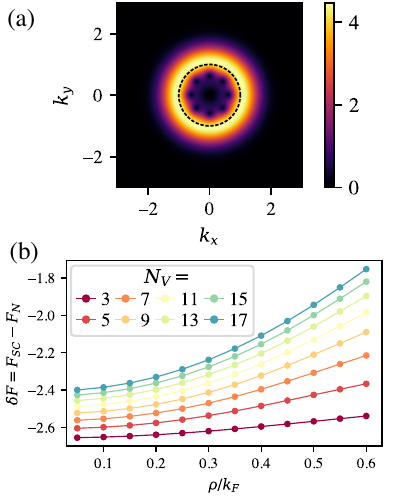}
    \caption{(a) $|\Delta_{\mathbf{k}}|$ as recreated by the ansatz in Eq.~\eqref{eq:vortex_ansatz_app} where we have chosen $N_V=9$, $\xi_l=0.3$, $\Delta_0=5$, $\beta=0.4$, $\ell_i=+1$. The Fermi surface is highlighted by the dashed line and $k_F=1.0$. (b) Shows the variation of the condensation energy with the radius of the momentum-space vortex ring $\rho$, calculated using the ansatz for the gap for different values of $N_V$.}
    \label{fig:ansatz_gap_and_condensation_energy}
\end{figure}

Fig.~\ref{fig:ansatz_gap_and_condensation_energy}(a) shows qualitatively how the gap function can be approximately captured by this ansatz with $N_V=9$ vortices. We may use this ansatz to try to recreate the trend in the condensation energy as the radius of the vortex ring contracts.

We show this trend in Fig.~\ref{fig:ansatz_gap_and_condensation_energy}(b) where we observe that as the radius of the vortex ring controlled by $\rho$ shrinks, the condensation energy decreases qualitatively matching the trend of the numerical solution of the gap equation presented in the main text. Additionally, the condensation energy grows with increasing number of vortices $N_V$, as the presence of more vortices drives the gap to have a value closer to zero over a larger area leading to an increase of the condensation energy.

\section{Suppression of the gap by the quantum metric}\label{sec_app:suppression_of_gap}
In this section, we show that the trace of the quantum metric $\text{tr}[g_{\mu\nu}(\mathbf{k})]$, can suppress the gap function in the limit of $\sigma^2\gg1$, following the derivation of Ref.~\cite{patri2025}. In this limit following Ref.~\cite{patri2025} the zero temperature gap equation can be expressed as
\begin{gather}
    |\Delta({\mathbf{k}})|=U\sum_{\mathbf{k'}}e^{-(\sigma^2\delta_{\mu\nu}+g_{\mu\nu}(\mathbf{k'})(\mathbf{k-k'})^{\mu}(\mathbf{k-k'})^{\nu})}\frac{|\Delta({\mathbf{k'}})|}{2E_{\mathbf{k'}}}
\end{gather}
Since $\sigma\gg1$, the integral is heavily suppressed away from $\mathbf{k}$. Therefore, for small momentum transfer $\mathbf{k}-\mathbf{k'}=\mathbf{q}$ with $|\mathbf{q}|\ll 1/\sigma$, we can write $\Delta({\mathbf{k}})\approx\Delta({\mathbf{k'}})$, $E_{\mathbf{k}}\approx E_{\mathbf{k'}}$, $g_{\mu\nu}(\mathbf{k})\approx g_{\mu\nu}(\mathbf{k'})$. This enables us to write the gap equation as
\begin{gather}
 2E_{\mathbf{k'}}=U\sum_{\mathbf{q}}e^{-(\sigma^2\delta_{\mu\nu}+g_{\mu\nu}(\mathbf{k'})q^{\mu}q^{\nu})}
\end{gather}
We can perform the Gaussian integral to obtain
\begin{gather}
    E^2_{\mathbf{k'}}=\left(\frac{U}{8\pi}\right)^2\frac{1}{\sigma^4+\sigma^2 \text{tr}[g_{\mu\nu}(\mathbf{k'})]+\text{det}[g_{\mu\nu}(\mathbf{k'})]}
\end{gather}
Expanding in the limit of $\sigma\gg1$, we can write
\begin{gather}
    |\Delta({\mathbf{k'}})|^2\approx \left(\frac{U}{8\pi \sigma^2}\right)^2\left(1-\frac{\text{tr}[g_{\mu\nu}(\mathbf{k'})]}{\sigma^2} \right)-\xi_{\mathbf{k'}}^2,
\end{gather}
giving us Eq.~\eqref{eq:metric_suppressing_gap} of the main text. Therefore, we observe that that in this limit of $\sigma^2\gg1$, we can analytically see that the trace of the quantum metric suppress the gap function.

\section{Additional data from solving the gap-equation numerically}\label{sec_app:NLGE_additional_data}
Here we present additional data obtained from the solving the full-gap equation. We show variation of the gap function with changes to interaction strength $U$, interaction length $\sigma$, and Fermi surface topology.

\subsection{Variation with interaction strength $U$}\label{app:var_U}
\begin{figure}
    \centering
    \hspace{-0.5cm}
    \includegraphics[scale=0.8]{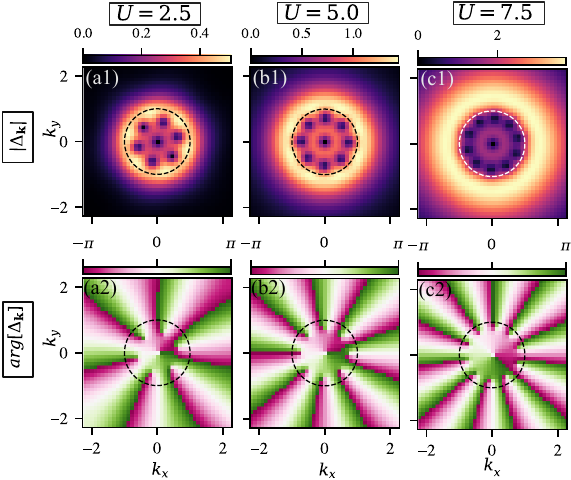}
    \caption{Gap structure as a function of increasing interaction strength $U$. With increasing $U$, both the number of nucleated vortices and the magnitude of the gap function increase. The model parameters are set at $\sigma^2=1$, $N=8$, $\mathcal{B}=8$. The dashed line indicates the Fermi surface.}
    \label{fig:vortices_increasing_U}
\end{figure}

In the main text, we assumed a constant value of the interaction strength $U$, and was set to $U=5$, to enable all electrons in the Fermi surface to participate in pairing. Below, we present additional data exploring variations in $U$. For a constant value of $N$, $\mathcal{B}$, and $\sigma$, an increase of $U$ enhances vortex formation as shown in Fig. \ref{fig:vortices_increasing_U}. Furthermore, as expected physically from Eqn. \eqref{eq:metric_suppressing_gap}, increasing $U$ leads to a larger gap magnitude, $\text{max}[\Delta_{\mathbf{k}}]$ (see Fig. \ref{fig:vortices_increasing_U}). For stronger interactions, electrons across the entire Fermi sea may participate in pairing, and therefore the Berry curvature exerts a greater influence, suppressing gap formation in regions where the Berry curvature is large. 

\begin{figure}
    \centering
    \hspace{-0.3cm}
    \includegraphics[scale=1.0]{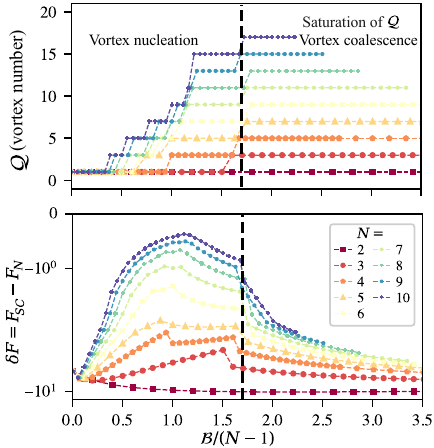}
    \caption{Top panel shows that vortex number $\mathcal{Q}$ with the x-axis is scaled as $\mathcal{B}/(N-1)$ for $U=5$. The bottom panel shows the condensation energy on the same scale. The vortex nucleation region is separated from the vortex number saturation by the vertical dashed line, and this is independent of $N$. With increasing $\mathcal{B}$ in the saturation regime the vortices tend to coalesce, as was discussed in the main text.}
    \label{fig:vortex_nucleation_coalescence_log_scale}
\end{figure}

\begin{figure}
    \centering
    \hspace{-0.3cm}
    \includegraphics[scale=1.0]{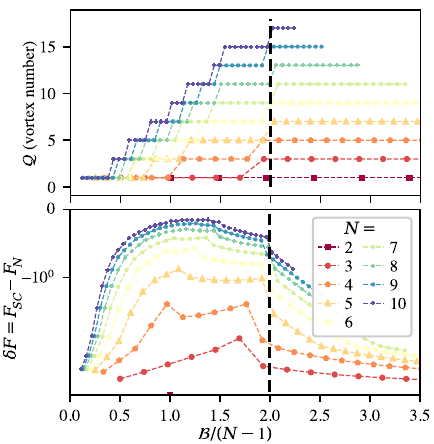}
    \caption{Same as Fig.~\ref{fig:vortex_nucleation_coalescence_log_scale} but with $U=3.5$. The vertical dashed line at $\mathcal{B}_{\text{crit}}/(N-1)\sim2.0$ separates the vortex nucleation and vortex number saturation regimes.
    }
    \label{fig:vortex_nucleation_coalescence_log_scale_3.5}
\end{figure}

Having identified the regions of vortex nucleation and vortex saturation in the main text, here we observe a general trend independent of $N$ which separates these two regions. In Fig. \ref{fig:vortex_nucleation_coalescence_log_scale} we plot the vortex number $\mathcal{Q}$ for different values of $N$, where the $x$-axis is scaled as $\mathcal{B}/(N-1)$. We observe that $\mathcal{B}/(N-1)\sim 1.7$ separates the two regions for $U=5$. This critical value  of $\mathcal{B}/(N-1)$ is dependent on the interaction strength as shown in Fig.~\ref{fig:vortex_nucleation_coalescence_log_scale}; for $U=3.5$, we find that $\mathcal{B}/(N-1) \sim 2.0$ separates the two regions. Therefore, for a smaller values of $U$, this critical threshold of $\mathcal{B}/(N-1)$ is higher as a larger $\mathcal{B}$ is required to achieve vortex saturation for smaller interaction strengths. The condensation energy in Figs.~\ref{fig:vortex_nucleation_coalescence_log_scale}, \ref{fig:vortex_nucleation_coalescence_log_scale_3.5}, reveals a dome-like feature. It initially increases with $\mathcal{B}$, and then tapers off as vortices are nucleated in the gap. Upon entering the vortex number saturation regime, it decreases rapidly to approach a value independent of $N$.

\begin{figure*}
    \centering
    \hspace{-0.5cm}
    \includegraphics[scale=0.85]{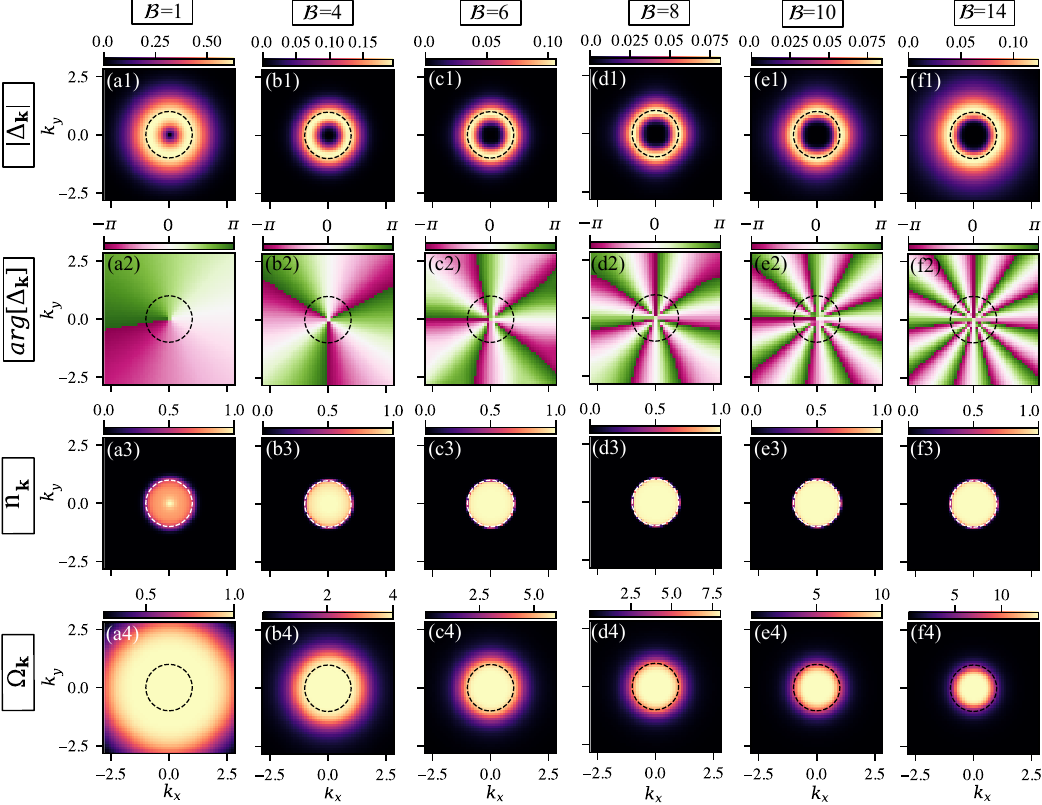}
    \caption{All panels are similar to Fig. \ref{fig:trend_gap_N_8}, with the distinction that the interaction strength is $U=1$. Similar to Fig. \ref{fig:trend_gap_N_8}, the parameters are $N=8$, $\sigma^2=1$. As the interaction strength is smaller compared to Fig. \ref{fig:trend_gap_N_8}, the gap is more localized near the Fermi surface.}
    \label{fig:trend_increasing_N_8_b}
\end{figure*}

If the interaction strength is small enough so that only electrons close to the Fermi surface participate in pairing (see Fig.~\ref{fig:trend_increasing_N_8_b}), then the gap is confined close to the Fermi surface and the momentum-space vortices vortices are confined close to the origin. However, as seen in Fig.~\ref{fig:trend_increasing_N_8_b}, more vortices are nucleated with increasing $\mathcal{B}$, a trend that is independent of the interaction strength.


\subsection{Variation with interaction length $\sigma$}\label{app:var_sigma}
\begin{figure}[h!]
    \centering
    \hspace{-0.5cm}
    \includegraphics[scale=0.8]{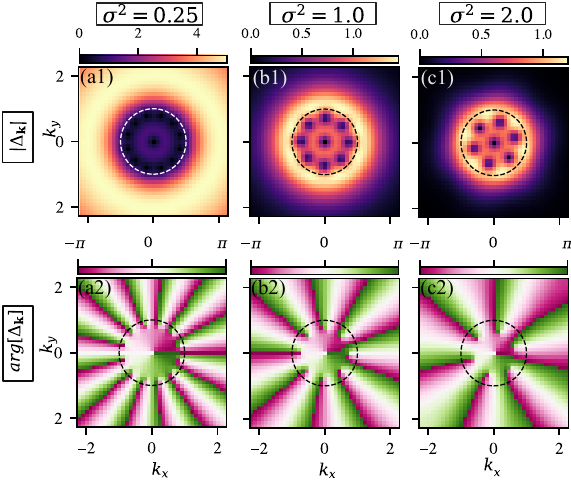}
    \caption{The variation of the gap structure with increasing the interaction length $\sigma^2$. Increasing $\sigma^2$ suppresses vortex nucleation, and the magnitude of the gap function. The model parameters are set at $U=5$, $N=8$, $\mathcal{B}=8$. The dashed line indicates the Fermi surface.}
    \label{fig:vortices_increasing_sigma}
\end{figure}

In the main text, the interaction length $\sigma$ was set to a constant ($\sigma=1$). We find from the gap equation that the interaction range controls the spatial extent of the gap in momentum space. In Fig. \ref{fig:vortices_increasing_sigma}, we show the variation of the gap function with $\sigma$, for a constant $U$, $N$, and $\mathcal{B}$. We find that a decrease of $\sigma$ enhances the nucleation of vortices, as is clear from Fig. \ref{fig:vortices_increasing_sigma}. Although Eq.\eqref{eq:metric_suppressing_gap} is derived in the limit $\sigma^2\gg1$, from the same equation we expect that decreasing $\sigma$ renders the gap more sensitive to the presence of Berry curvature, while simultaneously enhancing its maximum value. Therefore, $U$ and $\sigma$ play opposing roles with respect to both vortex nucleation and the gap magnitude.

%\bibliography{ref}
%\begin{comment}

%\end{comment}

\end{document}